\documentclass[longauth]{aa} 

\usepackage{graphicx}
\usepackage{soul}
\usepackage{indentfirst}
\usepackage{booktabs,multirow}
\usepackage{float}
\usepackage{txfonts}
\usepackage[colorlinks=True,linkcolor=blue,citecolor=blue,filecolor=blue,urlcolor=blue]{hyperref}
\newcommand{\ewlya}{\ifmmode W_{\rm Ly\alpha} \else $W_{\rm Ly\alpha}$\fi}
\newcommand{\ewhi}{\ifmmode W_{\rm \hi} \else $W_{\rm \hi}$\fi}
\newcommand{\rhi}{\ifmmode {\rm R(\hi)} \else ${\rm R(\hi)}$\fi}
\newcommand{\cfhi}{\ifmmode C_f({\rm \hi}) \else $C_f(\mathrm{\hi})$\fi}
\newcommand{\ewlis}{\ifmmode W_{\rm LIS} \else $W_{\rm LIS}$\fi}
\newcommand{\rlis}{\ifmmode {\rm R(LIS)} \else ${\rm R(LIS)}$\fi}
\newcommand{\cflis}{\ifmmode C_f({\rm LIS}) \else $C_f({\rm LIS})$\fi}

\newcommand{\fescabs}{\ifmmode f_{\rm esc}^{\rm abs} \else $f_{\rm esc}^{\rm abs}$\fi}
\newcommand{\feschi}{\ifmmode f\mathrm{_{esc}^{~abs,~HI}} \else $f\mathrm{_{esc}^{abs,~HI}}$\fi}
\newcommand{\fesclis}{\ifmmode f\mathrm{_{esc}^{~abs,~LIS}} \else $f\mathrm{_{esc}^{abs,~LIS}}$\fi}

\newcommand{\chion}{\ifmmode \xi_{\rm ion} \else $\xi_{\rm ion}$\fi}

\newcommand{\mfuv}{\ifmmode M_{\rm 1500}^{\rm int} \else $M_{\rm 1500}^{\rm int}$\fi}
\newcommand{\mobs}{\ifmmode M_{\rm 1500}^{\rm obs} \else $M_{\rm 1500}^{\rm obs}$\fi}
\newcommand{\auv}{\ifmmode A_{\rm UV} \else $A_{\rm UV}$\fi}
\newcommand{\ebv}{\ifmmode E_{\rm B-V} \else $E_{\rm B-V}$\fi}
\newcommand{\av}{\ifmmode A_{\rm V} \else $A_{\rm V}$\fi}
\newcommand{\oh}{\ifmmode 12 + \log({\rm O/H}) \else$12 + \log({\rm O/H})$\fi}
\newcommand{\Zsun}{\ifmmode {\rm Z_{\odot}} \else ${\rm Z_{\odot}}$\fi}

\newcommand{\hi}{\ion{H}{i}}
\newcommand{\si}{\ion{Si}{ii}}

\newcommand{\lya}{Ly$\alpha$}
\newcommand{\lyb}{\ifmmode \ion{H}{i}\lambda1026 \else \ion{H}{i}$\lambda$1026\fi}
\newcommand{\lyg}{\ifmmode \ion{H}{i}\lambda973 \else \ion{H}{i}$\lambda$973\fi}
\newcommand{\lyd}{\ifmmode \ion{H}{i}\lambda950 \else \ion{H}{i}$\lambda$950\fi}
\newcommand{\lye}{\ifmmode \ion{H}{i}\lambda938 \else \ion{H}{i}$\lambda$938\fi}
\newcommand{\lyz}{\ifmmode \ion{H}{i}\lambda930 \else \ion{H}{i}$\lambda$930\fi}

\newcommand{\siia}{\ifmmode \ion{Si}{ii}\lambda\lambda1190,1193 \else \ion{Si}{ii}$\lambda\lambda$1190,1193\fi}
\newcommand{\siib}{\ifmmode \ion{Si}{ii}\lambda1260 \else \ion{Si}{ii}$\lambda$1260\fi}
\newcommand{\siic}{\ifmmode \ion{Si}{ii}\lambda1020 \else \ion{Si}{ii}$\lambda$1020\fi}
\newcommand{\siid}{\ifmmode \ion{Si}{ii}\lambda989 \else \ion{Si}{ii}$\lambda$989\fi}
\newcommand{\oi}{\ifmmode \ion{O}{i}\lambda1302 \else \ion{O}{i}$\lambda$1302\fi}
\newcommand{\cii}{\ifmmode \ion{C}{ii}\lambda1334 \else \ion{C}{ii}$\lambda$1334\fi}
\newcommand{\mgii}{\ifmmode \ion{Mg}{ii}\lambda\lambda2796,2803 \else \ion{Mg}{ii}$\lambda\lambda$2796,2803\fi}
\newcommand{\oiii}{\ifmmode [\ion{O}{iii}]\lambda\lambda4959,5007 \else [\ion{O}{iii}]$\lambda\lambda$4959,5007\fi}
\newcommand{\oii}{\ifmmode [\ion{O}{ii}]\lambda\lambda3726,3729 \else [\ion{O}{ii}]$\lambda\lambda$3726,3729\fi}
\newcommand{\sii}{\ifmmode [\ion{S}{ii}]\lambda\lambda6717,6731 \else [\ion{S}{ii}]$\lambda\lambda$6717,6731\fi}

\usepackage{color}
\usepackage{amsbsy}

\begin{document} 

\title{The Low-Redshift Lyman Continuum Survey}

\subtitle{Unveiling the ISM properties of low-$z$ Lyman-continuum emitters}

\author{
    Alberto Saldana-Lopez\inst{1}\fnmsep\thanks{\email{alberto.saldanalopez@unige.ch}}
    \and
    Daniel Schaerer\inst{1,2}
    \and
    John Chisholm\inst{3}
    \and 
    Sophia R. Flury\inst{4}
    \and 
    Anne E. Jaskot\inst{5}
    \and \\
    G\'abor Worseck\inst{6}
    \and
    Kirill Makan\inst{6}
    \and
    Simon Gazagnes\inst{7}
    \and 
    Valentin Mauerhofer\inst{1}
    \and
    Anne Verhamme\inst{1}
    \and \\
    Ricardo O. Amor\'in\inst{8}
    \and
    Harry C. Ferguson\inst{9}
    \and
    Mauro Giavalisco\inst{4}
    \and
    Andrea Grazian\inst{10}
    \and
    Matthew J. Hayes\inst{11}
    \and
    Timothy M. Heckman\inst{12}
    \and
    Alaina Henry\inst{9,12}
    \and
    Zhiyuan Ji\inst{4}
    \and
    Rui Marques-Chaves\inst{1}
    \and 
    Stephan R. McCandliss\inst{12}
    \and
    M. S. Oey\inst{13}
    \and
    Göran Östlin\inst{11}
    \and
    Laura Pentericci\inst{14}
    \and 
    Trinh X. Thuan\inst{15}
    \and 
    Maxime Trebitsch\inst{7}
    \and 
    Eros Vanzella\inst{16}
    \and
    Xinfeng Xu\inst{12}
}

\institute{
    Department of Astronomy, University of Geneva, 51 Chemin Pegasi, 1290 Versoix, Switzerland
    \and
    CNRS, IRAP, 14 Avenue E. Belin, 31400 Toulouse, France
    \and
    Department of Astronomy, The University of Texas at Austin, 2515 Speedway, Stop C1400, Austin, TX 78712-1205, USA
    \and 
    Department of Astronomy, University of Massachusetts, Amherst, MA 01003, USA
    \and
    Astronomy Department, Williams College, Williamstown, MA 01267, USA
    \and 
    Institut für Physik und Astronomie, Universität Potsdam, Karl-Liebknecht-Str. 24/25, D-14476 Potsdam, Germany
    \and
    Kapteyn Astronomical Institute, University of Groningen, PO Box 800, 9700 AV Groningen, The Netherlands
    \and
    Instituto de Investigaci\'on Multidisciplinar en Ciencia y Tecnolog\'ia, Departamento de F\'isica y Astronom\'ia, Universidad de La Serena, Avda. Juan Cisternas 1200, La Serena, Chile
    \and 
    Space Telescope Science Institute; 3700 San Martin Drive, Baltimore, MD, 21218, USA
    \and
    INAF-Osservatorio Astronomico di Padova, Vicolo dell’Osservatorio, 5, 35122 Padova, Italy
    \and
    Department of Physics and Astronomy, Johns Hopkins University, 3400 North Charles Street, Baltimore, MD 21218, USA
    \and 
    Department of Astronomy, Oskar Klein Centre; Stockholm University; SE-106 91 Stockholm, Sweden
    \and
    University of Michigan, Department of Astronomy, 323 West Hall, 1085 S. University Ave, Ann Arbor, MI 48109
    \and
    INAF-Osservatorio Astronomico di Roma, via Frascati 33, 00078, Monteporzio Catone, Italy
    \and
    Astronomy Department, University of Virginia, P.O. Box 400325, Charlottesville, VA 22904-4325
    \and
    INAF-Osservatorio di Astrofisica e Scienza dello Spazio di Bologna, via Gobetti 93/3, 40129 Bologna, Italy
}

\authorrunning{A.\ Saldana-Lopez et al.}
\titlerunning{LzLCS -- Unveiling the ISM properties of low-$z$ Lyman continuum emitters}

\date{Received 17 November 2021; accepted 16 December 2021}



\abstract
    {}
    {Combining 66 ultraviolet (UV) spectra and ancillary data from the recent Low-Redshift Lyman Continuum Survey (LzLCS) and 23 LyC observations by earlier studies, we form a statistical sample of star-forming galaxies at $z \sim 0.2-0.4$ with which we study the role of cold interstellar medium (ISM) gas in the leakage of ionizing radiation. We also aim to establish empirical relations between the $\protect\hi$ neutral and low-ionization state (LIS) absorption lines with different galaxy properties.}
    {We first constrain the massive star content (stellar ages and metallicities) and UV attenuation by fitting the stellar continuum with a combination of simple stellar population models. The models, together with accurate LyC flux measurements, allow us to determine the absolute LyC photon escape fraction for each galaxy ($\protect\fescabs$). We then measure the equivalent widths and residual fluxes of multiple $\hi$ and LIS lines, and the geometrical covering fraction of the UV emission, adopting the picket-fence model.}
    {The LyC escape fraction spans a wide range, with a median $\protect\fescabs$ (0.16, 0.84 quantiles) of 0.04 (0.02, 0.20), and 50 out of the 89 galaxies detected in the LyC ($1\sigma$ upper limits of $\protect\fescabs \protect\la 0.01$ for non-detections, typically). The $\protect\hi$ and LIS line equivalent widths scale with the UV luminosity and attenuation, and inversely with the residual flux of these lines. Additionally, \lya\ equivalent widths scale with both the $\protect\hi$ and LIS residual fluxes, but anti-correlate with the corresponding $\protect\hi$ or LIS equivalent widths. The $\protect\hi$ and LIS residual fluxes are correlated, indicating that the neutral gas is spatially traced by the low-ionization transitions. We find that the observed trends of the absorption lines and the UV attenuation are primarily driven by the geometric covering fraction of the gas. The observed nonuniform gas coverage also demonstrates that LyC photons escape through low-column-density channels in the ISM. The equivalent widths and residual fluxes of both the $\protect\hi$ and LIS lines strongly correlate with $\protect\fescabs$: strong LyC leakers (highest $\protect\fescabs$) show weak absorption lines, low UV attenuation, and large \lya\ equivalent widths. We provide several empirical calibrations to estimate $\protect\fescabs$ from UV absorption lines. Finally, we show that simultaneous UV absorption line and dust attenuation measurements can, in general, predict the escape fraction of galaxies. We apply our method to available measurements of UV LIS lines of 15 star-forming galaxies at $z \sim 4-6$ (plus 3 high-$z$ galaxy composites), finding that these high-redshift, UV-bright galaxies ($M_{\rm UV} \protect\la -21$) may have low escape fractions, $\protect\fescabs \protect\la 0.1$.}
    {UV absorption lines trace the cold ISM gas of galaxies, which governs the physics of the LyC escape. We show that, with some assumptions, the absolute LyC escape can be statistically predicted using UV absorption lines, and the method can be applied to study galaxies across a wide redshift range, including in the epoch of cosmic reionization.}
{}


\maketitle

\defcitealias{R16}{R16}

\section{Introduction}\label{sec:intro}
Since the discovery of the Gunn-Peterson effect \citep{GP1965, Becker2001}, understanding how the Universe was reionized around $z \approx 6 - 9$ \citep{Planck2016} requires not only identification of the sources that contributed to that change, but also knowledge of the physical mechanisms under the last major cosmic phase transition. There are two principal instigators of cosmic reionization. The first is massive stars that emit sufficient ionizing photons to keep hydrogen in an ionized state \citep{Robertson2015}, and the second is active galactic nuclei \citep[AGN,][]{Madau2015}. Depending on the principal source for reionization \citep[galaxies or AGN, or both; see][]{Dayal2020}, the thermal history of the late Universe changes accordingly \citep[e.g.,][]{Kulkarni2017}. For example, models primarily driven by star-forming (SF) galaxies generally have delayed reionization compared to AGN-dominated reionization, where the Universe is reionized at a higher redshift \citep[e.g., see][]{Finkelstein2019}. 

To establish the contribution of galaxies and AGN in terms of ionization rate, one needs to quantify three main properties: (1) the ionizing photon production efficiency (\chion), (2) the ionizing photon escape fraction (\fescabs), and the number density of such galaxies or AGN at high $z$. The ionizing production efficiency (\chion) is the intrinsic number of ionizing photons emitted by the stars per unit UV luminosity, while \fescabs\ is the fraction of these photons escaping from them to the intergalactic medium (IGM). 

Although always powerful in terms of ionizing emissivity \citep[$\fescabs \sim 1,$][]{Stevans2014, Lusso2015}, AGNs appear to be very rare at high redshifts \citep{Fontanot2014, Matsuoka2018} and with lower accretion rates and emissivities than in the late Universe \citep{Dayal2020}; although there is still the possibility of having an extending low-luminosity AGN population that is numerous enough to provide a substantial contribution to the total ionizing budget \citep{Giallongo2015, Grazian2020}. Star-forming galaxies are much more numerous at high-$z$, and the question now becomes whether more massive or compact low-mass SF galaxies contribute the most. On the one hand, the less numerous but massive SF galaxies with enormous gas and dust reservoirs at high-$z$  can intrinsically generate a large amount of Lyman continuum (LyC) photons \citep{Naidu2020}, but these photons can be absorbed and attenuated within the gaseous reservoirs of their host galaxies at the same time. On the other hand, compact low-mass SF galaxies are more numerous at high $z$ \citep{Rosdahl2018, Finkelstein2019}. Additionally, they take advantage of their weak gravitational potential and feedback effects to clean the interstellar medium (ISM) from gas and dust, increasing the fraction of LyC photons escaping the galaxy \citep{Trebitsch2017}. The true contribution to reionization from different galaxy types is still under debate, but models require a threshold of $\fescabs=5-20\%$ for $M_{\rm{UV}} \lesssim -15$ galaxies to drive reionization around $z \sim 7$ \citep{Robertson2013,Robertson2015}.

Over the last decade, several campaigns have searched for LyC emitters at low and intermediate redshifts. Using the {\it Far Ultraviolet Spectroscopic Explorer} (FUSE), \citet{Bergvall2006} and \citet{Leitet2013} reported the first LyC detections, although later re-observations revealed that some of them were affected by foreground scattered light. Using the {\it Hubble Space Telescope} (HST), about $30$ new \emph{leakers} were later discovered at $z \sim 0.3$ \citep{Leitherer2016, Borthakur2016, Izotov16a, Izotov16b, Puschnig2017, Izotov18a, Izotov18b, Wang2019, Malkan2021, Izotov2021}, with \fescabs\ values spanning from a few to $70\%$. The majority consist of compact SF galaxies with relatively high star formation rates (SFR), strong optical emission lines, and ionizing photon production efficiencies comparable to their high-$z$ analogs \citep{Schaerer2016}. Nevertheless, it remains challenging to robustly detect faint LyC flux at the sensitivity limit of  
HST (see e.g., the re-analysis of \citet{Chisholm2017} versus \citet{Leitherer2016} and \citet{Puschnig2017}).

At intermediate redshifts, investigating the LyC leakage is more difficult, as LyC photons are absorbed by Lyman limit systems in the IGM, which makes the detection extremely challenging beyond $z\geq4$. The presence of low-redshift interlopers also renders potential LyC detections at high $z$ difficult \citep{Vanzella2010, Inoue2014, Naidu2018}. \citet{RT2019} recently reported LyC emission from the Sunburst Arc, a lensed compact dwarf galaxy at redshift $z \sim 2.2$, with an escape fraction of $\fescabs = 32\%$ averaged over 11 different gravitationally lensed HST images. Similar works are \citet{Bian2017} and the Cosmic Horseshoe in \citet{Vasei2016}. The \emph{Keck Lyman Continuum Survey} \citep[KLCS,][see also \citet{Shapley2016}]{Steidel2018} spectroscopically confirmed 13 individual LyC detections around $z \sim 3.1$ in 9 independent fields \citep[see the follow-up HST observations in][]{Mostardi2015, Pahl2021}. Using HST/F336W and narrow-band photometry, another 12 Lyman break galaxies (LBGs) and AGNs ---enclosed in the SSA22 proto-cluster field
\citep[cf.][]{Shapley2006, Micheva2017a, Micheva2017b}--- were discovered to be leaking in the \emph{LymAn Continuum Escape Survey} (LACES) by \citet{Fletcher2019}. In \citet{Vanzella2015, Vanzella2016} and \citet{deBarros2016}, the authors used the CANDELS $U$-band and HST/F336W imaging to select LyC leaker candidates in the GOODS-S field, resulting in the robust detection of the \emph{Ion2} galaxy at $z=3.2$. With a similar methodology, \emph{Ion1} was discovered by \citet{Vanzella2012} and \emph{Ion3} found by \citet{Vanzella2018}, which currently holds the record as the most distant SF galaxy observed to leak LyC photons ($z \sim 4$). The recent re-analysis of \emph{Ion1} by \citet{Ji2020} suggests a 5\%-10\% escape, although this was originally measured at 60\% \citep{Vanzella2012}. At $z \sim 3$, three individual detections were performed by \citet{Grazian2016, Grazian2017} in the CANDELS/ GOODS-N, EGS, and COSMOS fields thanks to deep ground-based observations with the \emph{Large Binocular Telescope} (LBT). Finally, the recently launched AstroSAT telescope discovered a remarkable $z = 1.42$ SF galaxy that emits ionizing photons at a rest-frame of 650\AA\ \citep{Saha2020}. Some other tentative detections have been conducted by \citet{Siana2015} and \citet{Smith2020}. 

So far, low-$z$ galaxy studies have shown that individual SF galaxies can have \fescabs\ larger than the 10\%-20\% threshold for galaxies to reionize the Universe. However, at high $z$, the actual number of individual LyC detections is not only low but biased to the sightlines with the highest IGM transmissivity. As the number density and the properties of such high-$z$ leakers are not yet known (and we cannot extrapolate them from the few clear detections), stacking analyses have become the most popular method in reionization studies, and are used to obtain the average \fescabs\ of the SF galaxy population. Deep broad-band \citep{Grazian2016, Grazian2017, Rutkowski2017, Japelj2017, Naidu2018}, narrow-band imaging \citep{Guaita2016}, and spectral stacking \citep{Marchi2018Lyalpha-Lyman-c, Mestric2021} have yielded relatively low upper limits for the escape fraction of the SF population at $z=2-4$ \citep[$<$5\%-10\% for $M_{\rm{UV}} \lesssim -19$ galaxies, cf.,][]{Naidu2018,Alavi2020}. 

In order to detect and quantify LyC emission at high $z$ ($z \ga 4$), including in the Epoch of Reionization, the development of indirect techniques that estimate the LyC escape are required. Low-$z$ galaxies are the most suitable laboratories for testing indirect LyC indicators, because the LyC, the far-UV (FUV) stellar, and the interstellar absorption spectra, as well as the rest-frame optical nebular lines, can be observed at the same time and with better S/N than at high-$z$, allowing spatially resolved studies. 

Starting from the closest to the Lyman limit (912\AA), the depth of the Lyman series lines directly links the neutral gas column density to the geometrical distribution of the same gas along the line of sight. The residual flux of the high-order Lyman absorption lines has been related with the neutral gas porosity \citep[see][]{Reddy2016b, Reddy2021}. Properly accounting for dust attenuation, these lines have proved to be an effective LyC tracer \citep{Gazagnes2018, Chisholm2018, Steidel2018,Gazagnes2020}, although simulations show different results \citep{Mauerhofer2020}. Very prominent when present, the \lya$\lambda$1216 line ---accessible from the ground at $z \gtrsim 2$--- can also be used because \lya\ and LyC photons interact with the same neutral gas. \citet{Jaskot2014} and \citet{Verhamme2015} proposed that the \lya\ peak separation can trace the presence of low-density gas: the physical mechanism driving the leakage is expected to be connected for both \lya\ and LyC \citep{RiberaThorsen2015, Verhamme2015, Dijkstra2016}. Indeed, the separation and shape of double- and triple-peaked \lya\ profiles have been associated with LyC leakage in observations \citep{Ribera-Thorsen2017, Verhamme2017, Vanzella2020, Izotov2020, Izotov2021}. However, the Lyman series are still efficiently suppressed by IGM absorption at high-$z$ and are therefore almost impossible to observe during the Epoch of Reionization. The use of indirect LyC indicators redwards \lya\ is essential.

The most promising indirect tracers of LyC escape to date are the UV low-ionization state (LIS) absorption lines \citep{Erb2015, Chisholm2018} and the \mgii\ emission doublet \citep{Henry2018, Chisholm2020}. On the one hand, measurements of some UV metal absorption line depths such as \siib ~\citep{Gazagnes2018} and \cii\ \citep{Grimes2009} have been shown to uniquely trace the neutral gas covering fraction \citep{Reddy2016b, Reddy2021}. The observed depth of a given absorption transition depends on the optical depth and the covering fraction. If the optical depth is high enough that the lines are saturated, then the covering fraction can be inferred from the observed spectral line depths. On the other hand, the intrinsically bright \ion{Mg}{II} nebular transitions have an ionization potential similar to that of warm \hi. This makes the \ion{Mg}{II} and LyC transitions optically thick at similar \hi\ column densities (for typical Mg/H abundances), making \mgii\  a good tracer of LyC escape as well.

In the rest-frame optical range, some previous studies suggest that both low-$z$ \citep{JaskotOey2013, Izotov18b} and high-$z$ leakers \citep{Nakajima2016} have high \oiii\ versus \oii\ ratios \citep[O$_{32}$, see also][]{Faisst2016}. Subsequent efforts showed that this quantity is not a good predictor of the LyC escape \citep{Basset2019, Nakajima2020}, mainly because O$_{32}$ alone cannot probe the outer extent of the ionization regions, the region that is most sensitive to the optical depth and thus chiefly affects the escape of LyC radiation. Different kinematic effects can also impact the LyC leakage, complicating the use of emission lines in the diagnostics \citep{McKinney2019, Hogarth2020}. Worth mentioning is also the \sii\ doublet \citep{Wang2019, Ramambason2020}, which shows a deficit in leakers and was recently investigated more thoroughly by \citet{Wang2021}.

Finally, other indicators have been investigated in the literature, such as the SFR surface density \citep{Heckman2001, ClarkeOey2002, Izotov18b, Naidu2020, Cen2020, Izotov2021}, and UV/optical lines such as \ion{He}{i}$\lambda3889$ \citep{Izotov2017, Guseva2020} or \ion{O}{i}$\lambda6300$ \citep{Ramambason2020}, but these require further calibrations to be used as robust predictors of LyC escape. A detailed analysis of the various indirect LyC diagnostics is presented in \citet{FluryII}.

All these indicators are sensitive to the ISM properties, and we can therefore expect correlations between different ISM-related quantities and LyC escape. Revealing the ISM properties of leakers versus nonleakers, one could illustrate the limitations and strengths of the previous tracers. In summary, two scenarios describing the ISM structure compete in the escape of LyC radiation \citep{Zackarisson2013}. The first, called density-bounded regions \citep{JaskotOey2013, Nakajima2014}, envision a massive stellar population surrounded by a homogeneous and low-density neutral gas media, so that photons can leak from the galaxy at a rate that is inversely proportional to the hydrogen column density. Alternatively, in the ``picket-fence'' model \citep{Heckman2001, Heckman2011, Vasei2016, Reddy2016b, Gazagnes2018, Steidel2018, Reddy2021}, stars are surrounded by an optically thick ISM and the ionizing radiation can only escape through low-column-density channels: the ISM is not homogeneous but patchy. So far, local observations of leakers \citep{Gazagnes2018, Gazagnes2020} and also simulations \citep{Trebitsch2017, Mauerhofer2020} favor a patchy ISM, and only the galaxies with the highest \fescabs\ have a mostly ionized ISM consistent with a density-bounded scenario \citep{Jaskot2019, Ramambason2020,Kakiichi2021}. On top of the geometric distribution, other aspects like photo-ionized outflows and feedback effects \citep{Chisholm2017, Hogarth2020, Carr2021} also change the ISM structure, affecting the profile of emission/absorption lines and the escape of LyC radiation. 

In this paper, we statistically explore the impact of the neutral and low-ionized ISM absorption on LyC escape, taking a novel approach and using a large set of low-$z$ LyC galaxy observations. We use spectra from the HST \emph{Low-$z$ Lyman Continuum Survey} (LzLCS, see Sect.\ \ref{sec:data}), a set of 66 SF galaxies of which 35 have a secure LyC detection. The methodology for modelling the stellar continuum of the galaxies in the sample, the adopted absolute LyC photon escape fraction, and the absorption line measurements are presented in Sect.\ \ref{sec:methods}. The absorption line results are shown in the same section, and the possible inclusion of systematic errors is studied through absorption line simulations. In Sect.\ \ref{sec:fit_results}, we discuss the fitting results, secondary products, and limitations of the modelling. The main empirical relations between \hi\ and LIS lines and different galaxy properties are then given in Sects.\ \ref{sec:results_abslines} and \ref{sec:results_lya_cf_ebv_fesc}. A physical interpretation of the results and literature comparisons are presented in Sect.\ \ref{sec:discussion}. We finally conclude in Sect.\ \ref{sec:conclusions}, also highlighting the relevance of these results for future high-$z$ searches.

Throughout this paper, a standard $\Lambda$CDM cosmology is used, with a matter density parameter $\Omega_{\rm M}$ = 0.3, a vacuum energy density parameter $\Omega_{\Lambda}$ = 0.7, and Hubble constant of $H_0$ = 70~km s$^{-1}$ Mpc$^{-1}$. All magnitudes are in the AB system, and we adopt a solar metallicity value of $12 + \log({\rm O/H})_{\odot}=8.69$ ($\Zsun = 1$).

\section{Data}\label{sec:data}
In this study, we use UV spectra in the observed frame wavelength range 800-1950\AA\ from the {\em Low-Redshift Lyman Continuum Survey} \citep[LzLCS,][]{FluryI}, the most comprehensive spectroscopic campaign so far to trace the LyC emission of galaxies in the nearby Universe ($z \sim 0.3$). In addition, we include and reanalyze previous LyC emitters from the studies of \citet{Izotov16a, Izotov16b, Izotov18a, Izotov18b, Izotov2021} and \citet{Wang2019} as a comparison sample. For all the sources, we also use ancillary data derived in a homogeneous fashion by the LzLCS team (see \ref{sub:data_othervalues}).

\subsection{The Low-$z$ Lyman Continuum Survey}\label{sub:data_lzlcs}
The {\em Low-z Lyman Continuum Survey} \citep[LzLCS, see][]{FluryI} is composed of 66 SF galaxies at $z \sim 0.3$ whose UV spectra were observed by the \emph{Cosmic Origin Spectrograph} (COS) on board the HST. At $z \sim 0.3$, the COS/G140L grating not only probes the LyC but also traces an entire window of wavelengths redder than \lya\ up to 1500\AA\ rest-frame.

The parent sample of LyC-emitter candidates was drawn from Sloan Digital Sky Survey sources \citep[SDSS Data Release 15,][]{SDSS15} with \emph{GALEX} counterparts in both the FUV (0.1530$~\mu$m) and near-UV (NUV; 0.2310$~\mu$m) photometric bands \citep{Morrissey2007}. Applying the methods described in \citet{Jaskot2019}, the equivalent widths and fluxes of different optical emission lines were measured in the SDSS spectra, and the SF galaxies in the redshift range $0.22 \leq z < 0.45$ were selected by applying usual BPT \citep{BPT1981} diagram diagnostics (HST/COS traces the LyC region at $z > 0.22$). Star-formation rate surface densities ($\Sigma_{SFR}$) were measured from the H$\alpha$ and H$\beta$ fluxes in the SDSS spectra and the SDSS $u$-band half-light radius, and UV $\beta-$indices for this subsample were computed using \emph{GALEX} photometry. These measurements were later improved with the COS data in hand.

The final sample was built by selecting the galaxies that fulfill one or more of the following criteria \citep[see][]{Izotov16a}: high [\ion{O}{III}]5007/[\ion{O}{II}]3727 flux ratios (O$_{32}>3$), high SFR densities ($\Sigma_{SFR}>0.1 $~M$_{\odot}$yr$^{-1}$kpc$^{-2}$), and blue UV continuum slopes ($\beta<-2$), resulting in 66 galaxies. Spectra of each galaxy were taken under the GO 15626 HST large program (P.I. Jaskot), with around two orbits per object on average. COS acquired each object via NUV imaging and centered its $2.5$" diameter spectroscopic aperture on the peak of the NUV emission. The G140L grating was then used to cover the 800-1950\AA\ observed wavelength range, with a nominal full width at half maximum (FWHM) resolution of $R \equiv \lambda / \Delta \lambda \sim$1000 at 1100\AA, increasing up to $R \sim$1500 at 1900\AA.

Science spectra were reduced following the methods described in \citet{FluryI} using the standard \textsc{calcos} pipeline (v3.3.9). The tackling of the systematic errors and the LyC signal-to-noise ratios (S/Ns) was improved thanks to the use of the \textsc{FaintCOS} custom software \citep{Makan2021}, which properly estimates the dark current and scattered geo-coronal \lya, co-adding spectra from multiple visits to improve the S/N. Finally, each spectrum was binned to 0.5621\AA\ to yield a sampling of $\gtrsim 2$ pixels per resolution element. The spectra were corrected for Milky Way extinction using the Galactic \ebv\ estimates from the \citet{Green2018} dust maps and the \citet{F99} attenuation law. 

The LyC flux ($f^{LyC}$) was measured as the mean over a 20\AA\ window as close as possible to but shortward of 900\AA\ in the rest-frame (typically 850-900\AA), while avoiding wavelengths redward of the observed frame of 1180\AA\ to reduce the impact of geo-coronal \lya\ emission. Over the 66 galaxies in the LzLCS, 35 were significantly detected (LyC flux S/N$\geq$2) and 31 were not detected at LyC wavelengths. In the detected galaxies, LyC luminosities span from $10^{38}$ to $10^{40}$ erg/s. In the undetected sample, the average upper uncertainty on the LyC flux density was reported as the 1$\sigma$ upper limit. See \citet{FluryI} for details.

\subsection{Low-z LyC emitters in the literature}\label{sub:data_literature}
In addition to the new LzLCS sources, we add 20 SF galaxies with LyC observations from the sample of \citet{Izotov16a}, \citet{Izotov16b}, \citet{Izotov18a}, \citet{Izotov18b}, and \citet{Izotov2021}. Some of these sources are among the strongest leakers in the literature at $z \sim 0.3$ (i.e., with the highest ionizing escape fractions). 

Originally, the \citet{Izotov16a, Izotov16b, Izotov18a, Izotov18b} galaxies were selected from SDSS in a similar way to some of the LzLCS in terms of compactness, FUV brightness, and ionizing features in the optical spectra (EW(H$\beta)>100\AA$ and O$_{32} \geq 5$). The selected galaxies exhibit narrow ($\lesssim 130$~km/s) double-peaked \lya\ emission and high (up to 60\%) \lya\ escape fractions in most cases, low metallicities (1/10-1/3~Z$_{\odot}$), low stellar masses (0.15-6$\times$10$^9$~M$_{\odot}$), and high SFRs (14-80~M$_{\odot}$~yr$^{-1}$) and SFR densities  (2-35~M$_{\odot}$yr$^{-1}$kpc$^{-2}$). The resulting LyC escape in those galaxies extends from a few up to $\sim 70\%$. Here, we used these Izotov et al. SF galaxies as a reference sample for the strongest leakers in the new LzLCS. 

Similarly, the \citet{Izotov2021} sample consists of 9 dust-poor and low-mass SF galaxies ($\lesssim 10^8 M_{\odot}$), with very high specific SFRs ($\sim$ 150-630~Gyr$^{-1}$), showing a wide range in \lya\ peak velocity separations ($\sim 200-500$km/s). Their published LyC escape fractions reach $\sim 35 \%$.

In order to increase the dynamic range for of some the galaxy properties in our working sample, we finally also included three more SF galaxies observed by \cite{Wang2019}, which were selected in terms of their [\ion{S}{ii}] deficiency (SDSS spectra), compactness ($\le 0''.5$ in the SDSS $u$-band), and relatively high GALEX FUV fluxes. These galaxies are more massive ($\sim 3 \times$10$^{10}$~M$_{\odot}$), more metal-rich (0.9-1~Z$_{\odot}$), dustier, and less extreme in terms of ionization properties than the Izotov et al. sample, but also have exceptionally high SFR densities (35-700~M$_{\odot}$yr$^{-1}$kpc$^{-2}$). Their estimated escape fractions are significantly lower (up to 3\%). 

The LzLCS team re-processed and re-reduced these 23 COS/G140L spectra with the same methodology as that described above in order to maintain the homogeneity and consistency of the results \citep[see][]{FluryI}. This results in 15 additional detections and 8 nondetections out of a total of 23 additional galaxies. The same total sample is described and analyzed in depth in \citet{FluryI} and \citet{FluryII}.

\subsection{Other LzLCS science products}\label{sub:data_othervalues}
The LzLCS team measured galaxy properties of every source in the LzLCS and in the selected literature samples, either from the SDSS spectra and the GALEX photometry, or directly from the COS spectra. In particular, for this work we use:
\begin{enumerate}
    \item[$-$] Optical dust attenuation parameters: $E{\rm _{B-V}-Balmer}$.
    \item[$-$] Gas-phase metallicities: \oh.
    \item[$-$] \lya\ equivalent widths: \ewlya.
\end{enumerate}
Optical attenuations were computed using an iterative {\em Balmer decrement} methodology from the measured SDSS H${\beta}$ to H${\epsilon}$ fluxes \citep[following][see also \citet{Flury2020}]{Izotov1994}. The gas-phase metallicities (relative oxygen abundances) were obtained using a direct-method approach using the extinction-corrected hydrogen and oxygen optical lines. Finally, \lya\ equivalent widths were measured by integrating the COS spectral flux close to the \lya\ line (continuum linearly fitted within a 100\AA\ window). Errors on every quantity come from $10^4$ Monte-Carlo realizations of the corresponding spectra. See \citet{FluryI} for more details. The LzLCS team is also measuring UV morphological parameters from the COS acquisition images (Ji et al. 2021, in prep.).

\begin{figure*}
    \centering
    \includegraphics[width=0.98\textwidth]{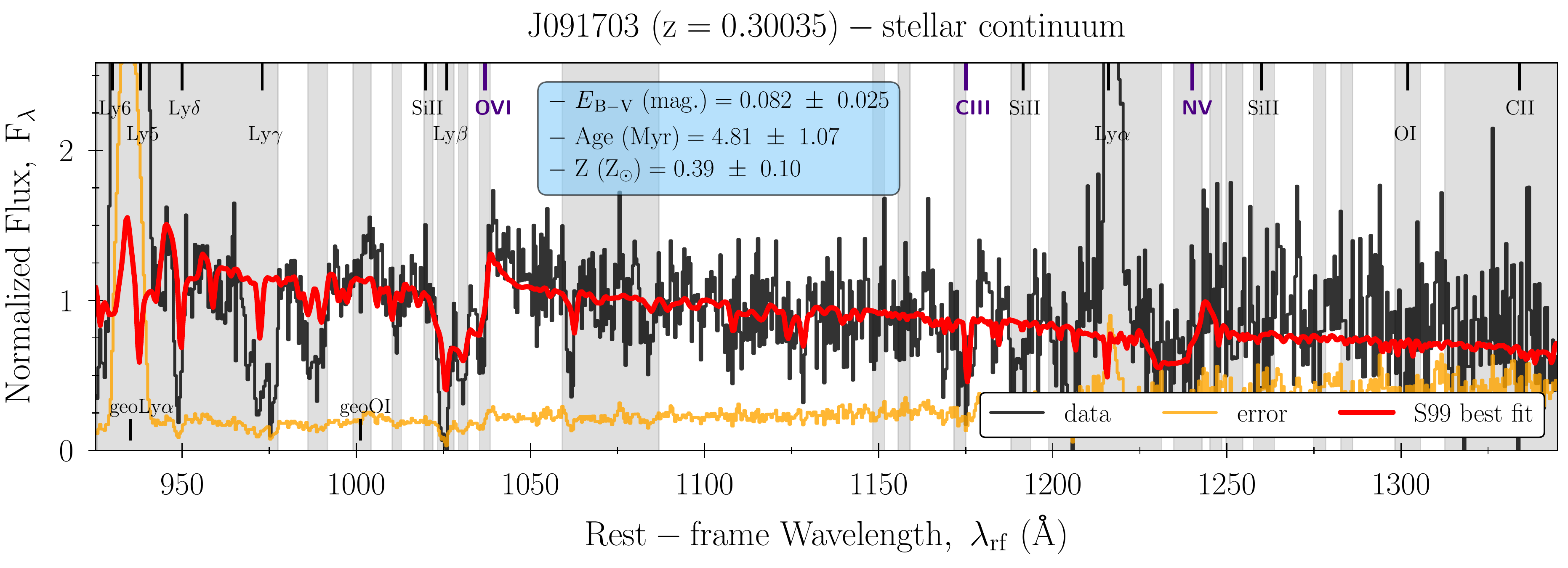}
\caption{Stellar continuum modelling for the galaxy J091703. Stellar continuum fit (red), observed spectrum (black), and error spectrum (yellow); spectral regions masked during the fit are shown in gray. The spectra, shown in $F_\lambda$ units, have been normalized over 1070-1100\AA. Different ISM absorption lines and stellar features are indicated with black and dark-blue vertical lines in the top part of the figure, respectively. Geo-coronal emissions are labeled at the bottom. The best-fit dust attenuation color-excess (\ebv, in mag., \citetalias{R16} law), light-weighted stellar age (Myr) and metallicity ($\Zsun$) inferred for this source are given in the inset. Uncertainties come from a consistent Monte-Carlo error propagation along the fitting routine.}
\label{fig:fits_hRsb99}
\end{figure*}

\section{Methods}\label{sec:methods}
Measuring the properties of the ISM features in the UV requires accurate determination of the stellar continuum. In this section, we describe our methodology for modelling the UV stellar continuum of extragalactic spectra \citep[\ref{sub:method_fitting}, following][]{Chisholm2019} and absolute LyC escape fractions in detail. The expressions used to determine the equivalent widths, residual fluxes, and corresponding escape fractions from the lines are also listed (\ref{sub:method_abs}).

\subsection{Stellar continuum modelling}\label{sub:method_fitting}
The stellar continuum modelling is achieved by fitting every COS spectrum with a linear combination of multiple bursts of single-age and single-metallicity stellar population models. We use the fully theoretical \textsc{starburst99} single-star models without stellar rotation \citep[S99,][]{Leitherer2011b, Leitherer2014} using the Geneva evolution models \citep{Meynet1994}, and computed with the \textsc{WM-Basic} method \citep{Pauldrach2001, Leitherer2010}. The S99 models use a Kroupa initial mass function \citep{Kroupa2001} with a high-(low-)mass exponent of 2.3 (1.3), and a high-mass cutoff at 100~M$_{\odot}$. The spectral resolution of the models is $R \equiv \lambda / \Delta \lambda \sim$ 2500, and remains constant at $\le 1500\AA$ rest-frame. 

A nebular continuum contribution was added to every single population by processing the stellar continuum models through the \textsc{cloudy v17.0} code \citep{Ferland2017}, assuming similar gas-phase and stellar metallicities, an ionization parameter of $\log(U)=-2.5$, and a volume hydrogen density of n$_{H} = 100~cm^{-3}$. The nebular continuum is negligible for the wavelengths that are typically fitted in this paper, but it can impact longer wavelength data ($\ga 1200\AA$ rest-frame).

We chose four different metallicities (0.05, 0.2, 0.4 and 1 $Z_{\odot}$) and seven ages for each metallicity (1, 2, 3, 4, 5, 8, and 10 Myr) as a representative set of 28 models for our low-$z$ UV spectra. In detail, ages are chosen to densely sample the stellar ages where the stellar continuum features appreciable change \citep{Chisholm2019}. 

The observed spectra are first placed into the rest-frame by dividing by $(1+z)$. Both the observed spectra and the models are then normalized by the median flux in the 1070--1100\AA\ rest-frame wavelength interval (free of stellar and ISM features), and all the fits are performed in the same wavelength range: 975--1345\AA\ (rest-frame). Finally, the models are convolved by a Gaussian kernel\footnote{The high nonGaussianity of the COS line spread function (LSF) does not have a strong impact on the current analysis.} to the COS/G140L nominal spectral resolution ($R \sim 1000$).

We first mask out the nonstellar features, the spectral regions that are affected by host-galaxy and Milky Way ISM absorptions, and geo-coronal emissions. We also mask bad pixels with zero flux, as well as pixels with S/N<1. We then fit the UV stellar continuum ($F^{\star}(\lambda)$) with a linear combination of multiple S99 models ($F^{S99}_{i, j}$) plus the nebular continuum. Adopting a uniform foreground dust-attenuation model for the galaxies (see Sect.\ \ref{sub:method_abs}), this yields: 
\begin{equation}
    \begin{aligned}
    F^{\star}(\lambda) &= 10^{-0.4 k_{\lambda} \ebv} \sum_{i,j} X_i F^{S99}_{i, j}, \\
    &i \equiv 1, 2, 3, 4, 5, 8, 10 ~Myr \\
    &j \equiv 0.05, 0.2, 0.4, 1 ~Z_{\odot}\\
    \end{aligned}
    \label{eq:s99_fit}
\end{equation}

\noindent where $10^{-0.4 A_{\lambda}}$ is the UV attenuation term, $A_{\lambda} = k_{\lambda} \ebv$, and $k_{\lambda}$ is given by the \citet{R16} attenuation law (\citetalias{R16} hereafter), {$k_{\lambda}=2.191+0.974/\lambda(\mu m)$}. We also tested the influence of different attenuation laws on the fitting results, in particular using the Small Magellanic Cloud (SMC) law instead \citep{SMC84, SMC85}. We discuss these results in the next Sect.\ \ref{sec:fit_results}. The light-fraction coefficients ($X_i$) determine the weight of the single-stellar population within the fit, and the best fit is chosen through a nonlinear $\chi^2$ minimization algorithm with respect to the observed data \citep[\textsc{lmfit}\footnote{A \textsc{python} version of the \textsc{lmfit} package can be found in \url{https://lmfit.github.io/lmfit-py/}.} package, ][]{lmfit}. Errors are derived in a Monte-Carlo way, varying observed pixel fluxes by a Gaussian distribution whose mean is zero and standard deviation is the 1$\sigma-$error of the flux in the same pixel, and re-fitting the continuum over 500 iterations per spectrum (enough realizations to sample the posterior continuum so that it approaches ``Gaussianity'' on each pixel). The adopted 1$\sigma-$errors on the spectra are approximated as the squared-root average of the original asymmetric spectral uncertainties \citep{FluryI}.

\begin{figure*}
    \centering
    \includegraphics[width=0.98\textwidth, page=4]{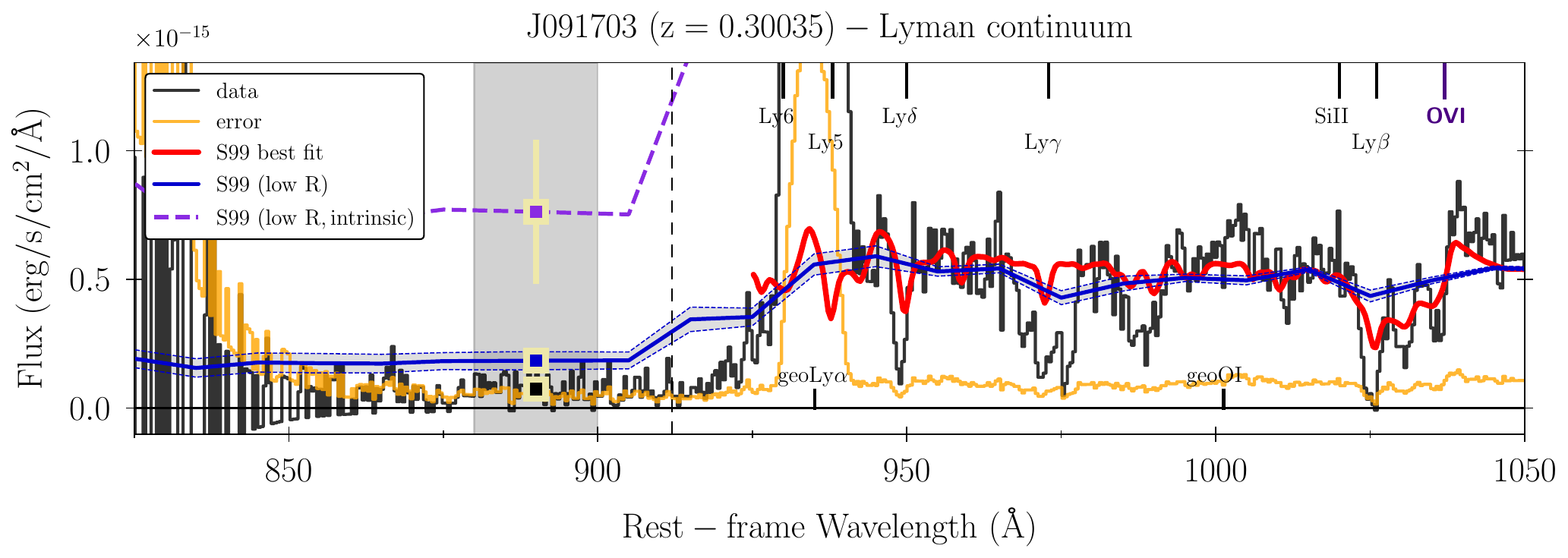}
\caption{Stellar continuum modelling in the LyC region for the J091703 galaxy. The high-resolution stellar continuum fit (down to 925\AA) and the observed J091703 flux and error are plotted in red, black, and yellow lines, respectively. The stellar continuum is now reproduced with models of lower resolution (blue line and shaded area), allowing us to go below the Lyman limit ($<912\AA$, dashed vertical line). The intrinsic (unreddened) flux is shown by the purple dashed line. In the LyC window (gray-shaded), the measured, modeled, and modeled-intrinsic mean LyC fluxes are indicated with filled squares and corresponding error bars. The ISM, stellar lines, and geo-coronal lines are displayed with the same color coding as in \autoref{fig:fits_hRsb99}.}
\label{fig:fits_lRsb99}
\end{figure*}

\autoref{fig:fits_hRsb99} shows one of the outputs of our fitting routine, the observed and fitted spectrum for J091703,
with the stellar continuum in red  overlaying the observed spectrum in black. The reduced-$\chi^2$ of the fit is $\chi^2_{\nu} = 0.93$, with a median of 1.03 for the whole sample. This galaxy is dominated by a young and low-metallicity population whose average light-weighted age is 4.81 Myr and light-weighted metallicity results in 0.39 \Zsun. J091703 is moderately attenuated, with a UV dust-attenuation parameter of $\ebv=0.082$~mag. We highlight with dark-blue labels the two main stellar features that help our algorithm to estimate the age and the metallicity of the stellar populations at those wavelengths: the \ion{O}{vi}$\lambda$1032 and \ion{N}{v}$\lambda$1240 P-Cygni stellar wind profiles (see also the \ion{C}{ii}$\lambda$1175 photospheric line). As, in general, the FUV spectrum is dominated by young massive stars, we do not expect large changes between the \textsc{starburst99} and other models including binaries \citep[e.g., \textsc{BPASS}, ][]{BPASS17}. Detailed results regarding the stellar populations will be discussed in a forthcoming paper (Chisholm et al., in prep.).

All the data sets are fitted using the same method. To conclude, the main derived quantities of interest for the present paper are the UV dust attenuation parameters (\ebv) and the fitted stellar spectral energy distributions (SEDs). The intrinsic SEDs (before UV attenuation) are used below to derive important quantities such as the absolute LyC escape fraction (\fescabs) and intrinsic UV luminosities. The UV-fit \fescabs\ values derived here are compared to other escape fraction estimates in \citet{FluryI} and \citet{FluryII}.

\subsubsection{Fiducial photon escape fractions}
Due to the fact that the high-resolution S99 models are only well defined above the Lyman limit (912\AA), we use the low-resolution S99 models ($R \sim 100$) to extend to the LyC region \citep{Leitherer2011a}. These low-resolution models are linearly combined with the same light-fractions and attenuation parameters as the previous fits. 

The absolute LyC escape fraction at $\sim 900\AA$ (denoted by \fescabs$^{\rm ,~900} \equiv$ \fescabs\ in the following) is then defined as the ratio between the modeled intrinsic LyC flux (i.e., the flux given by the low-resolution models) and the observed (measured, see Sect.\ \ref{sec:data}) LyC flux for every galaxy ($f^{LyC}$):
\begin{equation}
    \fescabs = \dfrac{f^{LyC}}{(\sum_{i,j} X_i F^{S99}_{i, j}) (LyC)}.
    \label{eq:fescS99}
\end{equation}

The modeled LyC flux is computed as the mean flux over the same 20\AA\ window as the observed LyC flux, which is typically at 850-900\AA,\, although varying from galaxy to galaxy \citep{FluryI}. For galaxies with nondetection in the LyC, we adopt the corresponding 1$\sigma$ COS/G140L upper limit. Error bars on \fescabs\ come from the simultaneous sampling of $f^{LyC}$ and the modeled flux. In the case of LyC detections, these error bars are dominated by the uncertainty in the modeled flux. \autoref{fig:fits_lRsb99} shows the low-resolution S99 modeled flux for J091703 (blue shaded region; see the figure caption) and emphasises this approach. The wavelength window over which the LyC flux is measured is shown by the gray region. Dividing the mean observed flux value in this window (black square) by the modeled intrinsic stellar flux (purple square) yields the absolute photon escape of this source ($\fescabs = 0.161^{+0.073}_{-0.055}$).

Throughout this paper, we use the definition of the photon escape given by Eq.\ \ref{eq:fescS99}. However, we note that similar trends and results are obtained if we use other methods to determine \fescabs\ \citep[see][]{FluryII}. The main quantities (\ebv\ attenuation, stellar age, metallicity, and LyC  escape fraction) derived for our sample are listed in Appendix\ \ref{appA}. 

Together with the LyC flux significance (i.e., the S/N in the LyC flux), the inclusion of the escape fraction makes it possible to distinguish between strong, weak, and nonleakers. As in \citet{FluryII}, we define strong leakers as those sources with secure LyC detection (significance>5) and $\fescabs \ge 0.05$. Weak leakers are considered to be galaxies with fair or marginal LyC detection (significance$\ge 2$), but not falling in the strong category. Finally, nonleakers are defined as undetected galaxies in the LyC (significance<2). The total sample of 89 galaxies is then divided into 20, 30, and 39 strong, weak, and nonleakers, respectively (see also Appendix\ \ref{appA}).

\subsection{Absorption lines}\label{sub:method_abs}
\subsubsection{Equivalent widths and residual fluxes}
In the rest-frame spectral range covered by our observations ($\sim$ 800-1500\AA), several lines of interest are present. Here we choose the following: five Lyman series lines of hydrogen, from Ly$\beta$ to Ly6 (\lyb, \lyg, \lyd, \lye, \lyz), and six low-ionization state (LIS) lines (\siid, \siic, \siia, \siib, \oi, \cii). Given the actual COS spectral resolution, some of the mentioned lines are blended with other lines at adjacent rest-frame wavelengths. In particular, the Lyman series lines are blended with \ion{O}{i} transitions, and \oi\ is also folded with the \ion{Si}{ii}$\lambda$1304\ line. \siid\ is blended with \ion{O}{i}$\lambda$988 and \ion{N}{iii}$\lambda$989.

Our goal is to measure the equivalent widths and residual fluxes of these different lines (when present) for all the galaxies in the LzLCS and literature samples. We first find the wavelength of the minimum depth of the line (minimum flux). We then define a $\pm$1250~km~s$^{-1}$ integration window in velocity space ($\Delta \lambda_i$, chosen by visual inspection) centered on this wavelength, over which the equivalent width $W_{\lambda_i}$ is computed as:
\begin{equation}
    W_{\lambda_i} = \int _{\Delta \lambda_i} \left(1 - f_\lambda/F^\star(\lambda)\right) d\lambda,
    \label{eq:ew}
\end{equation}

\noindent where $f_{\lambda}$ is the observed spectral flux density, and $F^{\star}(\lambda)$ is the modeled stellar continuum. We also compute the residual flux of the line, $R(\lambda_i) = \langle f_\lambda/F^{\star}\rangle$, as the median over an interval of $\pm$150~km/s around the minimum flux (to avoid biasing towards minimum depth pixels affected by noise). Errors on the observed spectra plus the error on the stellar continuum fitting are incorporated within every absorption line uncertainty. More precisely, we take the median and standard deviation of the $W_{\lambda_i}$ and $R_{\lambda_i}$ distributions (over the 500 realizations of the perturbed spectra and subsequent continuum). We consider these as the final measurement and error values of each line.

The drop in the COS/G140L sensitivity redward of 1600\AA\ in the observed frame reduces the S/N of the measured flux, thus providing poorer constraints on the stellar continuum and lowering the S/N of the LIS lines at wavelengths longer than \lya$\lambda$1216 ($\sim 1600\AA$ at $z = 0.3$). Then, and only for the LIS measurements, we prefer to locally model the stellar continuum by a linear function. Equivalent widths and residual fluxes for \siib, \oi,\ and \cii\ lines are measured assuming this linear continuum, and using the same previous equations. 

Ideally, we also want to determine the covering fraction of the lines. In general, the relation between the residual flux and the covering fraction depends on (1) the assumed gas and dust geometries, (2) the different velocity components of the gas clouds that potentially contribute to the line shape, and (3) the degree of saturation of the line, which we discuss in the forthcoming sections.

\subsubsection{The covering fraction: assumptions about the ISM geometry}
The escape of ionizing radiation and the UV absorption features of a galaxy can be explained by the so-called picket-fence model \citep[e.g.,][]{Heckman2001, Heckman2011, Vasei2016, Reddy2016b, Gazagnes2018, Steidel2018}. This model describes a stellar ionizing radiation field surrounded by a nonuniform or patchy ISM, where the neutral and metallic gas are distributed in high-column-density clouds, with low-column-density channels in between. Two gas and dust relative geometries can be assumed. In the first, a uniform foreground dust-screen envelops the whole galaxy (A) while in the second, dust is hosted in the discrete gas clouds of the ISM (B). 

A key parameter of the picket-fence model is the geometrical distribution of the gas parameterized by the covering fraction, $C_f({\lambda_i})$. This variable represents the fraction of sight lines that are covered by optically thick gas at wavelength $\lambda_i$, in contrast to sight lines with low amounts of gas or no gas at all. Observationally, the covering fraction can be related to the residual flux of individual absorption lines ($R(\lambda_i) = \langle f_\lambda/F^{\star}\rangle$). The functional form of this dependency changes with the gas and dust geometry \citep[e.g.,][]{Vasei2016, Gazagnes2018}, and always assumes absorption lines are saturated (i.e., the column density of the medium is high enough so that it is optically thick at the $\lambda_i$ transition). Furthermore, given the limited spectral resolution, we assume that the lines are described by a single gas component or, in other words, that all velocity components of the gas have the same covering fraction.

For geometry A, the gas covering is simply the complement of the residual flux, and can be measured from the line depths as
\begin{equation}
    R(\lambda_i) = 1 - C_f(\lambda_i),
    \label{eq:cfA}
\end{equation}

\noindent while for geometry B, the relation depends also on the dust attenuation \citep[see detailed derivation in][]{Vasei2016, Reddy2016b, Gazagnes2018}:
\begin{equation}
    R(\lambda_i) = \dfrac{1 - C_f(\lambda_i)}{C_f(\hi)\times 10^{-0.4 k_{\lambda_i} \ebv^B} + (1 - C_f(\hi))},
    \label{eq:cfB}
\end{equation}

\noindent where $\ebv^B$ is the dust attenuation measured according to geometry B, and \cfhi\ is the neutral gas covering fraction. The \hi\ and the dust have the same covering fraction according to this model. Both the stellar continuum and the residual flux of the lines must be determined consistently with the same geometry (under the same gas and dust distribution), but the actual S/N for most of the spectra does not allow a stellar continuum fitting using geometry B \citep{Gazagnes2018, Steidel2018}. Therefore, in this paper we adopt geometry A. However, our absorption line calculations will provide insights into the underlying gas and dust geometry and other ISM properties of low-z LyC-emitting galaxies (see Sect.\ \ref{sec:results_abslines}).

\subsubsection{Saturation regime for the \hi\ and LIS lines}
Applying Eq.\ \ref{eq:cfA} (or Eq.\ \ref{eq:cfB}) to compute the gas covering fraction intrinsically assumes that absorption lines are saturated. The condition of saturation is tested using the equivalent width ratios for lines of the same ion (see App.\ \ref{appB}). 

The saturation of \hi\ is studied through the ratio of the Lyman series lines Ly$\beta$ (\lyb) and Ly$\delta$ (\lyd). The Ly$\beta$-to-Ly$\delta$ equivalent width comparisons reveal that most of the sources fall in the optically thick (saturation) region of the diagram \citep[see][]{Draine}, while some of them may be compatible with an optically thin regime, although many of them have large uncertainties at the same time. We caution that some of the discrepancies in this analysis can be due to slight underestimations of the continuum level, or because of the \ion{O}{i} contribution, whose lines are blended with the Lyman series. For the metals, we use the \si\ as a representative ion, and compare the ratio of \siib\ to \siia. Nevertheless, this comparison is subject to large measurement errors and, although we can be statistically sure about the \hi\ saturation, we can neither discard nor assume saturation for the metallic lines.

Henceforth, we assume that all the \hi\ transitions and metal lines considered are saturated. A more detailed explanation of this reasoning is presented in Appendix\ \ref{appB}.

\subsubsection{Estimation of systematic errors}
We finally study the possible systematic errors on our measurements. When measuring absorption line properties from observed spectra, the instrumental resolution (R) tends to make the lines broader and so residual fluxes can be overestimated. In addition, the observed S/N of every pixel ---due to instrumental sensitivity, noise, etc.--- also impacts the line shape and can 
therefore affect our estimates.

To account for these effects, we simulate different absorption lines (Ly$\delta$, Ly$\beta$, \siib, \cii) with a standard Voigt profile \citep{Draine} and typical saturated column densities and Doppler broadening in SF galaxies, a single velocity component, and a fixed covering fraction of $C_f=0.85$ for all of them. A foreground dust-screen geometry of the picket-fence model is adopted. We then degrade to the COS/G140L instrumental resolution ($R \sim 1000$) and finally add a set of different random Gaussian S/N per pixel. After that, the residual flux is measured 500 times for every S/N value, and converted to a covering fraction using Eq.\ \ref{eq:cfA}. The median covering fraction is compared to the original input value to give the relative difference. We conclude that, for the typical S/N measured in our sample at the continuum level close to the lines in question (e.g., close to \lyd\ the continuum is S/N $\sim 4$) the systematic error on the covering fraction spans between 5\% and 20$\%$, with a bias towards lower coverings. However, this value does not dominate over original errors and is within the original error bars we are reporting for single covering fraction measurements (see Sect.\ \ref{sec:results_abslines}). In addition, given the actual instrumental resolution, systematic deviations do not change significantly with respect to $C_f=0.85$ when other input $C_f$ is chosen in the simulations. Even in the case when the residual flux of single lines could be slightly overestimated, the general trends in this study will hold, because all the spectra in the working sample have the same resolution. We therefore leave our initial residual flux measurements and errors unaltered. A more exhaustive explanation of the estimation of systematic errors is presented in Appendix\ \ref{appC}.

\subsection{Photon escape fraction from the Lyman series and LIS metallic lines}\label{sub:method_fesc}
In the above-described picket-fence model, the absolute LyC photon escape fraction is related to the \hi\ covering fraction of saturated lines and the dust attenuation \citep{Heckman2001, Heckman2011, Vasei2016, Reddy2016b, Gazagnes2018, Chisholm2018, Steidel2018, Gazagnes2020}. The expression of the predicted \fescabs\ depends on the assumed gas and dust geometry. Nonetheless, as demonstrated by \citet{Gazagnes2018, Chisholm2018}, although \ebv\ and \cfhi\ are model dependent, regardless of the use of geometry (A) or (B), the attenuation and covering fraction terms balance each other so that the predicted \fescabs\ remains almost unaltered (e.g., it does not strongly depend on the assumed geometry). This is because, for the same observed flux, model (B) would require larger dust obscuration but also higher gas covering fraction than model (A) in order to equally match the data.

Consequently, we decide to adopt a uniform dust screen model (A) consistently for the continuum and the \fescabs\ determination from the absorption lines. Covering fractions are therefore simply given by Eq.\ \ref{eq:cfA}, and finally the predicted photon escape from \hi\ is \citep[cf.][]{Gazagnes2018, Chisholm2018} 
\begin{equation}
    \feschi = 10^{-0.4k_{912}\ebv} \times (1 - \cfhi),
    \label{eq:fesc_Ly}
\end{equation}

\noindent where $k_{912}$ depends on the assumed attenuation law at 912\AA\ (e.g., $k_{912} \approx 12.87$ for \citetalias{R16}). \cfhi\ is the average \hi\ covering fraction. To compute this, we first use Eq.\ \ref{eq:cfA} to figure out individual covering fraction measurements for each line. Then, in order to improve the S/N in single measurements of $C_f(\lambda_i)$, we use an inverse-variance-weighted mean definition for the \hi\ covering fraction:
\begin{equation}
\begin{aligned}{}
    \cfhi & \pm \Delta{\cfhi} = \dfrac{\sum_{i=1}^{N} C_{fi} \times \frac{1}{\sigma_{C_i}^2}}{\sum_{i=1}^{N} \frac{1}{\sigma_{C_i}^2}} \pm \sqrt{\frac{1}{\sum_{i=1}^{N} \sigma_{C_i}^2}}, \\
    i & \equiv \lyz, \lye, \lyd, \lyg, \lyb 
\label{eq:cf_meanHI}
\end{aligned}
\end{equation}

\noindent where N$~\leq5~$ is the number of Lyman lines that can be measured on a single spectra, and $\sigma_{C_i}$ is the measurement error (from Monte-Carlo sampling of the spectra+continuum, see previous section). These methods are tested against the {\em fiducial} photon escape in Sect.\ \ref{sec:discussion}.

\begin{figure}[t]
    \centering
    \includegraphics[width=0.95\columnwidth, page=3]{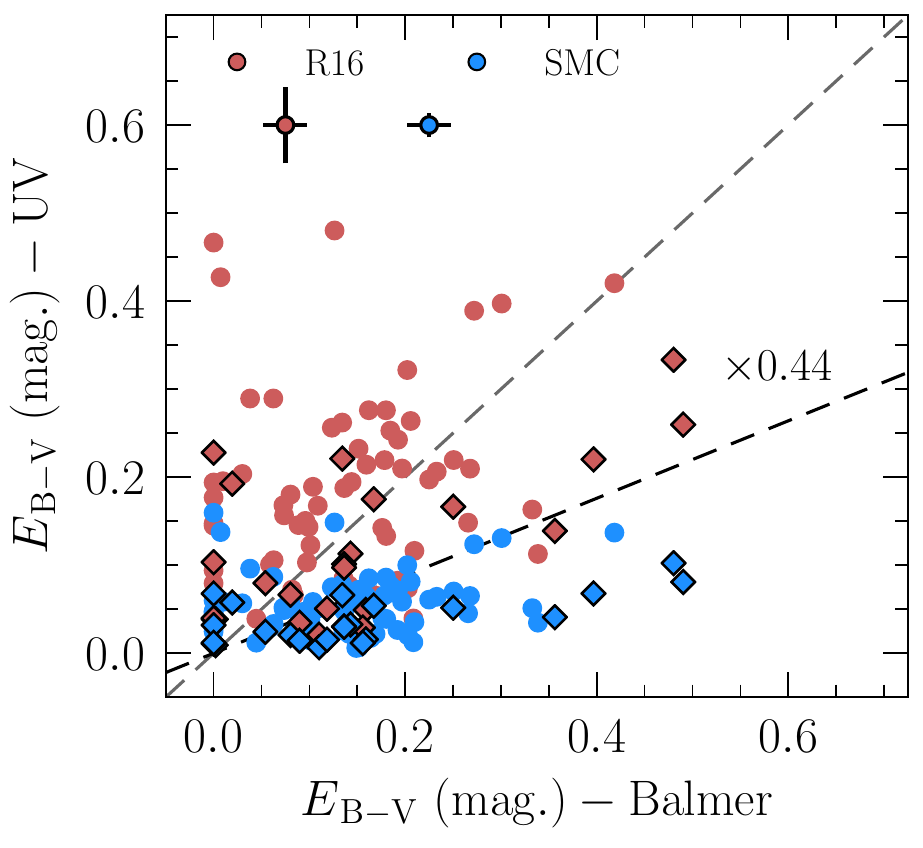}
\caption{Comparison between the optical attenuation parameter \ebv ~(color excess, in mag.) obtained from the Balmer decrements (CCM attenuation law) and the values derived from the UV--SED fits to the COS spectra. Here we explore two different attenuation laws: \citetalias{R16} (red) and SMC (blue symbols). Median error bars are plotted in the upper left. Circles represent the LzLCS galaxies while diamonds are the \citet{Izotov16a, Izotov16b, Izotov18a, Izotov18b, Izotov2021} and \citet{Wang2019} sources. The theoretical $\times0.44$ shift between the stellar and nebular attenuation, which assumes a foreground screen of dust \citep{Calzetti2000}, is also shown.}
\label{fig:EBV_comparison}
\end{figure}

\section{UV attenuation and absolute photon escape of the LzLCS galaxies}\label{sec:fit_results}
From the direct output parameters of the spectral fits we obtain information on the light-weighted stellar age and metallicity of the stellar population, and the UV dust attenuation. The UV dust attenuation significantly impacts the derived absolute escape fractions, and so here we mostly focus on the relation between these two variables. The stellar population properties of the LzLCS sample will be discussed in a separate paper (Chisholm et al., in prep.).

\begin{figure}
    \centering
    \includegraphics[width=0.95\columnwidth, page=4]{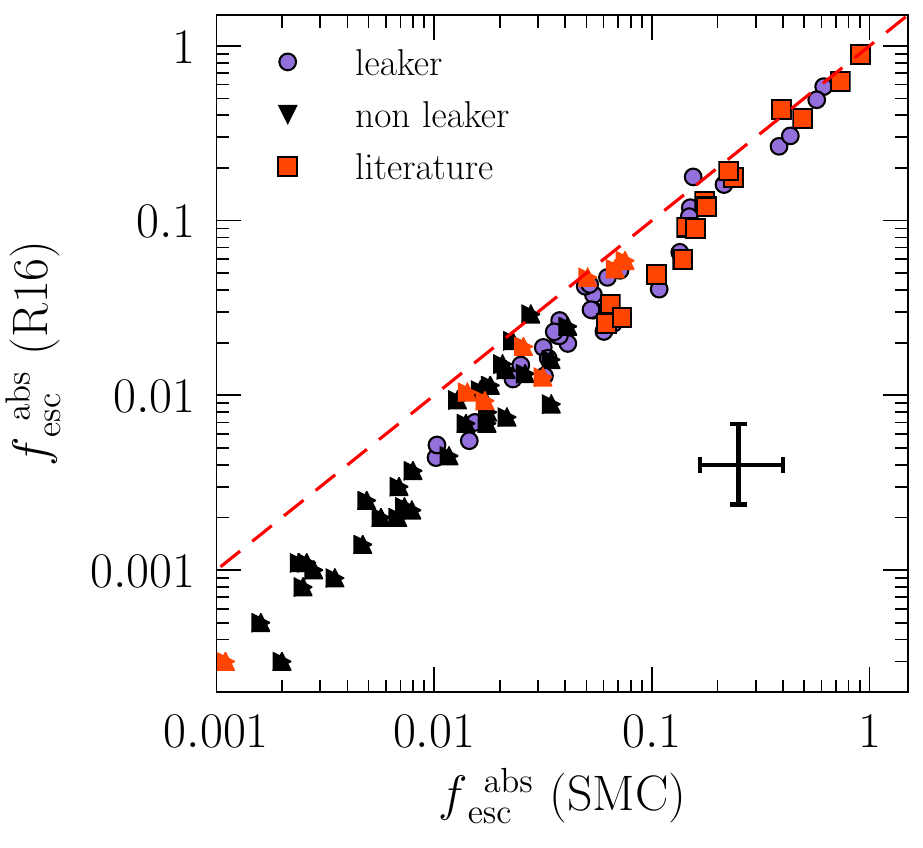}
\caption{1:1 comparison between the absolute LyC photon escape fractions (\fescabs) from the UV--SED fits using \citetalias{R16} and SMC dust-attenuation laws. LzLCS leakers and nonleakers are indicated by blue points and black arrows (upper limits), respectively. Red squares are the \citet{Izotov16a, Izotov16b, Izotov18a, Izotov18b, Izotov2021} and \citet{Wang2019} sources. Median relative error bars are shown in the lower right corner.}
\label{fig:FESC_comparison}
\end{figure}

The fitted \ebv\ ranges between $\sim$0.02 and 0.45, with a mean value of $\sim$ 0.15 (0.07 and 0.24 as the 0.16 and 0.64 quantiles). \autoref{fig:EBV_comparison}  shows the \ebv\ derived from the UV--SED fits and a comparison to the ones determined from the Balmer decrement measurements using the optical SDSS spectra and the \citet{CCM89} attenuation law (CCM). For the \ebv\ from the UV fits, we examined two different attenuation laws: \citetalias{R16} and SMC, chosen to bracket the expected UV attenuation in SF galaxies\footnote{Using the dust-attenuation law of \citet{Calzetti2000}, our results do not change significantly with respect to the \citetalias{R16} law, because both attenuation laws have a similar wavelength dependency of $k_\lambda$, implying a similar absolute UV attenuation.} \citep{Shivaei2020}. A direct comparison between \ebv\ values obtained from SMC and CCM is straightforward, because both laws have identical $k_{\lambda}$ at optical wavelengths. In the case of \citetalias{R16}, we used the \citet{Reddy2015} parametrization at $\lambda>1500\AA$ converted to CCM law via the usual Balmer Decrement formulae\footnote{$\dfrac{\ebv^{R16}}{\ebv^{CCM}}=\dfrac{(k_{\lambda_1}-k_{\lambda_2})^{R16}}{(k_{\lambda_1}-k_{\lambda_2})^{CCM}}=0.93, ~\lambda_1=H\beta \lambda4863, \lambda_2=H\gamma \lambda4370$.}. When \citetalias{R16} is used, the overall \ebv\ derived from the UV spectra is higher than that from the CCM law typically by a factor of $\sim 1.2$, while it is a factor of $\sim 2.6$ lower when SMC is chosen. 

A systematic shift between the nebular (obtained from emission line ratios) and the stellar attenuation (from the slope of the continuum) may be expected when a foreground dust-screen distribution is in play \citep[see][]{Calzetti2000}. The stellar attenuation is typically a factor of 0.44 lower than the nebular attenuation (see \autoref{fig:EBV_comparison}). We only find this trend when the SMC law is used; \citetalias{R16} derived values agree better with the \ebv\ from the Balmer decrements. Moreover, an SMC-like attenuation law has appeared more suitable for high-$z$ SF galaxies \citep{Salim2018} and low-metallicity starburst \citep{Shivaei2020}, requiring even steeper laws for leakers \citep{Izotov16b}. 

Distinguishing the attenuation law from the data used here is not possible, because the fit quality (reduced-$\chi^2$) is similar in both cases. Therefore, and as we mentioned in Sect.\ \ref{sec:methods}, we use the \citetalias{R16} attenuation law as a default in this study, motivated by the fact that this law is defined below 1000\AA\ by a significant number of galaxies. This translates to a factor  two lower attenuation at 900\AA\ than estimates derived from extrapolations of dust attenuation curves such as the SMC, and will impact the determination of the escape of ionizing photons from SED fitting, as we explore below \citep[see also][]{Reddy2016b, Chisholm2019}. Finally, to avoid a strong dependence on the assumed attenuation law, we refer to the UV attenuation defined as $A_{\lambda} = k_{\lambda}~\ebv$ (Eq.\ \ref{eq:s99_fit}) instead of \ebv. Still, differences in $A_{\lambda}$ are found for the two laws at all wavelengths, being systematically higher for \citetalias{R16}. This reveals that the fitted stellar parameters ($X_i$) are slightly lower for the SMC law, meaning that both fits can match the observations similarly. In any case, both approaches yield the same light-averaged ages and metallicities (further investigations in Chisholm et al., in prep.).

\begin{figure*}
    \centering
    \includegraphics[width=0.95\textwidth, page=2]{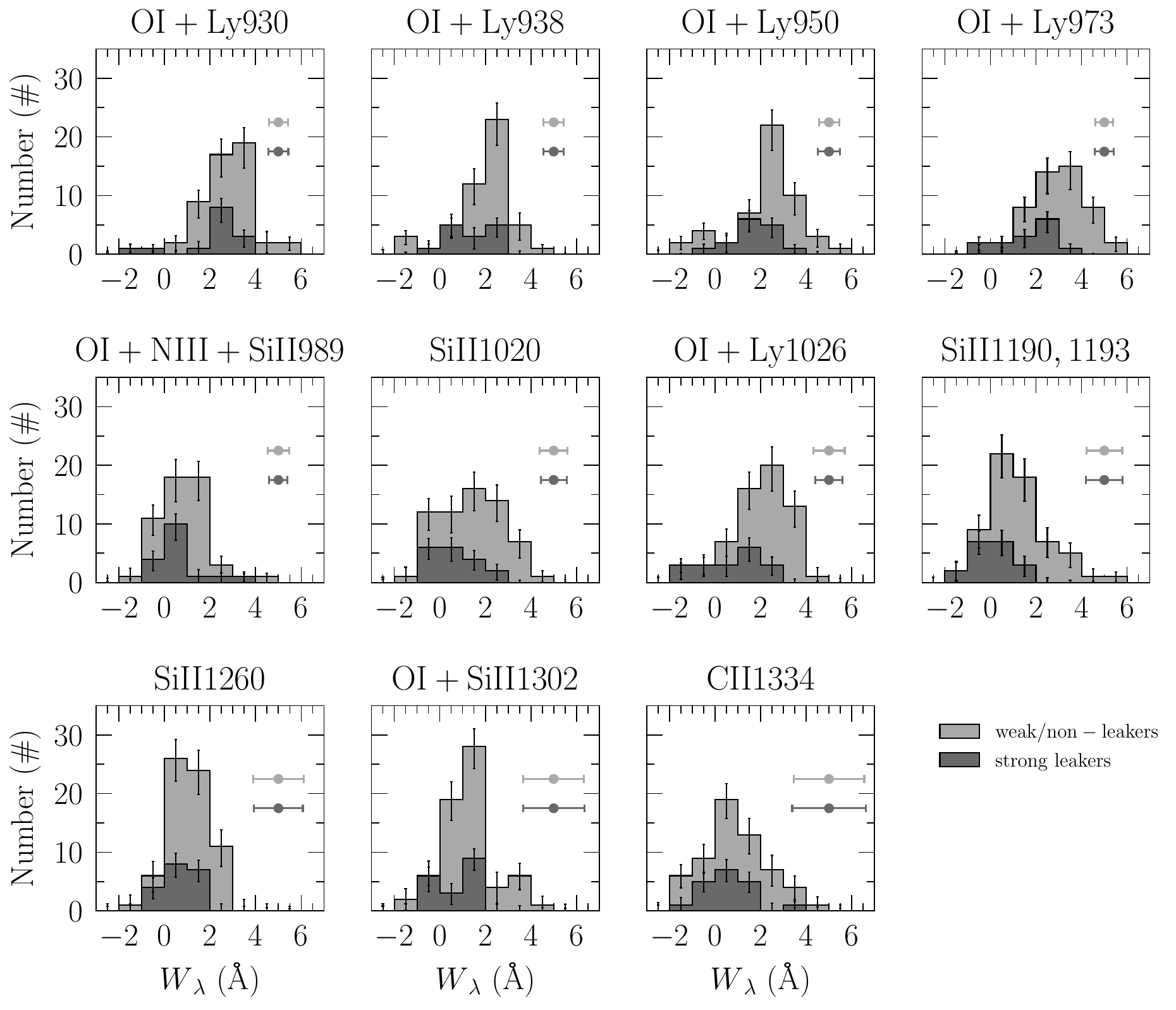}
\caption{Equivalent width distributions ($W_{\lambda}$, in \AA) for the measured absorption lines. Dark-gray histograms correspond to the strongest leakers (20) in the LzLCS plus published joint samples, while light-gray histograms represent the weak leakers (30) and nondetections in the LyC window (39), out of the total of 89 galaxies. Strong leakers are defined as galaxies with significant detection in the LyC and $\fescabs \geq 0.05$. Weak leakers
also have significant LyC detection but $\fescabs < 0.05$. Nonleakers are not detected in the LyC. Median equivalent-width errors are indicated in all the panels and for each source type.}
\label{fig:abslines_ew}
\end{figure*}

\begin{figure}
    \centering
    \includegraphics[width=0.95\columnwidth, page=7]{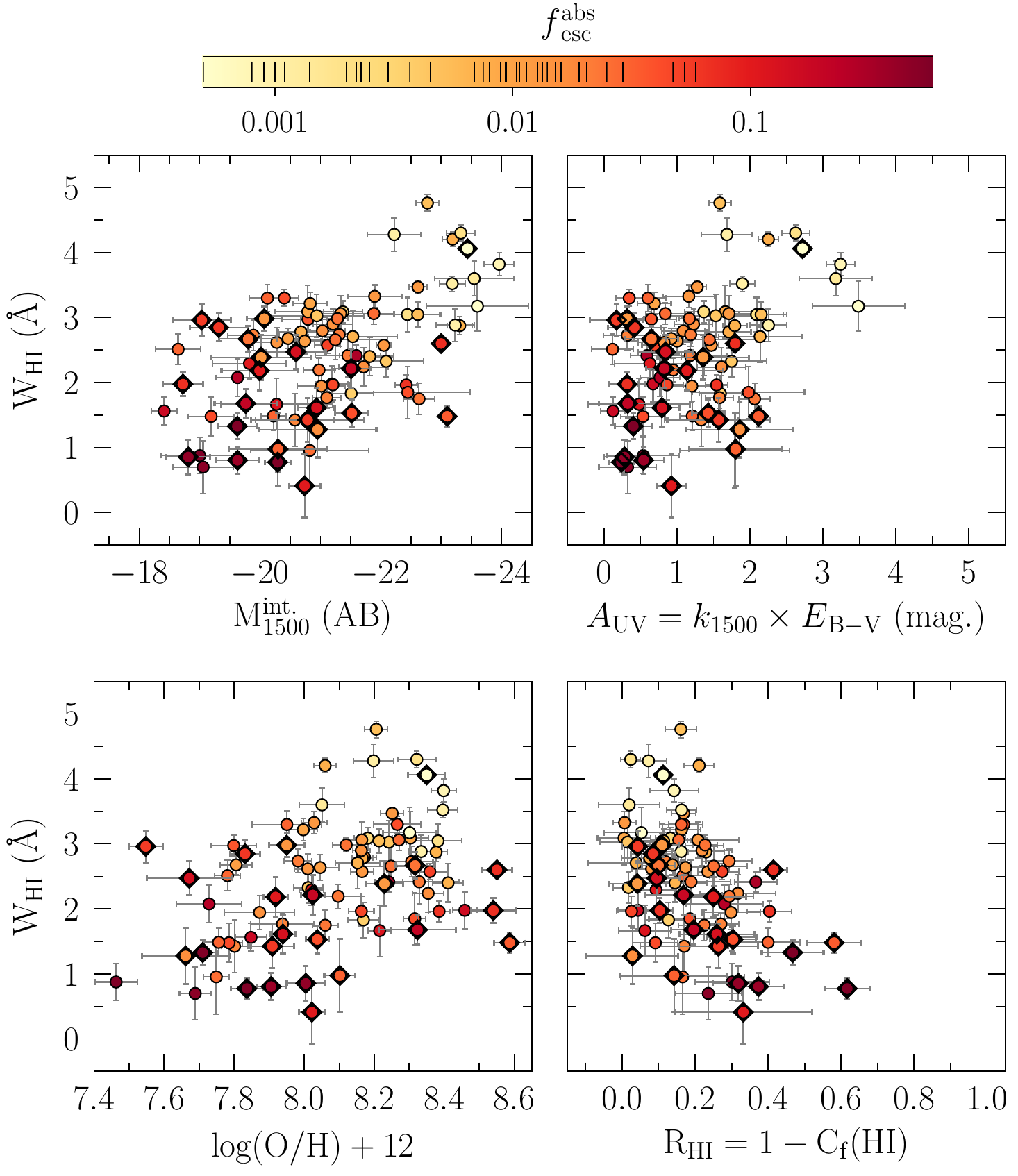}
\caption{Empirical trends between the \hi\ equivalent width (\ewhi) and different observational properties (from top to bottom and left to right): intrinsic absolute magnitude at 1500\AA ~(\mfuv), UV attenuation (\auv) at 1500\AA, measured gas-phase metallicity (\oh) and the average \hi\ residual flux (\rhi). Points are color-coded by their absolute LyC photon escape fraction, \fescabs\ (and upper limits values on \fescabs\ are indicated inside the color bar for reference). The \citet{Izotov16a, Izotov16b, Izotov18a, Izotov18b, Izotov2021} and \citet{Wang2019} sources are displayed with diamonds and also considered in the correlations (see text).}
\label{fig:ewHI_props}
\end{figure}

\begin{figure}
    \centering
    \includegraphics[width=0.95\columnwidth, page=8]{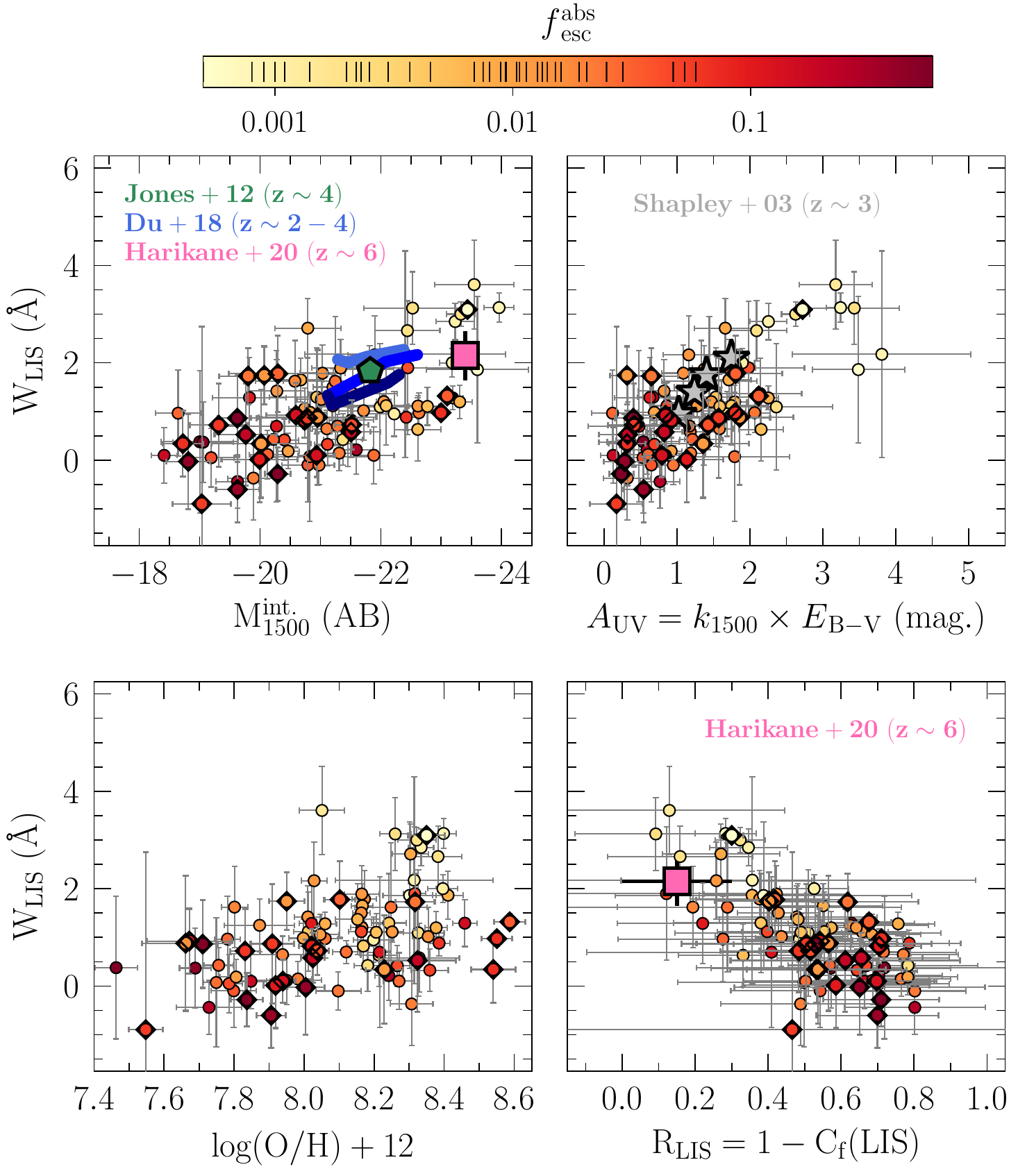}
\caption{Empirical trends between the LIS equivalent width (\ewlis) and different observational properties: \mfuv, \auv, \oh\ and \rlis, with sources displayed and color-coded as in \autoref{fig:ewHI_props}. Additionally, results from \citet[][$z \sim 3$]{Shapley2003} and \citet[][$z \sim 6$]{Harikane2020} composites are overplotted. The light-blue, blue, and dark-blue lines correspond to the different redshifts between $z = 2-4$ from \citet{Du2018}.}
\label{fig:ewLIS_props}
\end{figure}

In \autoref{fig:FESC_comparison} we show the derived \fescabs\ values (Eq.\ \ref{eq:fescS99}), assuming both the \citetalias{R16} and the SMC law \citep[see also][]{FluryI}. The derived LyC escape fractions vary over a wide range, with values up to $60\%$~($85\%$) for the LzLCS (literature) sample. Typically, upper limits are below 0.01. Clearly, the choice of the attenuation law has some impact on the derived \fescabs: the relative difference between the two \fescabs\ versions ranges from 2\%\ to 40$\%$ $\times \fescabs$, with systematically higher values obtained for the SMC law. This systematic shift is smaller for high values of the absolute escape fraction, which are found in galaxies with relatively low UV attenuation (Sect.\ \ref{sub:results_ebv_fesc}). The shift to higher \fescabs\ for the SMC law is due to fact that this law is steeper than \citetalias{R16}, which means that a smaller UV attenuation is needed to reproduce the overall, reddened observed spectral shape. Therefore, the intrinsic stellar flux (before attenuation by dust) is lower with the SMC law, and therefore \fescabs\ is higher.

Finally, our fitting algorithm is not devoid of degeneracies between stellar age, metallicity, and the \ebv\ \citep[see discussion in][]{Gazagnes2020}. The S/N of the original spectra impacts the ability of the fitting code at the time of solving this degeneracy, where only moderate (and high) S/N spectra will give a reliable estimation of the stellar population properties \citep{Chisholm2019}.

\begin{table}
\centering
\small
\caption{Median line equivalent widths (in \AA) for the whole working sample (all, 89 galaxies), strong leakers only (20 galaxies), and weak+nonleakers (69 galaxies). Leakers have lower median equivalent widths than nonleaking galaxies in most cases.}
\begin{tabular}{cccc}
\toprule
Ion $+$ & $W_{\lambda}~(\AA)$ & $W_{\lambda}~(\AA)$ & $W_{\lambda}~(\AA)$ \\
$\lambda_{rf}~(\AA)$ & all (89) & strong leakers (20) & weak+nonleakers (69) \\
\midrule
OI+HI930 & 2.75 & 2.05 & 2.94 \\
OI+HI938 & 1.71 & 1.41 & 1.80 \\
OI+HI950 & 2.09 & 1.62 & 2.24 \\
OI+HI973 & 2.78 & 1.58 & 3.15 \\
OI+NIII+SiII989 & 0.77 & 0.60 & 0.83 \\
SiII1020 & 1.13 & 0.57 & 1.32 \\
OI+HI1026 & 1.64 & 0.58 & 1.97 \\
SiII1190,1193 & 0.96 & 0.17 & 1.21 \\
SiII1260 & 0.96 & 0.56 & 1.08 \\
OI+SiII1302 & 1.08 & 0.64 & 1.21 \\
CII1334 & 0.79 & 0.52 & 0.88 \\
\bottomrule
\end{tabular}
\label{tab:lines_median}
\end{table}

\section{Empirical results and inferences from UV absorption lines}\label{sec:results_abslines}
We now search for the empirical relations between the \hi/LIS absorption lines, their equivalent widths (\ref{sub:results_lowews}), residual fluxes, and some galaxy properties (\ref{sub:results_HILIS}). We also compare the observed trends with those in high-$z$ galaxies (\ref{sub:results_highz_lines}).

\subsection{LyC leakers have weak absorption lines} \label{sub:results_lowews}
The equivalent width ($W_{\lambda}$) distributions of the major hydrogen and metal absorption lines are shown in \autoref{fig:abslines_ew}, where the strong leakers (dark gray) are separated from the weak and nonleaker distributions (light gray). The error bars on the histograms come from a proper estimation of the Poissonian confidence intervals for every $W_{\lambda}$ bin, taking into account the uncertainty on single measurements. 

Overall, the equivalent widths reach up to 4-6\AA\ for the strongest lines. The number of detections (defined here as value/error$~>2$) out of the total of 89 galaxies in the sample are 62, 51, 54, 56, and 50 for the Lyman series, and 32, 35, 23, 20, 13, and 9 for the LIS lines, respectively. The main cause for nondetection is the contamination by geo-coronal emission in the Lyman series range and the decrease in S/N towards longer wavelengths. The presence of negative equivalent widths can be due to different reasons. For some \hi\ lines, such as \lyb, the continuum near the line is complex to model because of contributions from a stellar \ion{O}{vi} P-Cyngi profile as well as numerous adjacent ISM absorption lines (in particular \ion{O}{vi}, \ion{C}{ii}, \ion{O}{i}). Slight underestimations of the continuum level would lead to negative equivalent widths. In the case of the LIS lines, the low S/N and overlap of some fluorescence transitions within the integration window would also produce some negative $W_{\lambda}$ values.

The equivalent width distributions of leakers peaks at lower values than the nonleaking distributions in general, suggesting that leakers have weaker \hi\ and LIS absorption lines than nonleakers. \autoref{tab:lines_median} shows the median of the $W_{\lambda}$ distributions for the  entire sample (89 galaxies), strong leakers (20), and weak+nonleakers only (70). For example, for \lyb\ (\siia), the median equivalent width value for leakers is 0.58\AA\ (0.17\AA), while nonleakers show median equivalent widths of 1.97\AA\ (1.21\AA), respectively. To test whether the two distributions differ or not, one can perform the {Anderson-Darling} test. For the previous lines, the null hypothesis that assumes leaker and nonleaker populations to come from the same parent distribution can be rejected at 97.5\% level, with a significance of $p_{val.} \le 10^{-2}$. Based on smaller samples \citet{Chisholm2017}  already noted that LyC emitters show weaker \hi\ and LIS lines \citep[see also][in simulations]{Mauerhofer2020}.

\subsection{HI and LIS absorption line strengths and UV attenuation are driven by the gas covering fraction} \label{sub:results_HILIS}
The observed differences in the \hi\ and LIS lines between LyC emitters and nonemitters suggests a connection between the mechanisms that favor the LyC leakage and those that lead to changes in the absorption lines profiles. Therefore, we explore several empirical correlations between \hi\ and LIS equivalent widths and other galaxy properties, such as the intrinsic absolute magnitude at 1500\AA ~(\mfuv), the UV attenuation (\auv), the measured gas-phase metallicity (\oh), and the average \hi\ and LIS residual flux (\rhi, \rlis). \mfuv\ is produced by taking the mean of the dust-free S99 modeled flux over a 100\AA\ window around 1500\AA, and converting this into an AB absolute magnitude given the redshift and the distance modulus. \auv\ is taken from the fits, $\auv = k_{\rm UV} \times \ebv$ --at 1500\AA--, and finally residual fluxes are computed using Eq.\ \ref{eq:cfA} and subsequently Eq.\ \ref{eq:cf_meanHI} for \hi\ and similarly for the LIS lines (\siia, \siib, \oi, \cii).

For the equivalent widths, we take the inverse-variance-weighted average of the single \hi\ and LIS line widths (\ewhi, \ewlis):
\begin{equation}
\begin{aligned}{}
    W_{K} & \pm \Delta{W_{K}} = \dfrac{\sum_{i=1}^{N} W_{\lambda i} \times \frac{1}{\sigma_{W_i}^2}}{\sum_{i=1}^{N} \frac{1}{\sigma_{W_i}^2}} \pm \sqrt{\frac{1}{\sum_{i=1}^{N} \sigma_{W_i}^2}}, \\
    K \equiv \hi \\
    i & \equiv \lyz, \lye, \lyd, \lyg, \lyb \\
    K \equiv LIS \\
    i & \equiv \siia, \siib, \oi, \cii,
\label{eq:ew_meanHI}
\end{aligned}
\end{equation}

\noindent where $N$ is the number of \hi\ or LIS lines that can be measured on a single spectra and $\sigma_{W_i}$ represents the error on the individual equivalent widths measurements (Monte-Carlo sampling). At this point it becomes necessary to take into account the following aspects. Some lines, such as \lyg\ and \siid,\ are contaminated by geo-coronal emission in many cases (those values are removed from the average). We exclude \siic\ from the average because this line has a very low oscillator strength and is usually very weak and likely optically thin. We also exclude \siid\ because of the geo-coronal contamination and the blending with the high-ionization-state \ion{N}{III}$\lambda$989 line.

Averaging the \hi\ and LIS lines of the same ion is reasonable owing to the fact that most of those lines have similar oscillator strengths or are likely saturated, and so their equivalent widths are similar. Even LIS transitions for very different ions (e.g., \si\ versus \ion{O}{i}) will only differ by a factor of two in SF galaxies \citep[e.g.,][]{Chisholm2016}. Averaging LIS line measurements has also frequently been reported for high-redshift galaxies \citep[e.g.,][]{Shapley2003,Du2018,Trainor2019,Du2021} to improve the S/N of single line measurements. As mentioned above, in our case the average is strictly necessary for the LIS lines redder than 1200\AA\ rest-frame (or below 1600\AA\ in the observed frame), where the S/N is usually very low due to the decrease in the COS instrument sensitivity.

In \autoref{fig:ewHI_props} and \ref{fig:ewLIS_props}, the \hi\ and LIS equivalent widths are plotted against the four quantities described above: \mfuv, \auv, \oh, \rhi,\ and \rlis. LzLCS sources are represented by circles while the published sample is shown using diamonds. The sources are color coded by our \fescabs\ determined from the UV spectral fits. Overall, strong absorption lines are only found at high \auv, high metallicity, and in UV-luminous sources (where \mfuv\ is dust corrected). In other words, \ewhi\ and \ewlis\ both correlate with the UV attenuation and metallicity, but anti-correlate with the intrinsic UV absolute magnitude and residual fluxes. High \fescabs\ sources are located at the lowest \hi/LIS equivalent-width regime (see also Sect.\ \ref{sub:results_cfs}). The variations in the absorption line properties with the LyC escape fraction are discussed further below.

In order to quantify the above-mentioned trends, we computed the {\em Kendall} correlation coefficient for every pair of variables described above ($\tau$, see \autoref{tab:kendall_ew}). For a sample of 89 objects, we consider correlations to be significant if $p_{val} \lesssim 2.275 \times 10^{-2}$ (i.e., 2$\sigma$) and strong if $\left| \tau \right| \gtrsim 0.2$. For \ewhi, the strongest and most significant correlation is found to be with the UV intrinsic absolute magnitude \mfuv. On the other hand, \ewlis\ shows the strongest trend with the LIS residual flux.

Generally, three different mechanisms can change the \hi/LIS equivalent widths ---if we assume the lines are formed in single gas clouds, with a single velocity and gas column density: (1) the \hi/metal column densities and metallicity, (2) the Doppler broadening (velocity dispersion of the cloud), and (3) the \hi/LIS covering fraction. Doppler broadening (2) is challenging to address, given our spectral resolution. This said, and based on visual inspection, we do not see any evidence for the highest equivalent widths being among the widest lines, but they are among the deepest. The effect of the covering fraction (3) is clearly visible in the bottom right panels of \autoref{fig:ewHI_props} and \ref{fig:ewLIS_props}, where the residual flux is linked to the covering fraction ---for a uniform dust geometry of the picket-fence model--- and so the equivalent width is directly related to the line depth. It remains to be seen whether or not other physical effects can also play a role.

For example, in a study of nearby SF galaxies, \citet{Chisholm2017} suggested that the decrease in the \ewlis\ is set by a combination of small \hi\ column densities and low metallicity. The relation between the \hi\ column density ($N_{\hi}$) and the column density of metals ($N_Z$) scales with metallicity ($Z$) as $N_Z \sim Z \times N_{\hi}$. A decrease in metallicity (Z) would produce a weaker \ewlis, leading to a natural $\ewlis - \auv$ dependency. However, a change in metallicity  cannot alter the \hi\ line widths and thus explain the observed $\ewhi - \auv$ trend. Furthermore, as the \hi\ lines must be saturated ---based on the ratio of equivalent widths of different \hi\ and \si\ transitions (Appendix\ \ref{appB})--- column density must be secondary. The observations of the Lyman series lines and the fact that they behave in a very similar fashion to those of the LIS lines therefore exclude option (1) to explain the trends shown in \autoref{fig:ewHI_props} and \ref{fig:ewLIS_props}.

\begin{figure}
    \centering
    \includegraphics[width=0.90\columnwidth, page=9]{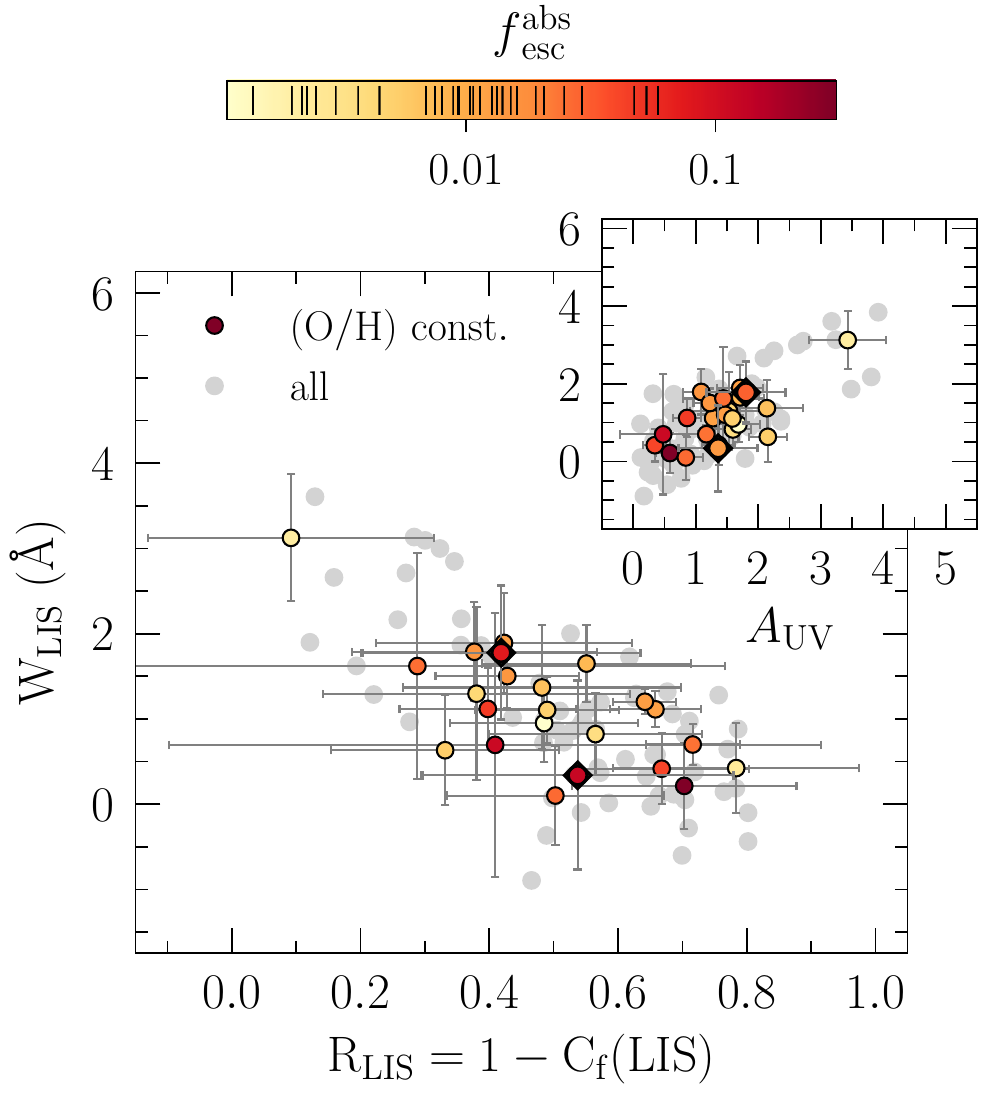}
\caption{LIS equivalent widths (\ewlis) versus residual fluxes (\rlis, {\em main panel}) and UV attenuation (\auv, {\em inset}), when the working sample is restricted to a {constant} metallicity of $\oh = 8.2 \pm 0.1$ (color-coded as in \autoref{fig:ewHI_props}). The global trend including all the sources is shown with gray background symbols.}
\label{fig:OHconst}
\end{figure}

The effect of metallicity is examined in more detail by restricting our analysis to the sources at a constant gas phase metallicity of $\oh = 8.2 \pm 0.1$ (25 galaxies). Doing so, the trends seen in \autoref{fig:ewHI_props} and \ref{fig:ewLIS_props}  remain the same. As an example, in \autoref{fig:OHconst} we show the resulting $\ewlis-\rlis$ behavior when the sample is limited in metallicity (colored points) against the whole LzLCS+literature sample (gray points in the backgorund). Quantitatively, a {\em Kendall} test performed over the $\ewlis-\auv$ relation
for example (see inset in \autoref{fig:OHconst}), still reveals a similar strong and significant correlation when the sample is restricted in metallicity ($\tau \sim 0.25, p_{val.} \sim 10^{-3}$). On the other hand, when restricting to constant \rlis, the statistical connection between \ewlis\ and \auv\ disappears, and similarly for the other variables. This excludes the metallicity (1) as the parameter driving the LIS correlations. We therefore conclude that variations of the geometric covering fraction (3) are the most likely parameter driving the observed trends  both for the \hi\ and LIS lines, and also for the UV attenuation (with a possible cloud-to-cloud velocity dispersion, and the different column densities in front of the stars contributing to the scatter). These findings are similar to the conclusions reached in \citep{Du2021} from the analysis of UV absorption lines and attenuation in LBGs at $z \sim 2-3$.  \citet{Du2021} find that metallicity only affects the LIS equivalent widths when the \ewlya\ is fixed at a constant value or, in other words, when the \hi\ covering fraction is also roughly constant.

\subsection{Comparison with high-redshift galaxies}
\label{sub:results_highz_lines}
Few studies have explicitly shown high-order Lyman series trends with other observables \citep{Henry2015, Reddy2016b, Gazagnes2018, Steidel2018, McKinney2019, Gazagnes2020}, because these lines are in the extreme FUV and hard to detect, often only accessible in the nearby Universe \citep[e.g., see][using the FUSE satellite]{LdE2004, Lebouteiller2006, Grimes2009}. In general, our \hi\ and LIS equivalent width lines correlate similarly with the other galaxy properties, supporting a direct link between neutral gas and metallic gas distributions (see next Sect.\ \ref{sub:results_cfs}), and the claim that the neutral gas content of galaxies is well traced by LIS species \citep{Shapley2003}. The similarity in the behavior of \hi\ and LIS lines also suggests that metallicity cannot be the major contributor to these trends, whose effect would no longer be visible if the lines were saturated. 

In \autoref{fig:ewLIS_props}, we compare the observed \ewlis\ and UV magnitudes with results from \citet{Jones2012} and \citet{Harikane2020}, who used LBG composite spectra at $z \sim 4$ and $z \sim 6$, respectively. Here, \citet{Jones2012} and \citet{Harikane2020} reported values are not dust-corrected, i.e., they are observed AB absolute magnitudes. As seen in \autoref{fig:ewLIS_props} (top left), the agreement is fair with the high-$z$ literature results, suggesting that the ISM structure and properties of the low-$z$ galaxies might be similar to those at high redshift. 

Continuing, similar empirical relations between \ewlis\ and \auv\ (or \ebv) were found by \citet{Shapley2003, Du2018} and \citet{Pahl2020}, although the relations are generally shifted to lower reddening values. In the same \autoref{fig:ewLIS_props}, we also show the results from \citet{Shapley2003}\footnote{For a proper comparison we adopted the appropriate $k_{UV}$ values, depending on the attenuation law used in each study.}. The continuum sample selection in \citet{Shapley2003} against our continuum plus emission line selected sample could be a possible reason for higher $W_{\lambda}$ at fixed \auv. Moreover, the use of a broad-band SED fitting approach to infer the reddening or the spectral stacking could lead to the current differences and, less importantly, the use of different attenuation laws. \citet{Leitherer2011a} and \citet{Grimes2009} show similar trends with the \oh, but shifted to higher metallicities (typical of local starburst galaxies). The trend between the \ewlis\ and LIS residual fluxes (bottom right panel) is compatible with the results of \citet{Harikane2020} who use a $z \sim 6$ galaxy composite. Finally, another possible reason for the small systematic differences in the observed equivalent widths at low- and high$-z$ could be the different spectral resolution and aperture effects, which would systematically lead to stronger observed lines for the low-$z$ galaxies.

\begin{figure}
    \centering
    \includegraphics[width=0.95\columnwidth, page=10]{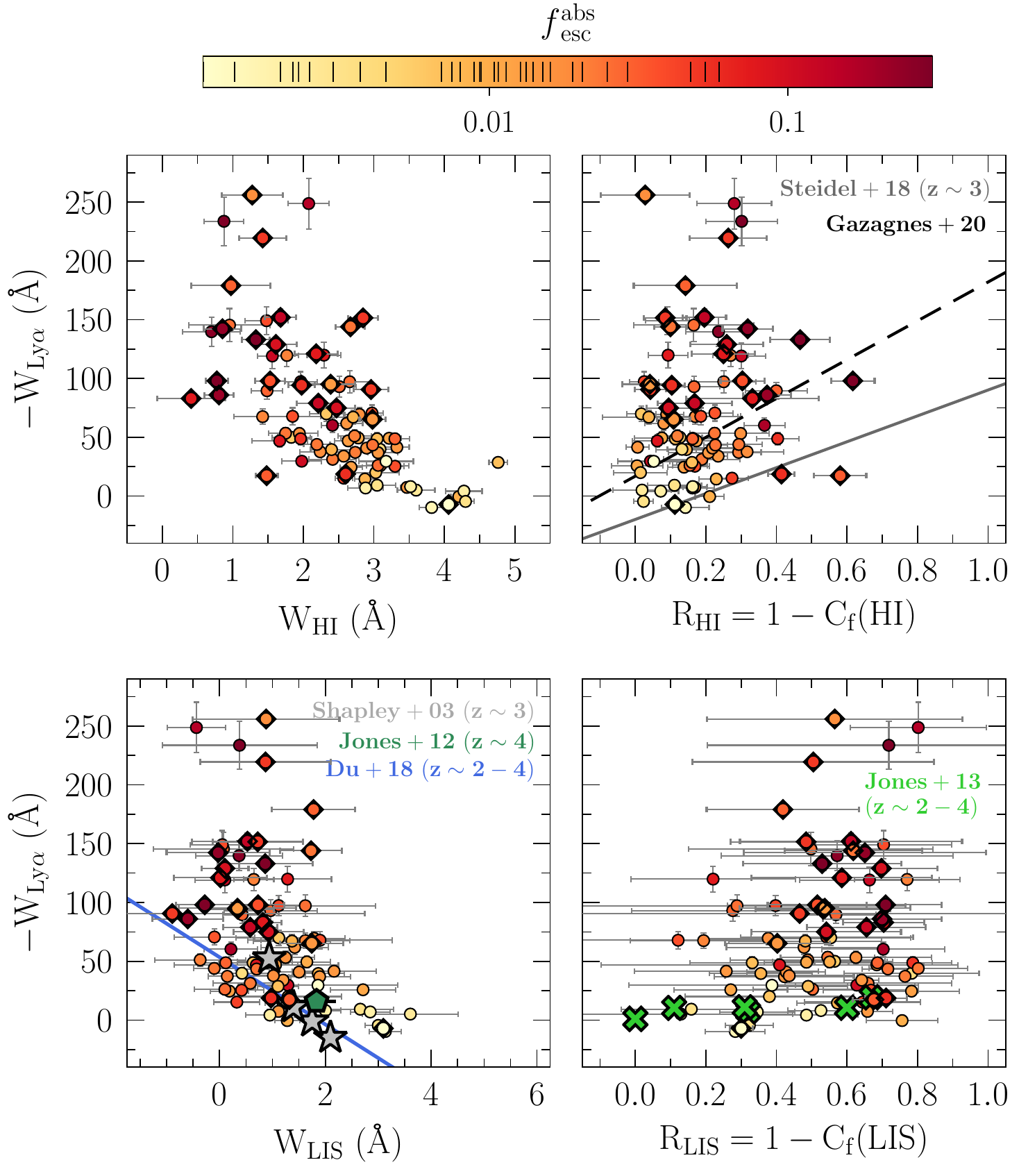}
\caption{Relation between the \lya\ (\ewlya), \hi\ ({\em top}), and LIS line ({\em bottom}) equivalent widths and residual fluxes, respectively. Sources are displayed and color-coded as in \autoref{fig:ewHI_props}. Solid and dashed lines represent the linear fits to the relations found in \citet{Steidel2018}, \citet{Du2018}, and \citet{Gazagnes2020} for comparison. Gray stars, the green polygon, and green crosses correspond to \citet{Shapley2003}, \citet{Jones2012}, and \citet{Jones2013} results.}
\label{fig:lya_HILIS}
\end{figure}

\begin{table*}
\centering
\caption{{\em Kendall} ($\tau$) correlation coefficients$^1$ and $p-$values for \hi\ and LIS equivalent widths (\ewhi, \ewlis) versus diverse galaxy properties$^2$.}
\begin{tabular}{cccccc}
\toprule
\multicolumn{1}{c}{} & \multicolumn{2}{c}{\ewhi} & \multicolumn{1}{c}{} & \multicolumn{2}{c}{\ewlis} \\
\midrule
 & $\tau$ & $p_{val}$ & & $\tau$ & $p_{val}$ \\ 
\cmidrule{2-3} \cmidrule{5-6}
\mfuv\ & $-0.342$ & $4.123 \times 10^{-6}$ & & $-0.383$ & $1.493 \times 10^{-7}$ \\ 
\auv\ & $+0.220$ & $3.009 \times 10^{-3}$ & & $+0.447$ & $9.011 \times 10^{-10}$ \\ 
\oh\ & $+0.274$ & $2.284 \times 10^{-4}$ & & $+0.297$ & $4.641 \times 10^{-5}$ \\ 
\rhi\ & $-0.331$ & $8.576 \times 10^{-6}$ & & -- & -- \\ 
\rlis\ & -- & -- & & $-0.462$ & $2.431 \times 10^{-10}$ \\ 
$-$\ewlya\ & $-0.535$ & $5.819 \times 10^{-13}$ & & $-0.302$ & $4.745 \times 10^{-5}$ \\ 
\bottomrule
\end{tabular}
\label{tab:kendall_ew}
\tablefoot{$^1$ See \citet{ATS1996}. 
$^2$ Intrinsic (dust-free) absolute magnitude at 1500\AA ~(\mfuv), UV attenuation (\auv), measured gas-phase metallicity (\oh), the average \hi\ and LIS residual flux (\rhi, \rlis) and \lya\ equivalent widths (\ewlya).
}
\end{table*}

\section{Relations between the UV absorption lines, \lya, and LyC escape}
\label{sec:results_lya_cf_ebv_fesc}

Here, we present the empirical relations between the UV absorption line, the \lya\ emission line, and the Lyman continuum escape (\ref{sub:results_Lya}). We then investigate the connection between \hi\ and LIS covering fractions (\ref{sub:results_cfs}). Finally, we study the effect of the neutral and low-ionized gas covering fraction on the leakage of ionizing radiation (\ref{sub:results_cf_fesc}) and the influence of dust (\ref{sub:results_ebv_fesc}). 

\subsection{Empirical correlations between the HI and LIS absorption lines, Ly$\alpha$ and LyC}\label{sub:results_Lya}
Although \lya\ and LyC photons interact with the same neutral gas, gas kinematics can influence the \lya\ output, while it has almost no effect (directly) on LyC emission. The interplay between the \lya\ emission and the LyC leakage has been widely studied \citep{RiberaThorsen2015, Verhamme2015, Dijkstra2016, Ribera-Thorsen2017, Verhamme2017, Marchi2018Lyalpha-Lyman-c, Izotov2020, Gazagnes2020, Izotov2021}. Hence, it is a natural step to also study the behavior of the \hi/LIS lines with the \lya\ line. We use the \lya\ equivalent widths (\ewlya) measured by the LzLCS team from the COS spectra \citep{FluryI}. 

In \autoref{fig:lya_HILIS}, the \ewlya\ is compared to the different \hi\ and LIS line properties. Very strong and significant correlations (first column) are found between \ewlya\ and \ewhi, and with \ewlis\ (see \autoref{tab:kendall_ew}). \ewlya\ and \rhi\ are also moderately correlated (however they have larger uncertainties). In summary, larger \lya\ equivalent widths are found in galaxies where the weaker \hi\ and LIS absorption lines are, that is, \ewlya\ is anti-correlated with \ewhi\ and \ewlis\ but correlates with \rhi\ and \rlis. The color-coding demonstrates the correlation between \ewlya\ and \fescabs. Galaxies with higher \fescabs\ are statistically the strongest \lya\ emitters (LAEs), that is, they show large equivalent widths \citet{FluryII}. Additionally, larger scatter at the high-\ewlya\ end at fixed \ewhi\ and \ewlis\ may result from boosted (or at least variations in) intrinsic \lya\ production.

\citet{Steidel2018} and \citet{Gazagnes2020} studied the connection between the \lya\ equivalent width and the \hi\ residual flux. In \autoref{fig:lya_HILIS} (top right) we show how their fitted linear relations compare to our data (gray and black lines). Both correlations are less steep than ours, but a more prominent difference exists when comparing with \citet{Steidel2018}. As discussed in \citet{Gazagnes2020}, sample selection is likely the key to understand these differences. \citet{Steidel2018} photometrically selected LBGs with rest-frame UV bands, meaning that they are generally brighter, and have higher masses and/or higher SFRs than the low-redshift galaxies studied by \citet{Gazagnes2020} and here. Another possible explanation is the scattering of the blue side of the \lya\ line by the IGM, which would produce lower \ewlya\ in the \citet{Steidel2018} sample. Additionally, flux aperture losses would also lead to a decrease in the \ewlya\ of the former sample with respect to the LzLCS.

Finally, important to note is the bias introduced by the redshifts of the two samples: from nearby $z \sim 0.3$ to distant $z \sim 3$ galaxies, with intrinsic different properties. After all, the current sample may include sources with larger \ewlya\ than the bulk of the LBGs at $z \sim 3$, whereas in \citet{Steidel2018}, sources were selected to have brighter continua.

The \lya\ and LIS connection has been widely studied by stacking LBGs at $z \sim 3-5$ \citep{Shapley2003, Jones2012, Du2018, Pahl2020} and also with individual detections at slightly lower redshifts \citep[][at $z \sim 2-3$]{Trainor2019, Du2021}. For comparison, we include the linear regression  between the \lya\ and LIS equivalent widths found by \citet{Du2018} (\autoref{fig:lya_HILIS}, bottom left), and data from stacked spectra in \citet{Shapley2003} and \citet{Jones2012}. High-$z$ LBG measurements seem to be more consistent with low-\fescabs\ or nonleaking galaxies: strong leakers occupy a different portion of this plot, with the highest \ewlya. Similar trends between the \lya\ and metal equivalent widths were found in \citet{Jaskot2019} for individual Green Pea-like galaxies. 

The \ewlya\ correlation with the metal residual flux is more difficult to interpret ---mainly because of the larger error bars on \rlis--- but it is still moderately significant. This result may suggest that the trend between \lya\ and LIS absorption equivalent width is predominantly due to variations in the covering fraction of neutral hydrogen (see below), which modulates the \lya\ profile and is directly connected with the LIS covering fraction (see Sect.\ \ref{sub:results_cfs}). The analysis of \citet{Gazagnes2020} indicates that the gas kinematics could also explain part of the observed scatter, as well as the distribution of different gas column densities in front of the stars. For comparison, we include the \ewlya--\rlis\ correlation from the \citet{Jones2013} compilation of $z\sim2-4$ galaxies, although the latter sample lacks strong LAEs, as opposed to ours, which explains the differences.

As the LIS absorption lines are a probe of neutral gas, there is a direct link between the \lya\ profile and LIS lines \citep{Shapley2003}. From the apparent nonevolution of the $\ewlya-\ewlis$ relation with redshift, \citet{Du2018} and \citet{Pahl2020} suggest that \lya\ and the LIS absorption features are fundamentally related, indicating that the gas giving rise to LIS absorption modulates the radiative transfer of \lya\ photons. We now know that, rather than metallicity, variations of the line covering fractions are the main cause for changes in $W_{\lambda}$. In order to examine the role of the gas covering fraction in \ewlya\ from the low-redshift sample, we again restrict our sample to the sources at a constant gas phase metallicity of $\oh = 8.2 \pm 0.1$ (25 galaxies). With this restriction, we find the same trends as in \autoref{fig:lya_HILIS}, confirming that the observed  $\ewlya-\ewlis$ relation is not primarily due to metallicity variations, but it is ultimately driven by variations in the \hi\ covering fraction (with the different column densities in front of the ionizing sources contributing to the scatter). Given the locus of the strongest leakers in the $\ewlya\--\ewlis$ plane, having large \lya\ equivalent widths but weak LIS lines, this diagram can be used to select potential LyC emitters in high-redshift surveys.

\subsection{HI and metals are not evenly distributed, but trace each other}\label{sub:results_cfs}
Although with different ionization potentials, it is well-known that \hi\ and LIS absorption lines in SF galaxies generally show similar kinematics \citep[line profiles, central velocities, etc.; see ][]{Mirka2010, Trainor2015, Chisholm2016, Reddy2016b}. This indicates that the LIS and neutral gas are comoving and trace similar gas content. Here, we proceed to study the relation between the observed neutral and metallic gas geometrical distributions through the covering fraction. 

\begin{figure}
    \centering
    \includegraphics[width=1.\columnwidth, page=11]{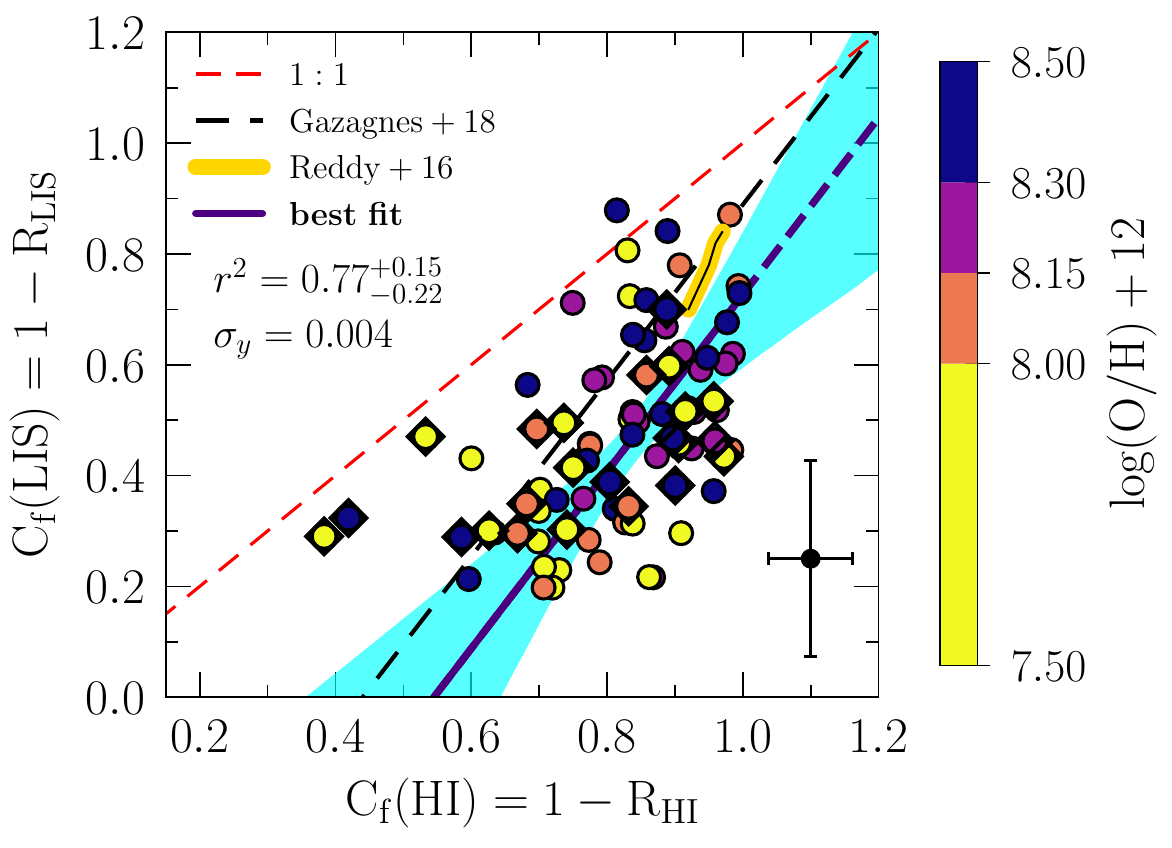}
\caption{Empirical correlation between the LIS and \hi\ covering fractions. Red, yellow, black, and dark blue lines correspond to the 1:1, \citet{Reddy2016b}, and \citet{Gazagnes2018} relations and our Bayesian linear fit (\textsc{linmix}), respectively. The points are color-coded according to the gas-phase metallicity of each source, \oh, and typical error bars are shown in the bottom-right corner. The blue shaded area represents the 2$\sigma$ confidence interval including the intrinsic scatter of the fit. The linear correlation coefficient ($r^2$) and intrinsic scatter over the fit ($\sigma_y$) are given. The \citet{Izotov16a, Izotov16b, Izotov18a, Izotov18b, Izotov2021} and \citet{Wang2019} sources are displayed in diamonds, and also included in the fit.}
\label{fig:cfHI_cfLIS}
\end{figure}

Covering fractions for every line were derived from the corresponding line residual fluxes using Eq.\ \ref{eq:cfA}, and applying Eq.\ \ref{eq:cf_meanHI} for \hi\ and similarly for the LIS lines. This assumes a uniform dust-screen geometry in the picket-fence model, a single velocity component, and saturated lines. In \autoref{fig:cfHI_cfLIS}, the LIS covering \cflis\  is compared to the neutral gas covering fraction obtained from the set of Lyman series lines \cfhi. Generally, the LIS covering fraction is lower, typically by a factor of two, compared to that of the \hi\ lines. The median covering fraction is 0.82 for \hi\ and 0.46 for the LIS. A linear relation between the two is clear from \autoref{fig:cfHI_cfLIS}, such that the LIS covering fraction increases with increasing \hi\ covering fraction. 

To quantify this, we fit a linear regression model to the data using the Bayesian fitting code \textsc{linmix}\footnote{A \textsc{python} version of the \textsc{linmix} package can be found in \url{https://linmix.readthedocs.io/en/latest/index.html}.} \citep{Kelly2007}, which accounts for errors on both variables, also including censored data (upper or lower limits). Further details are provided in Appendix\ \ref{appD}. We obtain
\begin{equation}
    \cfhi = (0.63 \pm 0.19) \times \cflis + (0.54 \pm 0.09).
    \label{eq:cfhi_cflis}
\end{equation}
The resulting fit is quite significant, with a linear correlation coefficient of $r^2=0.77~(+0.15,-0.22)$, and a negligible estimated intrinsic scatter of $\sigma_y=0.004$.

Our result is compared to the \citet{Gazagnes2018} fit, where a similar approach was followed but for a small subset of the LyC emitters (dashed black line). Both linear relations have similar slopes, but slightly different normalizations. Possible differences could be due to the use of the mean of different LIS lines ---in our case--- instead of single \siib\ line in the previous work, or due to the effect of the resolution,  which was corrected for by \citet{Gazagnes2018}. The measured residual fluxes can be affected by the spectral resolution, and therefore the actual linear relation also depends on the resolution. However, as discussed in Sect.\ \ref{sub:method_abs}, the resolution effect is subdominant in this sample. Our results are also compatible with those of \citet{Reddy2016b}, although spanning a wider range of $C_f$ values. As sketched in the previous paragraph, the strong leakers are located not only with the lowest \hi\ but also the lowest LIS covering fractions, having 0.77 (\hi) and 0.41 (LIS) as median values. 

Multiple explanations can lead to these covering fraction differences. As explained in \citet{Reddy2016b}, gas kinematics can influence the measured residual fluxes of metal lines. If most of the metals reside in narrow and unresolved absorption regions, the limited spectral resolution could underpredict the gas covering fraction estimated from the residual intensity of the lines. On the other hand, if the metal-enriched gas has a large velocity gradient in the line of sight, the derived covering fractions may be lower limits \citep[see discussion in][]{Mauerhofer2020}. Nevertheless, the similar blueshifted velocity and spread of \hi\ and LIS lines suggests that both phases are not separated, but coupled in the ISM, although once again, the limited spectral resolution does not allow us to confirm this definitely. Orientation effects for lines created in the circumgalactic medium (CGM) instead, where \hi\ and LIS are not particularly well-coupled, could also contribute to the scatter in the relation. Additionally, LIS lines could also be affected by scattering \citep[][in simulations]{Mauerhofer2020} and fluorescence of LIS emission in-filling \citep{Scarlata2015}, although this effect is very difficult to estimate.
 
\citet{Gazagnes2018} suggested that a different distribution of the metals, which would also depend on metallicity, could explain the current relation. In an ISM  where \hi\ is not fully mixed with the metals, \hi\ would probe high and low-metallicity regions, whereas metals would only probe the higher metallicity areas. Therefore, the \hi\ covering fraction would be higher than that of the metals. To test this scenario, we color-coded the points in \autoref{fig:cfHI_cfLIS} according to the 0.25, 0.5, and 0.75 gas-phase metallicity (\oh) quantiles. Overall, sources with low \hi\ and LIS covering fractions have lower metallicities, although the scatter and error bars blur this behavior. If we divide the sample into {high-} ($\oh > 8.1$) and {low-}metallicity ($\oh \le 8.1$) subsamples, we find average covering values of $\cfhi = 0.85~(0.78)$ and $\cflis = 0.52~(0.43)$ for high (low) metallicities, confirming the previous hypothesis. Nevertheless, high- and low-metallicity subsamples approximately follow the same linear relation predicted by Eq.\ \ref{eq:cfhi_cflis} 
(similar slopes and intercepts). This excludes a significant dependence of the previous $\cflis - \cfhi$ relation on metallicity, but as the measured metallicities are integrated over the whole galaxy, we cannot exclude the inhomogeneous mixing scenario.

Finally, there is still the possibility that the lower end of this $\cfhi-\cflis$ correlation is mainly driven by unsaturated lines. There is no simple way to address this question rather than the analysis already made in Appendix\ \ref{appB}. In there, we conclude that the number of unsaturated lines sources must be minimal, but without certainty for the LIS lines. For example, for galaxies with $\oh=7.8$ (among our lower metallicities) and assuming that \si\ roughly traces the \hi\ fraction, one has $\log({\rm Si/H}) \sim -5.7$, adopting the nearly constant\footnote{A constant Si/O ratio is expected because both Si and O are $\alpha$-elements sharing similar formation mechanisms.} observational relation of $\log({\rm Si/O}) \sim -1.5$  \citep[see][]{IzotovThuan1999}. According to a curve-of-growth analysis \citep{Draine}, the LIS lines are likely saturated at $\log (N_{\rm Si} / {\rm ~cm^{-2})} > 14$, roughly corresponding to $\log (N_{\rm \hi} / {\rm ~cm^{-2})} > 19.7$, which can be accomplished for most of our sources but not for all. Therefore, at neutral column densities below $\log (N_{\rm \hi} / {\rm ~cm^{-2})} \sim 19.7$, where \hi\ is still optically thick, some of the LIS lines could be optically thin. For these unsaturated metal lines, Eq.\ \ref{eq:cfA} no longer applies, and the covering fraction cannot be derived from the residual flux of the lines.

In any case, the \hi\ residual fluxes can still be roughly predicted from the LIS line residual fluxes using Eq.\ \ref{eq:cfhi_cflis}, a trend which is also found in simulations \citep{Mauerhofer2020}. This is important at high $z$, where the Lyman series is not easily observable due to the increase in IGM opacity, and so the \hi\ residual fluxes are not measurable and the \fescabs\ cannot be obtained from them. It should be finally considered that the relation between \hi\ and LIS residual fluxes (Eq.\ \ref{eq:cfhi_cflis}) may not be universal. However, so far the available data at high $z$ are compatible with our findings.

\begin{figure*}
    \centering
    \includegraphics[width=0.45\textwidth, page=15]{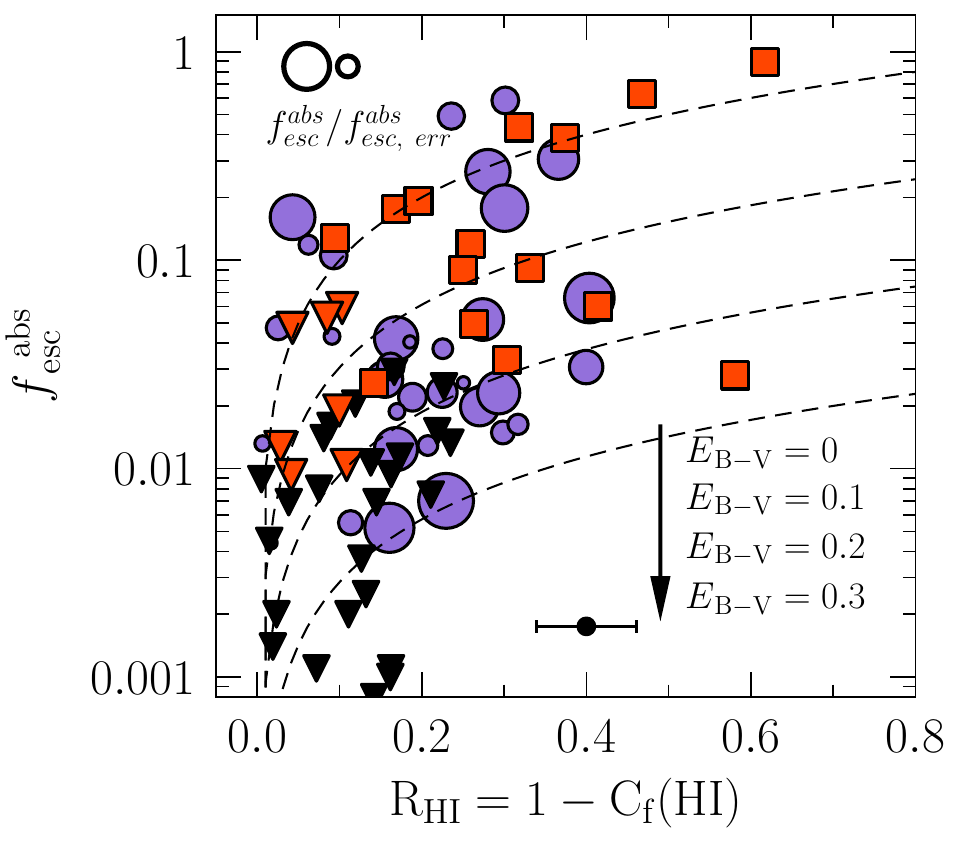}
    \hspace{0.5cm}
    \includegraphics[width=0.45\textwidth, page=16]{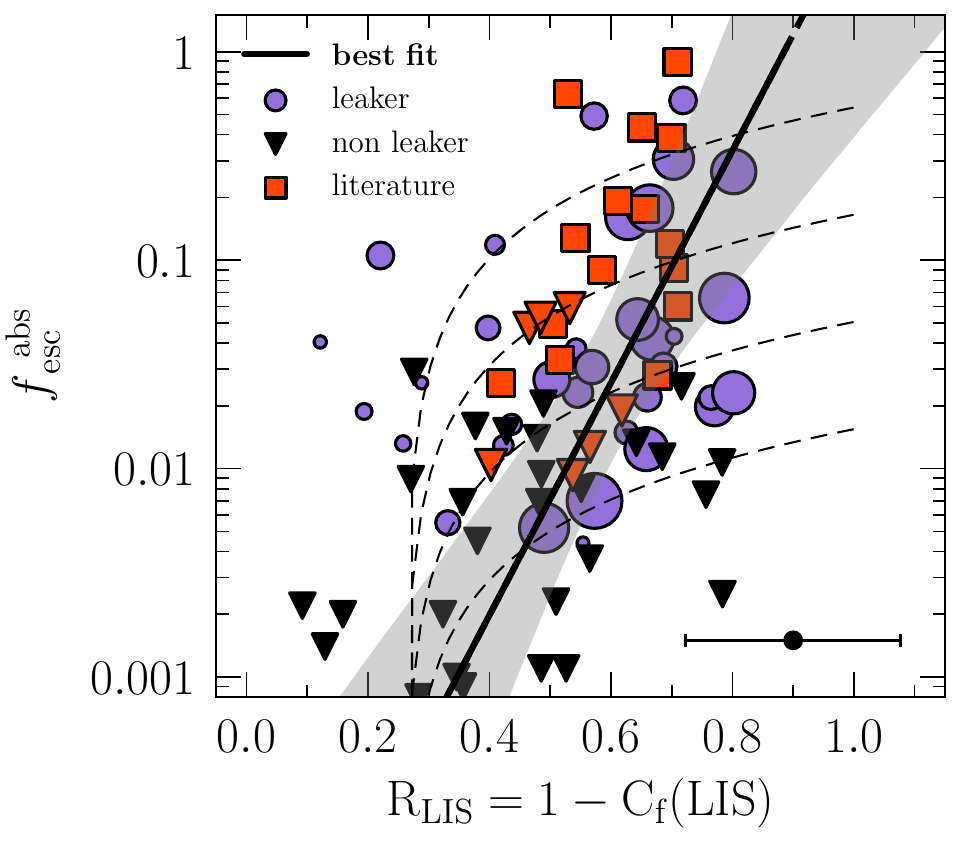}
\caption{Correlation between the LyC absolute photon escape fraction (\fescabs) and the \hi\ ({\em left}) and LIS line ({\em right}) residual fluxes. Purple circles represent the leakers, while the black triangles are the nondetections ---and for which the \fescabs\ is an upper limit. The size of the points corresponds to the relative uncertainty on the escape fraction ($\fescabs /f_{\rm esc}^{\rm abs,err}$), 
and the median 1$\sigma$ uncertainty of the residual flux is placed in the bottom-right corner of each panel. Red squares correspond to the \citet{Izotov16a, Izotov16b, Izotov18a, Izotov18b, Izotov2021} and \citet{Wang2019} galaxies (see legend). Overplotted are the curves predicted by Eq.\ \ref{eq:fesc_Ly} ({\em right}) and Eq.\ \ref{eq:fesc_LIS} ({\em left}) for $\ebv =$ 0, 0.1, 0.2, and 0.3 mag, respectively (dashed lines). The black solid line in the right panel corresponds to the best linear fit, and the gray-shaded area is the 2$\sigma$ confidence interval including intrinsic scatter (Eq.\ \ref{eq:fesc_RLIS}, see text).}
\label{fig:fesc_cfHILIS}
\end{figure*}

\begin{figure*}
    \centering
    \includegraphics[width=0.45\textwidth, page=12]{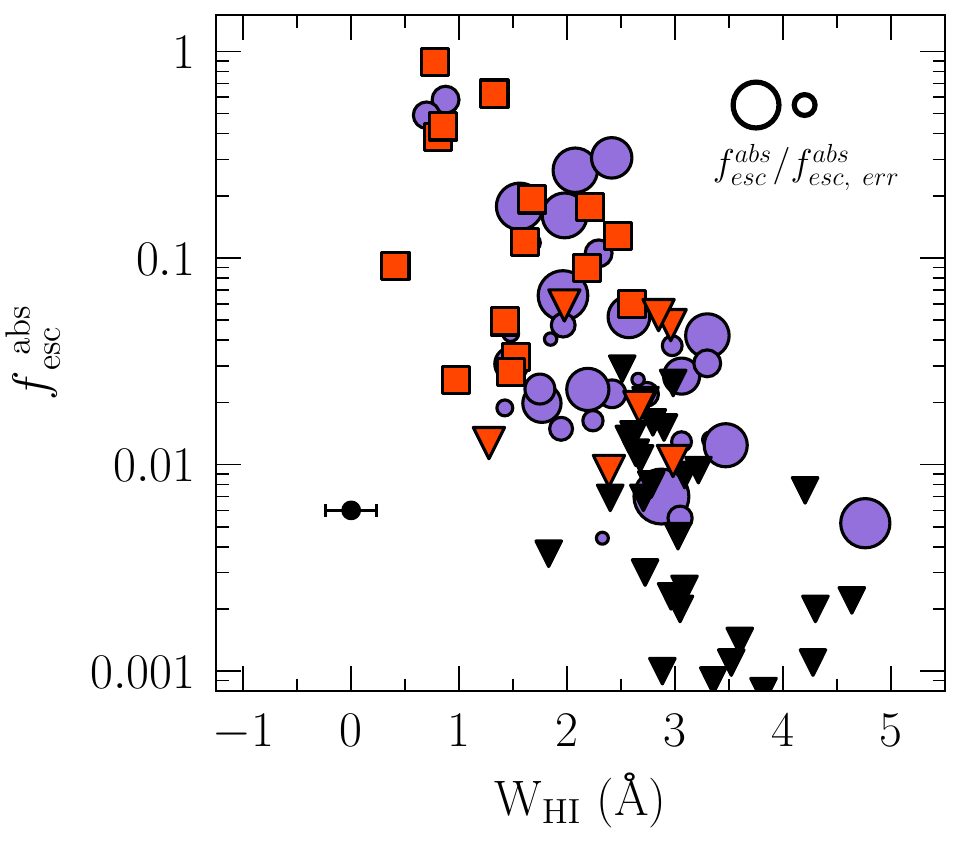}
    \hspace{0.5cm}
    \includegraphics[width=0.45\textwidth, page=13]{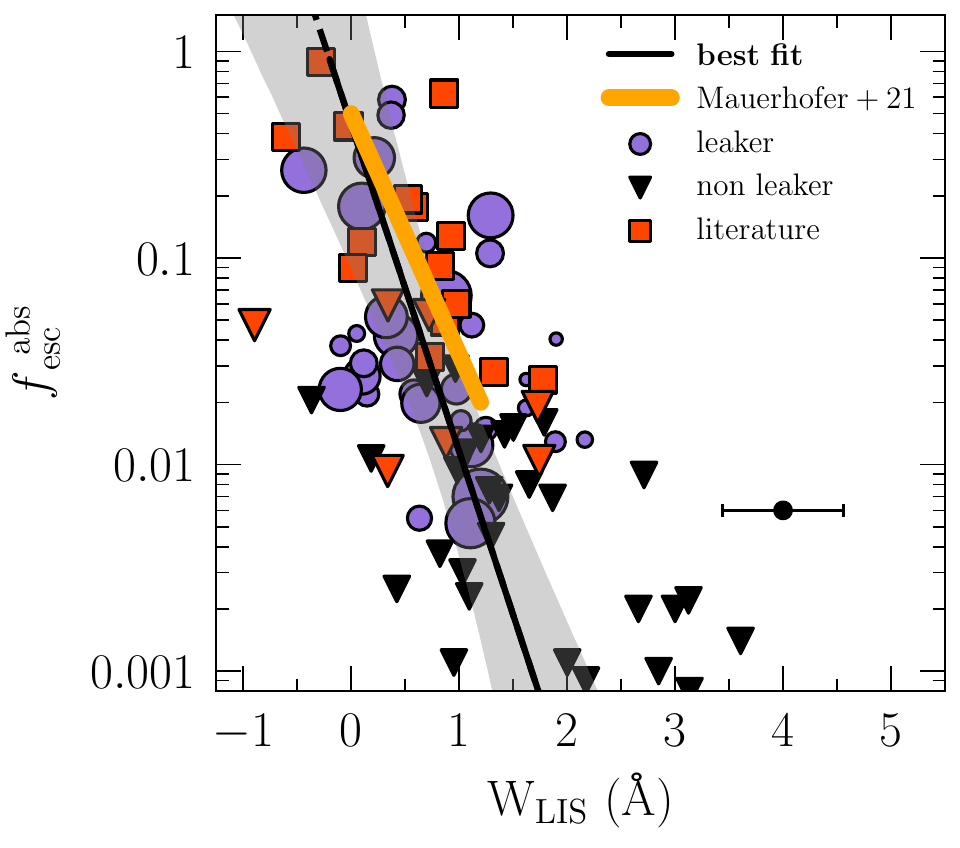}
\caption{The correlation between the LyC absolute photon escape fraction (\fescabs) and the \hi\ ({\em left}) and LIS lines ({\em right}) equivalent widths. The sizes and colors of the points have the same meaning as in \autoref{fig:fesc_cfHILIS}. The median 1$\sigma$ uncertainty on the equivalent width is placed at the bottom of every panel. The black solid line in the right panel corresponds to the best linear fit (Eq.\ \ref{eq:fesc_ewLIS}), and the gray-shaded area represents the 2$\sigma$ confidence interval, including intrinsic scatter. The upper limit prediction from \citet{Mauerhofer2020} using the \cii\ line is plotted in orange.}
\label{fig:fesc_ewHILIS}
\end{figure*}

\begin{figure}
    \centering
    \includegraphics[width=0.98\columnwidth, page=14]{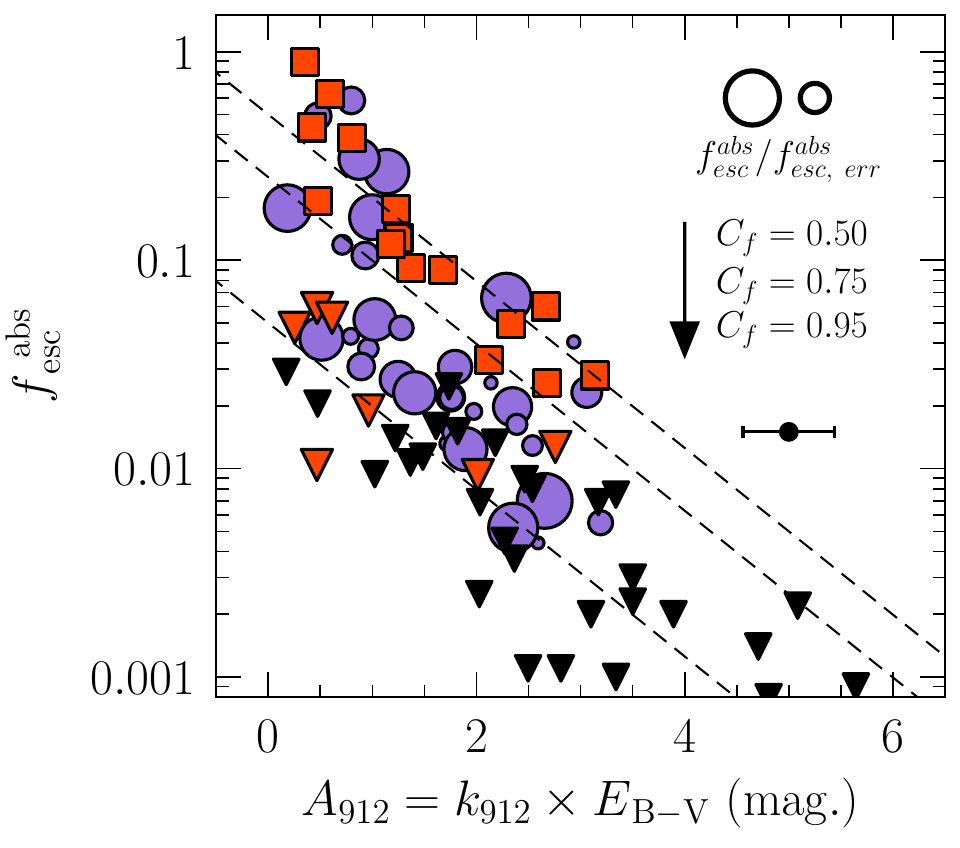}
\caption{Correlation between the LyC absolute photon escape fraction (\fescabs) and the UV dust-attenuation: $A_{\rm 912} = k_{\rm 912} \times \ebv$. Labeling is as in \autoref{fig:fesc_cfHILIS}. The median 1$\sigma$ uncertainty on \auv\ is plotted on the right of the plot. Overplotted are the three lines predicted by Eq.\ \ref{eq:fesc_Ly} for $\cfhi = 0.5, 0.75,$ and $0.95$, respectively (dashed lines), using $k_{912}=12.87$ (\citetalias{R16} law).}
\label{fig:fesc_EBV}
\end{figure}

\subsection{The porosity of the ISM favors the escape of LyC photons}\label{sub:results_cf_fesc}
As discussed in Sect.\ \ref{sub:method_abs}, \hi\ lines are likely saturated which means high neutral gas column densities of $\log (N_{\hi} / {\rm cm^{-2})} \gtrsim 15-16$. \citet{Gazagnes2018} measured the \hi\ column density of 6 out of 11 of our Izotov et al. leakers using the weak \ion{O}{i}$\lambda$1039 line and found column densities of $\log (N_{\hi} / {\rm cm^{-2})} \gtrsim 18$, which means that the neutral gas medium is also opaque to the LyC photons. Therefore, a density-bounded leaking scenario as proposed by \citet{Zackarisson2013, JaskotOey2013, Nakajima2014}, for example, can be excluded in principle. In contrast, the existence of saturated but nonblack \hi\ lines suggests a nonunity \hi\ covering fraction, where the radiation can escape through holes in the ISM, although these channels may not be free of gas and dust, and therefore may not be fully optically thin \citep{Gazagnes2018, Steidel2018}. Such a configuration can also explain the high ionization parameters and other properties associated with Green Pea galaxies \citep{Jaskot2019}, and is needed to explain the observed emission line ratios of LyC emitters \citep{Ramambason2020}.

The relationship ---already seen in previous figures--- between the LyC absolute escape fraction (\fescabs) and the residual flux, \rhi\ or \rlis, is shown in \autoref{fig:fesc_cfHILIS}. We performed a survival {\em Kendall} correlation analysis in order to take into account the upper limits on the escape fraction and the lower limits on the residual fluxes \citep[see][]{ATS1996, FluryII}. A strong ($\tau = -0.358$) and significant correlation ($p_{val.} = 1.428 \times 10^{-6}$) between \fescabs\ and \rhi\ is  found (see \autoref{tab:kendall_fesc}). The strongest leakers tend to have the highest residual fluxes or, conversely, the lowest \hi\ covering fractions. Such behavior is expected from the picket-fence model, where the neutral gas is distributed in clumps around the light sources.

Clearly, \hi\ residual flux overestimates the photon escape in most of the cases: $\fescabs \leq 1-\cfhi$. This is expected, because the presence of dust needs to be accounted for when inferring the \fescabs\ \citep{Gazagnes2018, Chisholm2018, Steidel2018}. In the same figure, we plot the predictions of the Eq.\ \ref{eq:fesc_Ly} for \ebv\ = 0, 0.1, 0.2, and 0.3 (\citetalias{R16} law). Visually, one can see how our data points follow the curves on top given by the picket-fence scenario (uniform dust-screen). At a given \rhi, variations in the LyC photon escape can be explained by different amounts of dust reddening.

The LyC photon escape fraction behaves similarly to \rlis\ ($\fescabs < 1-\cflis$), and shows a strong correlation too (\autoref{fig:fesc_cfHILIS}, $\tau=0.236, p_{\rm val.}=1.457 \times 10^{-3}$), where leakers have higher residual LIS flux than nonleakers. The differences between the two plots is principally introduced by the $\cfhi-\cflis$ relation (Sect.\ \ref{sub:results_cfs}). There are, on the other hand, several possible scenarios in which the escape of ionizing photons could differ from the measured residual flux. First, as soon as the ISM conditions move from optically thick (saturated lines) to optically thin regimes (linear), the optical depths of the gas clouds are close to one, and so the covering fraction is not well-defined. The presence of residual gas and dust in the low column density channels can also contribute to the same effect \citep[see discussion in][]{Gazagnes2020}. Secondly, the distribution of sources emitting ionizing photons and nonionizing UV continuum may not be the same.  From detailed radiation transfer simulations, \citet{Mauerhofer2020} found that around 15\%-30$\%$ of the light at for example 1020\AA\ is produced by stars that do not contribute to the budget of ionizing radiation. Hence, these continuum photons would be affected by screens of gas and/or dust that do not influence the LyC escape fraction. Thirdly, as already mentioned in the previous section, the gas kinematics and turbulence can spread the absorption line profiles, increasing the residual fluxes and causing the covering fractions to be underestimated \citep[also][]{Mauerhofer2020}. All these reasons would contribute to the scatter and the decreased ability to infer the \fescabs\ of galaxies (see Sect.\ \ref{sub:discussion_fescs}) using the \hi/LIS lines. More generally, a relation between the LyC escape and the residual fluxes of the  lines, interpreted as the escape fractions of the lines in question, can always be expected \citep{Mauerhofer2020}, even when \fescabs\ cannot be analytically predicted via the picket-fence model (see Sect.\ \ref{sub:caveats}).

Given the strong correlation between \fescabs\ and the residual line flux, we can also expect correlations with \hi\ or LIS equivalent widths. These relations are displayed in \autoref{fig:fesc_ewHILIS}. The \ewhi\ is significantly scattered (left panel), mainly driven by the nonleaker population ($\tau=-0.425, p_{val.}=1.082 \times 10^{-8}$). On the LIS side (right panel), a tight correlation between the \fescabs\ and the \ewlis\ exists, with $\tau=-0.331$ and a $p-$value of $p_{\rm val.}=8.273 \times 10^{-6}$.

It is of great interest to derive empirical relations using the LIS lines in order to predict the LyC escape fraction of galaxies. Nonetheless, caution is required, and all the previous limitations of the method and the current scatter in the correlations must be considered (see \autoref{fig:fesc_cfHILIS} and \ref{fig:fesc_ewHILIS}). For the LIS lines, a Bayesian linear fit to the data in log-linear space was performed using \textsc{linmix} \citep{Kelly2007}, which provides a proper treatment of the \fescabs\ upper limits. The resulting relation between the absolute LyC escape fraction and the LIS residual flux is
\begin{equation}
    \log \fescabs = (-4.93 \pm 0.65) + (5.57 \pm 1.11) \times \rlis, 
    \label{eq:fesc_RLIS}
\end{equation}

\noindent where the linear correlation coefficient of the fit is $r^2=0.69~(+0.18,-0.21)$, with an important intrinsic scatter of $\sigma_y=0.16$. For the \fescabs\ to \ewlis\ relation, and applying once again the \textsc{linmix} code to the log-linear data space, we get
\begin{equation}
    \log \fescabs = (-0.35 \pm 0.32) - (1.58 \pm 0.34) \times \ewlis / \AA,
    \label{eq:fesc_ewLIS}
\end{equation}

\noindent with $r^2=0.88~(+0.06,-0.09)$ and $\sigma_y=0.18$. In \autoref{fig:fesc_ewHILIS} (right), our results are compared to the ones derived by \citet{Mauerhofer2020} using the \cii\ line. These latter authors simulated a low-mass high-$z$ galaxy and generated mock \cii\ observations along different sight lines. The \cii\ equivalent width ($W_{\ion{C}{II}}$) was then computed together with the escape fraction for every line of sight. \citet{Mauerhofer2020} concluded that the equivalent width can be treated as an upper limit to the absolute photon escape as: $\fescabs < 0.5 - 0.4 \times W_{\ion{C}{ii}}$. Compared to our linear fit, we can see that both results are similar, although the \citet{Mauerhofer2020} relation follows a slightly different slope. In the latter work, the absorption line measurements were carried out in a similar way to that described in the present paper but only for a single line. Also, the shift could be due to the modeling of the lines, the dust location within the simulated galaxy (the density of dust is proportional to the metallicity located in the highest density \hi\ clouds), or the different intrinsic properties (SFR, ionizing emissivity, metallicity, etc.) between our low-$z$ sample and the $z \sim 3$ simulated galaxy. {\em Kendall} survival statistics quantifying the correlations of \fescabs\ versus \hi/LIS equivalent widths and residual fluxes are computed and presented in \autoref{tab:kendall_fesc}.

\subsection{Dust attenuation and LyC leakage}\label{sub:results_ebv_fesc}
The impact of dust extinction on the escape of ionizing photons is not well understood. Based on simulations, \citet{Kimm2019} and \citet{Mauerhofer2020} suggested that the absorption due to dust is subdominant compared to that due to neutral hydrogen. Physically, LyC photons are destroyed in the same way as nonionizing photons at slightly longer wavelengths, but if the dust resides only in regions of  high neutral gas density (usually assumed in the simulations), the high ionization cross-section and large abundance of the hydrogen atom is such that \hi\ likely dominates the absorption of LyC photons in galaxies more than dust does. However, if the dust is still present in the ionized regions or channels (as in the uniform dust-screen scenario), LyC photons would significantly attenuate before escaping the galaxy.

\citet{Chisholm2018} observationally demonstrated that dust needs to be accounted for in order to infer the absolute escape from the Lyman series. It is essential to correct for the UV attenuation in order to derive the intrinsic LyC photon production, which yields the absolute escape fraction by comparison to the observed LyC flux. Therefore, the ``effective'' UV attenuation \auv\ enters any derivation of the {absolute} escape fraction, whereas the {relative} escape fraction only accounts for the differential attenuation between 912\AA\ and nonionizing flux \citep[see e.g.,][]{Siana2015, Steidel2018}.

To investigate the role of the dust in the escape fraction in the LzLCS sample, in \autoref{fig:fesc_EBV} we plot \fescabs\ against the dust attenuation at 912\AA: $A_{\rm 912} = k_{\rm 912} \times \ebv$ (\citetalias{R16}). A strong correlation of $\tau=-0.367$ is found between the two variables ($p_{val.}=7.624\times 10^{-7}$). Most of the sources are below the $10^{-0.4 \auv}$ curve (i.e., $\fescabs \leq 10^{-0.4 \auv}, \cfhi=0$ following Eq.\ \ref{eq:fesc_Ly}) which indicates, as mentioned in the previous section, that the covering fraction and dust both overestimate the LyC photon escape of galaxies when considered independently \citep{Heckman2001, Heckman2011, Chisholm2018, Gazagnes2020}.

In \autoref{fig:fesc_EBV}, the results given by Eq.\ \ref{eq:fesc_Ly} for $\cfhi = 0.50, 0.75,$ and $0.95$ are also shown. Clearly, the spread in the \fescabs--dust plane can be explained by joint changes of \ebv\ and the \hi\ covering fraction of galaxies. We use this expression below to predict the \fescabs\ values of the LzLCS and high-$z$ galaxies. Simultaneously considering the neutral gas covering (given by the depths of the Lyman lines) and the dust attenuation (from the UV spectral fits) provides the most accurate photon escape prediction of this work. 

We note that a correlation between \fescabs\ and \ebv\ is expected because the \fescabs\ values come from de-reddening the S99 models (see Eq.\ \ref{eq:s99_fit} and \ref{eq:fescS99}) and comparing the predicted LyC flux to the observed one. Ideally, an independent measurement of the UV attenuation, which is here obtained from fitting the UV spectra, would test if this correlation is indeed as strong as shown here. We also note that the exact value of the UV attenuation (or \ebv) depends in principle on the assumed geometry and the attenuation law, for which we have only considered two simple cases (see Sect.\ \ref{sec:fit_results}). Probably the most direct verification of the robustness of these relations comes from the comparison of the absolute escape fractions derived using two independent methods, namely the one presented here, which uses the stellar UV spectrum, and the method used by \citet{FluryI} \citep[see also][]{Izotov16b}. In the latter study, \fescabs\ is computed from the observed LyC flux and the extinction-corrected H$\beta$ flux, together with H$\beta$ equivalent width which is used as an indicator of the age of the underlying ionizing stellar population. As shown in \citet{FluryI}, the overall agreement between the two methods is quite good, verifying the strong correlation that we see here between the escape and the dust-attenuation.

Finally, it is worth mentioning that the correlations between the absolute LyC escape fraction, residual fluxes, equivalent widths, and the UV attenuation shown in \autoref{fig:fesc_cfHILIS}, \ref{fig:fesc_ewHILIS}, and \ref{fig:fesc_EBV} also hold when one considers the observed flux ratio $f_{LyC}/f_{1100}$ instead of \fescabs. In particular, the fact that both $f_{LyC}/f_{1100}$ and \fescabs\ correlate with the UV attenuation indicates that dust is primarily in the clouds, as shown in Sect.\ \ref{sub:results_HILIS} (see also Sect.\ \ref{sub:caveats}), and hence the amount of attenuation increases with the covering fraction.

For easier comparison with future studies, and to better illustrate the behavior of the LyC escape with the previous parameters, a summary plot of \autoref{fig:fesc_cfHILIS}, \ref{fig:fesc_ewHILIS} and \ref{fig:fesc_EBV} can be seen in Appendix\ \ref{appE} (\autoref{fig:summary_plots}). To this end, we computed the median value of \rhi, \rlis, \ewhi, \ewlis\ and \ebv\ in three \fescabs\ intervals, specifically at $\fescabs \in [0-5\%), [5-10\%)$, and $\boldsymbol[10-100\%)$. \autoref{tab:summary_table} emphasizes these results, and a quantitative analysis of these trends, comprising the sample median values, will be further investigated in future LzLCS publications using stacked spectra.

\begin{table}
\centering
\caption{Survival {\em Kendall} ($\tau$) correlation test$^1$ results for \fescabs\ versus different galaxy properties$^2$.}
\begin{tabular}{ccc}
\toprule
\multicolumn{1}{c}{} & \multicolumn{2}{c}{\fescabs} \\
\midrule
 & $\tau$ & $p_{val}$ \\ 
\cmidrule{2-3} 
\rhi\ & $+0.358$ & $1.428 \times 10^{-6}$ \\
\rlis\ & $+0.236$ & $1.457 \times 10^{-3}$ \\
\ewhi\ & $-0.425$ & $1.082 \times 10^{-8}$ \\
\ewlis\ & $-0.331$ & $8.273 \times 10^{-6}$ \\
\auv\ & $-0.367$ & $7.624 \times 10^{-7}$ \\
\bottomrule
\end{tabular}
\label{tab:kendall_fesc}
\tablefoot{
$^1$See \citet{ATS1996}.
$^2$ \hi\ and LIS equivalent widths (\ewhi, \ewlis), \hi\ and LIS line residual fluxes (\rhi, \rlis) and UV attenuation (\auv) accounting for censored data (upper or lower limits).
}
\end{table}

\section{Discussion}\label{sec:discussion}
Having established relations between the UV absorption lines and LyC escape, we now examine the power of these lines to predict the ionizing photon escape fraction of galaxies (\ref{sub:discussion_fescs}). We also present a physical picture of the ISM of low-$z$ LyC emitters based on our UV absorption lines and \fescabs\ results as well as earlier results in the literature (\ref{sub:discussion_ism}). We finally discuss some implications and comparisons for the high-$z$ Universe (\ref{sub:discussion_highz}), and the caveats of our modeling approach (\ref{sub:caveats}).

\subsection{Predicting the Lyman-continuum photon escape with the Lyman series and LIS lines}\label{sub:discussion_fescs}
We want to test whether the \hi\ and, more particularly, measurements of the LIS lines could be used to effectively predict the \fescabs\ of galaxies. Under the assumption of a foreground screen of dust in the picket-fence model, \citet{Chisholm2018} determined that the parameters that contribute the most to predicting the photon escape are the dust attenuation and the \hi\ covering fraction, implying that the column density negligibly affects the line depths (if the lines are saturated). Therefore, the accurate determination of both \ebv\ and \cfhi\ is necessary. In this section, we carry out a similar exercise as in \citet{Chisholm2018} and \citet{Gazagnes2020}, but taking advantage of our larger LzLCS data set, and thus improving the robustness of the method and providing more statistical meaning. First, we take the derived \ebv\ from our SED fits as well as the weighted-averaged \hi\ covering fractions, \cfhi. The predicted photon escape from the \hi\ lines is computed from Eq.\ \ref{eq:fesc_Ly}, \feschi.

\begin{figure}
    \centering 
    \includegraphics[width=0.9\columnwidth,  page=17]{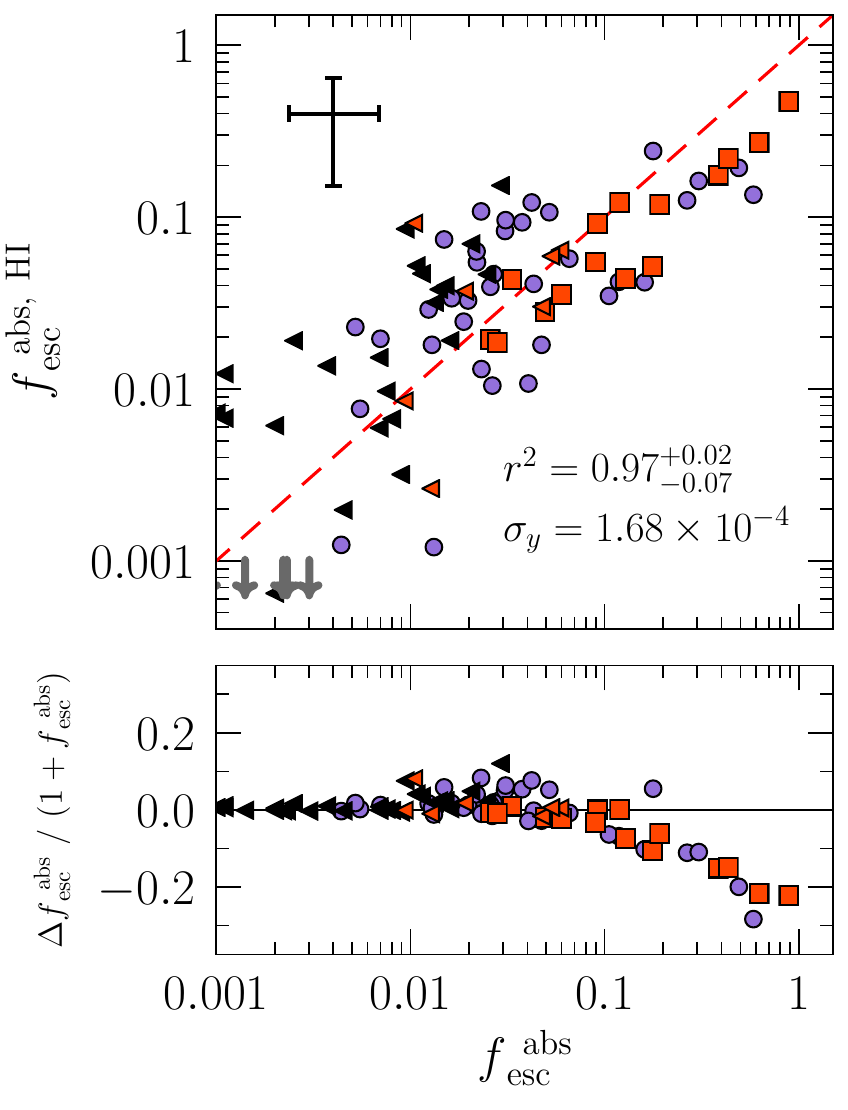}
    \includegraphics[width=0.9\columnwidth,  page=18]{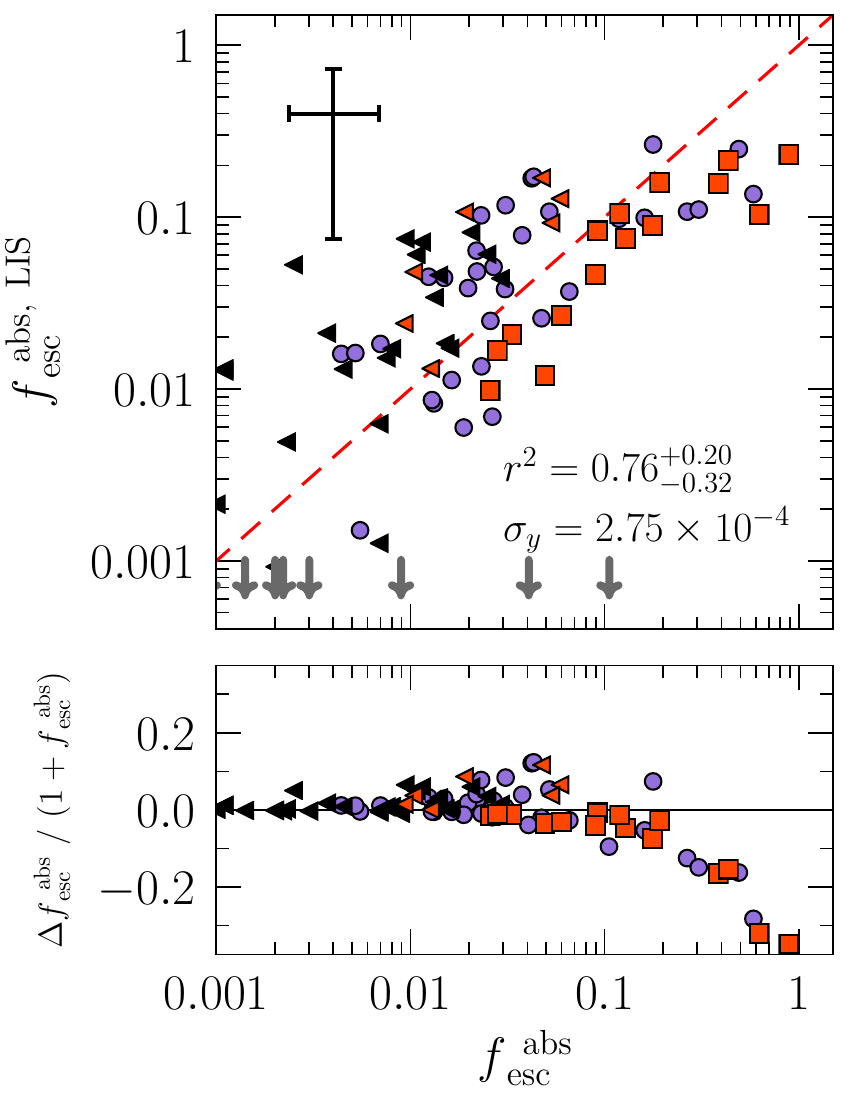}
\caption{Comparison of predictive methods for the LyC absolute escape fraction. \emph{Top:}  Lyman series photon escape (Eq.\ \ref{eq:fesc_Ly}) against the fiducial S99-derived values. \emph{Bottom:} Same as top panel but for the photon escape inferred using the LIS line covering fraction (Eq.\ \ref{eq:cfhi_cflis}). Purple points correspond to the LzLCS leakers while the red squares are the \citet{Izotov16a, Izotov16b, Izotov18a, Izotov18b, Izotov2021} and \citet{Wang2019} galaxies. Black triangles correspond to nonleakers. Median relative error bars are shown in the upper right corners of each panel. The subplots at the bottom illustrate the overall accuracy of the predicted \fescabs\ (i.e., $\Delta \fescabs / (1 + \fescabs)$) as a function of the fiducial \fescabs .}
\label{fig:fesc_comparison}
\end{figure}

Thus, the predicted \feschi\ value is compared to the fiducial value derived from the spectral UV fits (Sect.\ \ref{sub:method_fesc}) in \autoref{fig:fesc_comparison} (top). \textsc{linmix} determines a clear correlation between the two, with a correlation coefficient of $r^2=0.97~(+0.02,-0.07)$. The intrinsic scatter is negligible ($\sigma_y \sim 10^{-4}$), which means that the errors on the variables can fully explain the scatter in the 1:1 relation. The median of the predicted \fescabs\ for leakers does not change with respect to the fiducial \fescabs\ distribution, although a small systematic shift to higher predicted escape values exists for the $\fescabs < 10\%$ distribution, where a significant fraction of the sources are upper limits in \fescabs. In conclusion, \fescabs\ can be well recovered from the Lyman series and dust attenuation. 

In any case, our original purpose is to decipher whether or not the \fescabs\ can be predicted using the LIS lines. To approach this, we first take advantage of the intrinsic relation between the residual flux (or covering factor) of \hi\ and metals, which is empirically given by our Eq.\ \ref{eq:cfhi_cflis}. The \hi\ is then computed by substituting every measured \cflis\ in that expression, and then the predicted photon escape from LIS, \fesclis, is determined with a formula similar to that in Eq.\ \ref{eq:fesc_Ly}: 
\begin{equation}
    \fesclis = 10^{-0.4k_{912}\ebv} \times (1 - [a \times \cflis + b]),
    \label{eq:fesc_LIS}
\end{equation}

\noindent where the coefficients [$a, b$] $=$ [$(0.63 \pm 0.19), (0.54 \pm 0.09)$] are given by Eq.\ \ref{eq:cfhi_cflis}. The results are plotted against the fiducial \fescabs\ in \autoref{fig:fesc_comparison} (bottom). The gray arrows point towards $ f_{\rm esc}^{\rm abs, ~LIS} \approx 0$, for those sources with $\cflis \approx 1$ (zero residual flux) and/or very high \ebv, falling below the axis limits. Still, a strong linear correlation is present when using the LIS lines ($r^2=0.76~(+0.20,-0.32)$), with no significant intrinsic scatter.

As discussed earlier (Sect.\ \ref{sub:results_cf_fesc}), several mechanisms can lead to disagreement between the {real} covering fraction and the covering derived using the residual flux of the lines: ISM regions close to the optically thin limit, the mismatch between the distribution of UV ionizing and nonionizing sources, the gas kinematics and turbulence, and the ISM geometry can  all potentially contribute to the scatter in the relations shown in \autoref{fig:fesc_comparison} \citep[see][]{Mauerhofer2020}. 

A more detailed look into \autoref{fig:fesc_comparison} reveals that for the highest $\fescabs \geq 0.1$ sources, the Lyman series and low-ionization covering methods systematically underestimate the photon escape. The effect is more noticeable when one plots the overall accuracy of the predicted \fescabs\ ($\Delta \fescabs / (1 + \fescabs)$) as a function of the fiducial \fescabs\ itself (\autoref{fig:fesc_comparison}, bottom panels). This result, which was already noted by \citet{Gazagnes2020} and is supported by hydrodynamical simulations by \citet{Kakiichi2021}, suggests that the leakage could be partially due to a density-bounded ISM, where the LyC photons can escape through optically thin $\log (N_{\hi}) ~{\rm cm^{-2} \lesssim 10^{17.2}}$ regions to the LyC. 

Overall, we propose three different approaches to predict the LyC absolute photon escape from the LIS lines, and demonstrate the limitations on each: (1) using the residual flux of the absorption lines (Eq.\ \ref{eq:fesc_RLIS}), (2) their equivalent widths (Eq.\ \ref{eq:fesc_ewLIS}), and (3) a combination of residual flux and dust attenuation (Eq.\ \ref{eq:fesc_LIS}). When compared directly to \fescabs, the third method is ideally recommended in all the cases, as it presents the most stringent correlation coefficient of the fit and the least intrinsic scatter.

A similar methodology to (3) allowed \citet{Chisholm2018} to predict the \fescabs\ of five high-$z$ galaxies detected in LyC, and without Lyman-series observations. In particular, for the Cosmic Horseshoe \citep{Belokurov2007} at $z = 2.38$, these authors predicted a \fescabs\ of around 1\%, in agreement with the $\fescabs < 2\%$ limit imposed by HST imaging \citep{Vasei2016}. 

At high $z$, low-mass galaxies are characterized by hosting a metal- and dust-poor ISM similar to our low-$z$ analogs, where $\ebv < 0.1$ at $z \sim 3-6$ \citep{Schaerer2013}. Taking this attenuation as representative, and using the median \hi\ and metal LIS covering fraction for the leakers in our sample (0.70 and 0.38, respectively), this gives $\fescabs \geq 0.07$ for \hi\ and LIS predicted values. This is below but similar to the $10\%$ theoretical lower limit for faint SF galaxies to dominate reionization at $z \sim 6-9$ \citep{Robertson2013}\footnote{ The reader may notice here that various samples have different $M_{UV}$ ranges. For instance, \citet{Robertson2013} consider $M_{\rm{UV}} \lesssim -15$ for $z \sim 7$ galaxies, while our sample spans $-18 \leq M_{\rm{UV}} < -21$ at $z \sim 0.3$.}, and is in good agreement with the $9\%$ ($6\%$) average value derived in \citet{Steidel2018} \citep{Pahl2021} at the same redshifts; and also \citet{Rosdahl2018} in simulations. However, it should also be stressed that the predictions for \fescabs\ given by Eq.\ \ref{eq:fesc_LIS} are only reliable in a statistical sense. Given the scatter in the above-mentioned relations, predictions for any individual object are not very reliable.

We finally point out that higher resolution and higher S/N UV spectra would help to obtain a better measurement of the covering fractions and ionic column densities (fundamental to disentangling the role of the ISM structure and gas kinematics in LyC leakage), which would improve the \fescabs\ determination from the SED modeling (subject to quite large uncertainties in some cases), and to better constrain the stellar population and dust properties (\ebv).

\subsection{The absorbing ISM of LyC emitters}\label{sub:discussion_ism}
Low-metallicity (subsolar) galaxies typically have lower masses ($\lesssim 10^{10} {\rm M_{\odot}}$), higher gas fractions, and less dust than the typical local SF galaxy \citep{Trebitsch2017}. Low-mass systems have lower gravitational potentials, which increases the efficiency of stellar feedback and leads to more bursty star-formation episodes \citep{Muratov2015}. The low-metallicity and burstiness of their star-formation history means that these sources also have high ionizing production efficiencies (intrinsic number of ionizing photons per UV luminosity or mass, equivalently) and moderate ionizing escape fractions \citep{Ma2020}. This, together with their higher number density at all epochs (with respect to more massive galaxies), makes them the most promising analogs of the sources of cosmic reionization \citep{Finkelstein2019}.

Different mechanisms can remove the gas and dust from such galaxies, creating holes in the ISM with appreciably lower gas and dust column densities. Our observations suggest that LyC photons primarily escape through these channels. These holes are likely created by stellar winds and supernova explosions after starburst episodes, or they are photoionized channels induced by turbulence during the starburst period \citep{Kakiichi2021}. They could also be embedded in an intrinsically low-density ISM, which results from catastrophic cooling effects around low-metallicity star clusters \citep{Jaskot2019}. 

The picket-fence model describes an ionization-bounded galaxy \citep{Zackarisson2013, Vasei2016, Reddy2016b} where the ionizing photons can partially escape through these holes in the ISM. The fraction of sight lines covered by optically thick neutral gas clouds is the covering fraction. We now summarize a scenario where the \hi\ covering fraction, \cfhi, can explain most of the observed empirical trends found in the present study.

The presence of (likely) saturated but nonblack \hi\ and LIS absorption lines in our spectra (Sect.\ \ref{sub:results_HILIS}) implies the existence of low-column-density channels, and therefore a nonzero covering fraction for most of the galaxies. Measurements of the \hi\ column density of Green-Pea galaxies lead to similar conclusions in \citet{Gazagnes2018} and \citet{McKinney2019}, and in the recent, detailed study of the local galaxy Haro11 by \citet{Ostlin2021}. 

The LzLCS sample reveals many strong correlations between the UV absorption lines and other secondary galaxy parameters. We find that the measured \hi\ and LIS equivalent widths scale with (1) the intrinsic absolute magnitudes at 1500\AA, \mfuv, and with (2) the UV attenuation, \auv\ (or \ebv, equivalently), and scale inversely with  (3) the \hi\ and LIS residual fluxes (\autoref{fig:ewHI_props} and \ref{fig:ewLIS_props}). We also demonstrated that metallicity plays a minor role in these relations.

The inverse connection between the equivalent width of the lines and the residual flux (3) naturally comes from the definition of a saturated absorption line profile \citep{Draine}. In contrast, the $W_{\lambda}-\auv$ relation (2) can be explained by invoking a clumpy gas-to-dust geometry \citep{Shapley2003, Steidel2018, Du2018, Pahl2020, Du2021}, where the total reddening is related to the fraction of the total sight lines with optically thick neutral gas \citep{Vasei2016, Reddy2016b}. A galaxy with a larger fraction of sight lines covered by optically thick neutral gas (a higher \cfhi) also encounters more dust and the spectrum is more attenuated (\auv\ increases), because the dust resides in the same gas. In other words, the same gas which gives rise to \ewlis\ also reddens the continuum \citep{Jones2012}. 

Aside from the $W_{\lambda}-\auv$ connection, there are other lines of evidences for a clumpy gas-to-dust geometry in our sample. For instance, in a clumpy geometry, transitions for the same ion at different wavelengths will have different residual fluxes \citep[see Eq.\ \ref{eq:cfB} and][]{Gazagnes2018}. We tentatively see this trend when looking at the median residual flux of every Lyman line in the LzLCS. Values vary by as much as 0.71, 0.81, 0.83, and 0.89 from Ly5 to Ly$\beta$, but typical errors of $\sim 0.13$ obscure robust conclusions. The hypothesis of a clumpy distribution of gas and dust has been widely proposed in the literature, and it has special relevance for example when trying to explain the observed Ly$\alpha$-to-H$\alpha$ ratios in SF galaxies \citep[see][and references therein]{Scarlata2009, Jaskot2019}, and the \lya\ peak-velocity separation \citep{Gronke2016, Orlitova2018, Jaskot2019}.

When \lya\ varies from weak absorption to strong emission, the strengths of the \hi\ and LIS absorption lines decrease \citep[\autoref{fig:lya_HILIS} in Sect.\ \ref{sub:results_Lya}, see also][]{Shapley2003, Jones2012, RiberaThorsen2015, Du2018, Trainor2019, Pahl2020, Du2021}. Invoking our picket-fence model, the higher the covering fraction, the higher the fraction of \lya\ photons that are resonantly scattered out of the line of sight and absorbed by dust, resulting in a weaker \lya\ emission with stronger \hi\ and LIS profiles \citep{Steidel2018, Gazagnes2020}. Similarly, using $z \sim 3-5$ composites, \citet{Jones2012} demonstrated that other host properties, such as the luminosity, have little effect on the $\ewlya - \ewlis$ variations. The relationships between $\ewlya-\ewlis$ and $\ewlya-\auv$ are presumably invariant across cosmic time, as both are affected by variations in the same \hi\ covering \citep{Du2018}. This also suggests that the \lya\ production efficiency does not change drastically with redshift \citep{Trainor2019, Pahl2020}. 

Moreover, the existence of double- or triple-peaked \lya\ profiles in leakers \citep{Verhamme2017, Izotov2020, Izotov2021} can be associated with the characteristic bimodal distribution of \hi\ column densities in the patchy ISM \citep{Verhamme2015, Mauerhofer2020, Kakiichi2021}, or the presence of completely ionized channels \citep{Ribera-Thorsen2017}. This may lead to the observed correlations between the escape of \lya\ and LyC photons \citep[e.g.,][]{Verhamme2017,Marchi2018Lyalpha-Lyman-c,Steidel2018,Gazagnes2020,Pahl2021}.

\begin{figure}
    \centering
    \includegraphics[width=0.98\columnwidth, page=19]{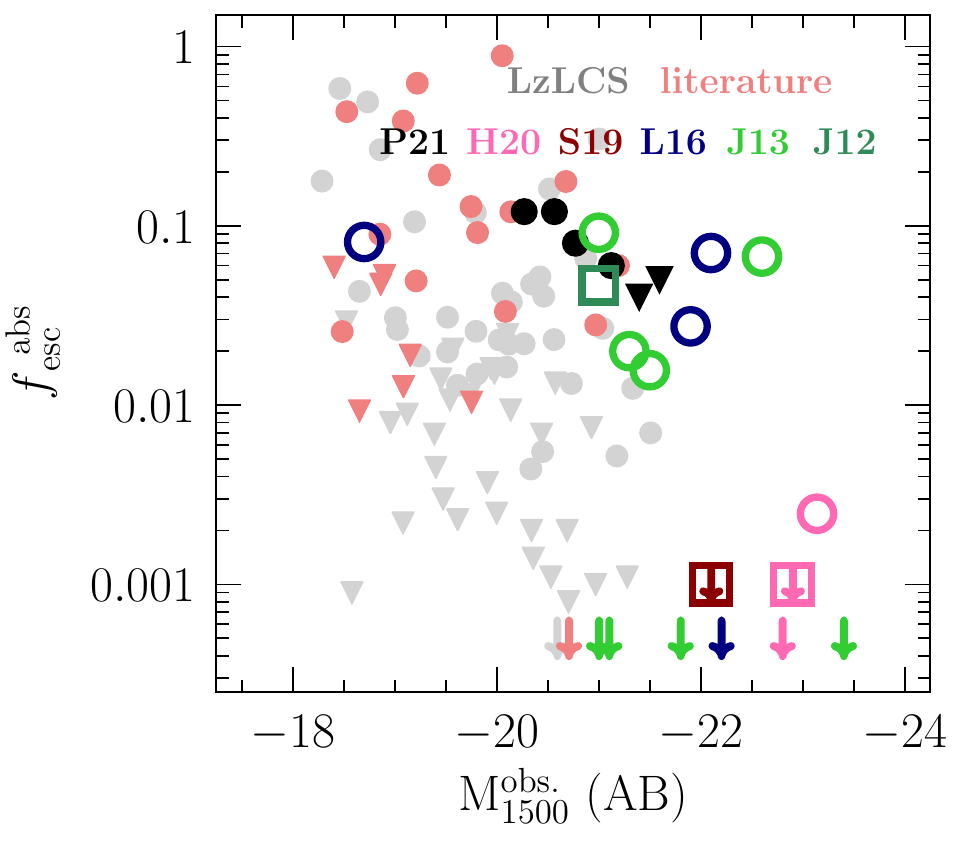}
\caption{Estimated and measured LyC absolute escape fractions of high- and low-$z$ galaxies as a function of the observed \mobs. Open circles and squares show the \fescabs\ predictions from the UV LIS lines for individual galaxies and stacks, respectively, taken from \citet[][]{Harikane2020} at $z \sim 6$, in pink;  \citet[][]{Sugahara2019} at $z \sim 5$, in dark red; \citet[][]{Leethochawalit2016} at $z \sim 4-5$, in blue; \citet[][]{Jones2013} at $z \sim 2-4$, in light green; and \citet[][]{Jones2012} at $z \sim 4$, in green. Direct \fescabs\ measurements from this paper (LzLCS and literature) are plotted in light gray and red; $z \sim 3$ measurements from \citet{Pahl2021} measurements are displayed with filled black symbols, for comparison. Downward-pointing arrows represent sources with estimated $\fescabs \approx 0$.}
\label{fig:fescMUV}
\end{figure}

While the picket-fence model allows us to predict  (at least on average) the LyC escape fraction relatively well, we find that the absorption lines tend to underestimate the escape fractions for the strongest leakers with \fescabs\ $\ga 0.1-0.2$ \citep[see also][]{Gazagnes2020}. This suggests that there could be a secondary scenario where LyC emission is regulated by a density-bounded ISM in these galaxies, as suggested by several previous studies \citep{Zackarisson2013, JaskotOey2013, Nakajima2014, Jaskot2019}. Indeed, from simulations, \citet{Kakiichi2021} show that a switch from an ionization-bounded regime to one that is density-bounded may naturally happen at evolved stages in the history of a galaxy. This transition occurs when turbulence ceases and the ionization radius then increases with time in a more static situation. Indications for the existence of the density-bounded regime for leakers with the highest \fescabs\ was presented by \citet{Gazagnes2020} and  \citet{Ramambason2020}, but whether this also quantitatively agrees with the present observations will be examined in future publications.

\begin{figure}
    \centering
    \includegraphics[width=0.98\columnwidth, page=20]{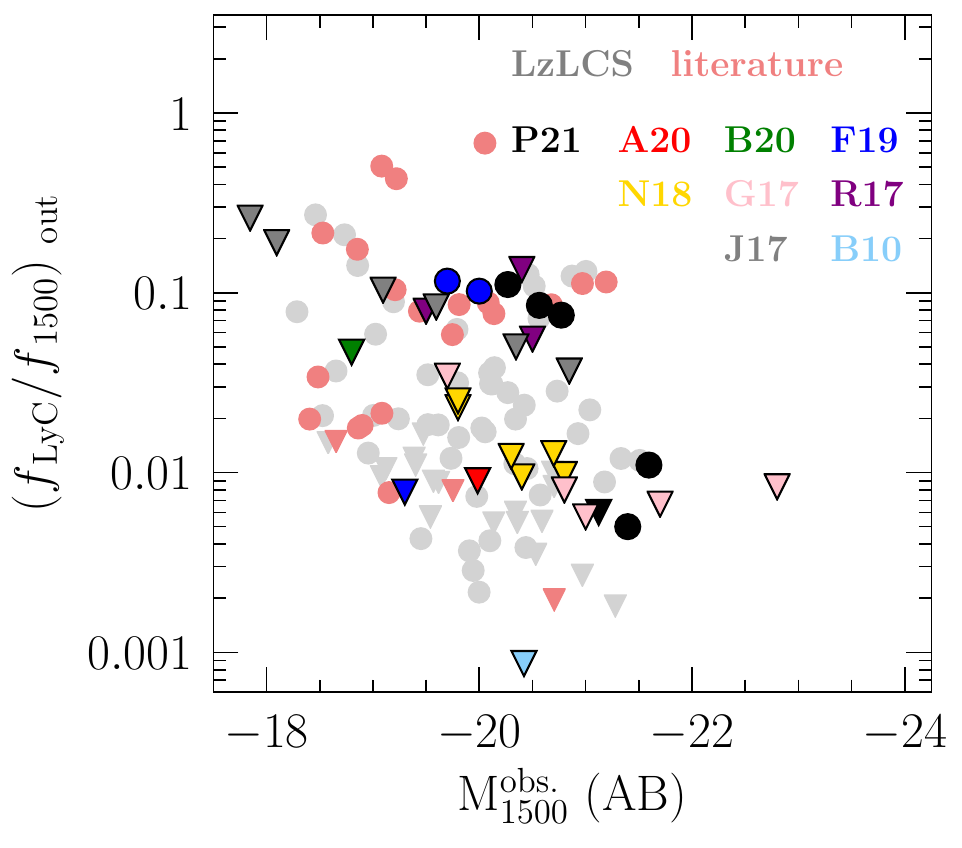}
\caption{Ionizing-to-nonionizing observed flux ratio $(f_{\rm ~LyC}/f_{1500})_{\rm ~out}$ in units of $F_\nu$ as a function of \mobs\ for high- and low-$z$ galaxies. At high redshift, we only show measurements determined from stacks. LyC detections from \citet{Pahl2021} and \citet{Fletcher2019} at $z \sim 3$ are plotted in black and blue symbols. Circles show detections and downward pointing triangles are 1$\sigma$ upper limits that are color coded by sample \citep{Pahl2021, Alavi2020, Bian2020, Fletcher2019, Naidu2018, Grazian2017, Rutkowski2017, Japelj2017}. The high-$z$ data probe $z \sim 2.5 - 4$, except for the study of \cite{Alavi2020} at $z \sim 1.3$, and \citet{Bridge2010} at $z \sim 0.7$. The low-$z$ data from the present study (LzLCS and literature) are plotted in light gray and red, as in the previous figure.}
\label{fig:frelMUV}
\end{figure}

\subsection{The LyC escape fraction of high-$z$ galaxies}\label{sub:discussion_highz}
With the empirically calibrated and theoretically motivated relations between the residual flux, UV attenuation, and the absolute LyC escape fraction (Eq.\ \ref{eq:fesc_LIS}), we can estimate \fescabs\ from literature measurements of the LIS absorption lines for high-$z$ galaxies, and compare them to local measurements from the LzLCS.

\autoref{fig:fescMUV} presents the result of this exercise, showing the predicted \fescabs\ from individual and stacked galaxy spectra reporting residual flux measurements of the UV LIS lines as a function of the observed absolute UV magnitude (the latter without a correction for UV attenuation). To compute the LyC escape fraction, we use the reported LIS residual flux,  and then Eq.\ \ref{eq:cfhi_cflis} to estimate the \cfhi, and the \citetalias{R16} attenuation law to estimate the \ebv\ from the average dust corrections given in \citet{Bouwens2020}, as a function of redshift\footnote{More precisely, we adopt $\ebv = 0.107, 0.095, 0.086, 0.058$ for $z = 3, 4, 5, 6$.}. We also show the \fescabs\ measurements at low $z$ from the present paper (gray, red circles and downward point triangles), and the latest update by \citet{Pahl2021} of the LyC escape fractions measured directly from stacks of  $z \sim 3$ galaxy spectra from the \emph{Keck Lyman Continuum Survey} \citep[KLCS, black circles; ][]{Steidel2018}. Downward arrows means predicted $\fescabs \approx 0,$ for those sources with $\cflis \approx 1$ (zero residual flux) and/or very high \ebv, falling below the axis limits.

As seen in \autoref{fig:fescMUV}, all our estimates yield absolute LyC escape fractions $\fescabs < 0.1$ at $z \ga 4$ for relatively bright $\mobs \la -21$ galaxies. Measurements from individual sources show large scatter, as also found at low $z$. For the highest redshift available ($z \sim 6$), the stack and the galaxy J0210-0523 of \cite{Harikane2020}, with $\mobs \sim -23$, and our method predict negligible escape fractions, $\fescabs \la 0.003$. From the $z \sim 4$ stack of \cite{Jones2012} with $\mobs \sim -21$, we estimate $\fescabs \sim 0.05$, which is comparable to the value measured at $z \sim 3$ for galaxies with similar UV magnitude \citep{Pahl2021}. We reiterate the fact that our \fescabs\ values of the high-$z$ galaxies are lower than those estimated in \cite{Jones2012, Harikane2020}, because we account for both the residual flux and the UV attenuation \citep{Gazagnes2018, Chisholm2018}. Escape fractions of $\sim 10$\% are found at $\mobs \sim -20$ from the KLCS stacks, and \cite{Pahl2021} find decreasing \fescabs\ towards UV-brighter galaxies, which overall appears consistent with our estimates. However, the available data are still scarce, and whether \fescabs\ increases further in fainter galaxies must be confirmed with future observations \citep[see][]{Bian2020}.

The overlap between the high- and low-$z$ data in observed UV magnitude is quite limited (\autoref{fig:fescMUV}) and meaningful sample averages remain to be obtained at low redshift. It is therefore difficult to draw conclusions on the possible redshift evolution of \fescabs\ from this figure, because possible trends may be a consequence of a general luminosity evolution or due to different detection limits within the high-$z$ samples. Furthermore, the adopted UV attenuations for the high-redshift samples are unconstrained, making the comparison of \fescabs\ as a function of the intrinsic UV magnitude also difficult to interpret. In any case, we note that both individual galaxies and stacks with $\fescabs \sim 5\%-10$\% are found over a range of four magnitudes (from $\mobs \sim -19$ to $-22$). Objects with $\fescabs > 0.1$ are (so far) found only at magnitudes fainter than $\mobs <-20$ at low $z$. However, some individual galaxies at $z > 3$ are bright ($\mobs < -21$) and have high LyC escape fractions \citep[e.g., {\em Ion2} in][]{Vanzella2016}. Thus, bright high-redshift galaxies may have markedly different LyC escape properties. Overall, the low-$z$ results concur, albeit with the low LyC escape fractions found in bright galaxies (\mobs $\la -20$) by many studies using stacks at high $z$ \citep[see e.g.,][]{Grazian2016,Guaita2016,Marchi2017,Japelj2017}.

Another quantity to be compared with high-$z$ studies is the observed ``Lyman decrement'' between the LyC and the nonionizing UV flux, that is, the commonly used $(f_{\rm ~LyC}/f_{1500})_{\rm ~out}$ corrected for the IGM attenuation. This quantity can directly lead to the ionizing emissivity of galaxies when integrating over the UV luminosity function \citep[see e.g.,][]{Pahl2021}. In \autoref{fig:frelMUV} we compile the available data from high-$z$ stacks, for which we plot the IGM-corrected measurements of $(f_{\rm ~LyC}/f_{1500})_{\rm ~out}$. We use stacked spectra to compensate for differences in the mean IGM transmission. The data comprise results from a variety of surveys mostly carried out at redshift $z \sim 2.5 - 4$, plus the study of \cite{Alavi2020} at $z \sim 1.3$, and \citet{Bridge2010} at $z \sim 0.7$. We have also added the data from the composite SEDs of the LAE sample of \cite{Fletcher2019} at $z=3.1$, who report LyC-detected and nondetected subsamples.

Interestingly, the Lyman decrement observed both from the KLCS \citep{Steidel2018, Pahl2021} and the LAE sample of \cite{Fletcher2019} at $z\sim 3$ are very similar, and comparable to the observed values $(f_{\rm ~LyC}/f_{1500})_{\rm ~out} \sim 0.1$ at $z \sim 0.3$ and at same UV magnitudes, $\mobs \sim -19.5$ to $-21$. The upper limits found for galaxies with $\mfuv \la -21$ indicate small Lyman decrements that are compatible with low-redshift observations. On the UV-faint end ($\mobs \ga -20$), the high- and low-$z$ observations also
appear fairly consistent with each other\footnote{The upper limit of $(f_{\rm ~LyC}/f_{1500})_{\rm ~out}$ at $\mobs = -19.3$ from \cite{Fletcher2019} is low by construction, because this represents the composite SED of LAEs with nondetections in the LyC.}. Further studies are required to better understand what determines the differences found between the various high- and low-$z$ samples and to determine representative population averages among low-redshift galaxies.
In summary, the high-$z$ observations show a consensus of decreasing Lyman decrement towards brighter UV magnitudes, although the overall correlation is weak. This trend is also tentatively shared by the LzLCS sample.

\subsection{Main assumptions, limitations, and caveats}\label{sub:caveats}
Among the possible sources of uncertainty in our analysis are (1) the choice of the dust-attenuation law, (2) the assumed gas and dust geometry, (3) the adopted picket-fence model, and (4) the definition and meaning of the LyC escape fraction.

\paragraph{(1) Assumptions on the dust-attenuation law.}
As previously described (Sect.\ \ref{sec:fit_results}), the choice of the  attenuation curve primarily impacts the derived stellar population parameters and UV attenuation. In particular, $X_i$ coefficients (see Eq.\ \ref{eq:s99_fit}) and the dust-attenuation parameters (\ebv) can both change, while average stellar ages and metallicities remain consistent \citep[same light-fractions, see ][]{Chisholm2019}.

A steeper attenuation law (like the SMC curve) yields a lower UV attenuation and hence systematically higher escape fractions \fescabs\ compared to the flatter \citetalias{R16} attenuation law adopted as default here. Typically, \fescabs\ is found to be higher by a factor of between $  1.05$ and 1.50 from high to low LyC escape fractions (see \autoref{fig:FESC_comparison}). The intrinsic, that is, extinction-corrected UV  magnitudes (\mfuv) are therefore also fainter for the SMC law. On the other hand, line equivalent widths and residual fluxes are essentially independent of the attenuation law, because the stellar continuum level is similarly modeled for the different laws. 

From the present UV data alone, it is difficult to distinguish between the two attenuation laws used here. A joint analysis of the full SED from the UV to the optical, including an independent attenuation measurement such as the Balmer decrement, would be necessary to constrain the attenuation law \citep[see e.g.,][]{Shivaei2020}. The offsets found between the color excess \ebv\ derived from the UV spectra and the Balmer decrement shown in \autoref{fig:EBV_comparison} indicate that an attenuation law steeper than the \citetalias{R16} curve is probably appropriate for a fraction of the galaxies in our sample. 
For the compact, high-O$_{32}$ LyC emitters detected prior to the LzLCS, an SMC-like or even steeper curve in some cases had been found by \cite{Izotov16b, Izotov18a, Izotov18b}.

Also, from the analysis of 218 $z=1.4-2.6$ SF galaxies with both UV rest-frame and Balmer decrement measurements plus known gas-phase metallicities, \cite{Shivaei2020} suggest that ``low-metallicity'' galaxies have a steeper, SMC-like attenuation curve, whereas galaxies with $\oh \ga 8.5$ are better described by the \citetalias{R16} law. In any case, no consensus exists so far as to the most appropriate UV attenuation law, or even as to the number of dust components that should be considered \citep{CF00}, and significant variations may exist between individual galaxies, different galaxy types, and so on.

We primarily use the \citetalias{R16} as the default here because it has been observationally constrained down to the Lyman limit, and thus avoids the need for extrapolations of the SMC law to the spectral range covered by the HST/COS spectra used here. Furthermore the \citetalias{R16} law is often used for high-redshift studies \citep[e.g.,][]{Reddy2016b,Steidel2018}, and may thus simplify comparisons.

\paragraph{(2) Assumptions on the gas and dust geometry.} 
The so-called picket-fence model (Sect.\ \ref{sec:methods}) proposes two simple ways to describe the relative distribution of dust with respect to the gas in the ISM. Either the dust is distributed in a homogeneous foreground layer (uniform dust screen, A), or it resides within the same clouds as the gas (clumpy geometry, B). 

Both UV attenuation and the geometrical covering fraction depend on the adopted geometry, as amply discussed in \cite{Vasei2016, Reddy2016b, Gazagnes2018, Steidel2018}. Overall, higher values of \ebv\ and $C_f$ are found in the clumpy geometry \citep[geometry B, referred to as the holes scenario by][]{Steidel2018}, as seen for example in \cite{Gazagnes2018, Chisholm2018, Steidel2018}. This is simply explained by the fact that, in this case, a fraction $(1-C_f$) of UV continuum light is {directly} transmitted, that is, unaltered, to the observer. To fit the observed UV slope and the depth of the absorption lines, higher \ebv\ and $C_f$ are therefore needed. 

Despite the differences for \ebv\ and $C_f$, \cite{Chisholm2018} and \cite{Gazagnes2020}  showed that, to first order, the determination of \fescabs\ from the UV absorption lines is independent of the assumed dust distribution between these two picket-fence geometries. However, for reasons that are not completely clear, \cite{Steidel2018} obtain somewhat lower average absolute LyC escape fractions when assuming the ``holes'' geometry.

Our analysis of the \hi\ and low-ionization UV absorption lines in Sect.\ \ref{sub:results_HILIS} clearly shows that the observed behavior favors a clumpy dust geometry, i.e., geometry B, as also suggested by earlier studies \citep[see e.g.,][]{Du2021}. However, for our spectral fits, we assume the foreground dust-screen scenario (A), which is not the favored scenario. This assumption is primarily driven by practical reasons and simplicity,because the fits of the stellar continuum (where ISM absorption lines are masked) can then be done independently of $C_f$, and the absorption lines can subsequently be analyzed. Indeed, in this geometry, the covering fraction is simply related to the residual flux of the (saturated) lines via $R=1-C_f$ \citep{Gazagnes2018}.

As long as the two assumptions yield almost identical LyC escape fractions, this apparent inconsistency is not relevant. However, it must be borne in mind that values such as the UV attenuation (quantified by \ebv) and the covering fraction depend on geometrical assumptions, and they should be handled with care in comparisons between different studies.

\paragraph{(3) Validity of the picket-fence model.}
In this paper, we make use of the picket-fence model to derive simple analytical solutions of the radiative transfer equations, which can be used to fit the observed galaxy spectra using SEDs of integrated stellar populations and ISM absorption lines \citep[cf.\ Sect.\ \ref{sec:methods},][]{Vasei2016, Reddy2016b}. With the assumptions that the ``channels'' between the pickets have a negligible column density and the absorption lines are saturated in the ISM clouds (pickets), this model then allows one to determine the geometric covering fraction of the absorbing gas from the residual flux of these lines.

Although this simple model works well to describe the observations and predict the 
average LyC continuum escape fractions, as shown in this paper and earlier studies \citep{Gazagnes2018,Steidel2018,Chisholm2018},
we remind the reader that this model is a drastic over-simplification of the complexity observed in resolved nearby galaxies and predicted by state-of-the-art high-resolution galaxy simulations.
In realistic situations, the geometry may differ from the simple assumptions of a single UV-emitting source behind a picket-fence ISM and a uniform dust screen made here. Different star-forming regions, whose emission contributes to the integrated spectra, may be ``covered'' by varying column densities of gas and dust, and each of those may also show substructure. Furthermore, the kinematics of the gas may be complex. How important these effects are for the diagnostics used in the present paper is nevertheless impossible to estimate. In addition, some galaxies, for example, the most compact ones, may also be better approximated by a picket-fence model than others.

The recent work of \citet{Mauerhofer2020}, who simulate a $z \sim 3$ low-mass galaxy at high resolution, and performed mock observations of UV absorption lines, may provide some guidance towards answering these questions.
Indeed, these simulations illustrate the nontrivial definition of the covering fraction and the complex relations between this quantity, the residual line fluxes, and the escape fraction of LyC photons. Despite this, the simulations reveal correlations between residual fluxes and LyC escape, also with significant scatter, similar to those found in our analysis. In any case, going beyond the analytical picket-fence model for the analysis of spectroscopic observations will require the development of new methods and data of higher quality (in S/N and resolution).

\paragraph{(4) Definition and meaning of the LyC escape fraction.}
As described in Sect.\ \ref{sec:methods}, the LyC escape fraction \fescabs\ derived in this paper relies on comparing the observed LyC flux in a limited window shortward of the Lyman edge to the expected intrinsic LyC flux. In our analysis, we measure the LyC typically from 850--900 \AA\ in the COS spectra \citep{FluryI}, and the intrinsic LyC flux is predicted by the fits to the nonionizing UV spectrum. This method is the most basic and most commonly used method for ``direct'' determination of the LyC escape fraction, that is,\ when a direct measurement of the LyC flux is available. It has been used by almost all existing studies of this kind, allowing for meaningful comparisons of \fescabs\ at all redshifts \citep{Izotov16a, Steidel2018}. 

The method can be applied to both spectroscopic observations and to imaging where two or more filters are used to probe the LyC flux and the nonionizing UV. When spectra are used, the LyC flux has so far always been measured over a small wavelength range just shortward of 912 \AA. Imaging may probe a wider spectral range in the LyC, depending on the filter widths used; for HST/WFC3 imaging of the strong LyC leaker {\it Ion2}, e.g., the LyC is probed over $\sim$ 730-890 \AA\ restframe \citep{Vanzella2016}. 

One question that presents itself in regard to this method is to what extent a measure of such a monochromatic escape fraction reflects the {total} escape fraction of all H-ionizing photons, including those at shorter wavelengths. For example, in their numerical simulations of escaping LyC radiation, \cite{Gnedin2008} find that the total escape fraction of ionizing photons below 912 \AA\ is approximately a factor of $1.25$ higher than the fraction escaping close to the Lyman edge. Furthermore, \cite{McCandliss2017} show the relation between these two escape fractions for slabs of variable \hi\ column density and fixed ionization fraction, finding fairly large enhancements (a factor of $2$ and higher for ``Lyman edge escape fractions'' $< 0.32$, comparable to our sample). However, such a simplified geometry and the transmission curves computed in their work cannot be used to predict the resulting spectrum of density-bounded \ion{H}{ii} regions \citep[cf.\ below, and][]{Inoue2010}.

Generally, if both {\em (i)} the escape fraction of LyC photons is wavelength-independent and {\em (ii)} the only emission process at the measured spectral range is stellar emission, the ratio of the observed to intrinsic stellar SED will provide a correct measure of the total escape fraction of LyC photons. Therefore, both of these conditions need to be examined. In fact, the escape fraction is to first order wavelength-independent if the ISM can be described by a picket-fence model, where the ``leaking'' regions have a negligible \hi\ column density and the total escape fraction is governed by the geometrical covering factor \citep[see also][]{Inoue2010}. The detailed analyses in this paper and earlier studies of LyC-emitting galaxies \citep{Gazagnes2018,Steidel2018} have shown clear evidence in favor of the picket-fence model, although a low residual column density in the ``escape channels'' cannot be excluded. The second condition {\em (ii)} might not always be validated, because free-bound emission of hydrogen in ionized nebulae emits shortward of the Lyman edge, and may therefore contribute a non-negligible additional emission source close to the Lyman edge, as discussed for example by \citet{Inoue2010}. Theoretically, the strength of this emission primarily depends on the mean energy of the ionizing photons (i.e.,\ on the detailed SED) and on the electron temperature of the ionized gas, and might lead to a ``Lyman bump'' in extreme cases \citep{Inoue2010}\footnote{We note that free-bound emission could also be expected if the escape channels in the ISM contain a low \hi\ column density, as shown by the \textsc{cloudy} simulations of \citet{Inoue2010}.} To the best of our knowledge, this emission has so far not been identified, although it is expected. If present, our monochromatic determinations of \fescabs\ could overestimate the true total escape fraction.

A second, independent method to determine the LyC escape fraction has also been used for the LzLCS survey and the literature samples; see \citet{FluryI} and \citet{FluryII}. This method uses the de-reddened H$\beta$ flux, which is proportional to the total number of ionizing photons absorbed (i.e.,\ nonescaping) in the galaxy, and compares it to the observed LyC flux, using theoretical SEDs to relate the total ionizing photon flux to the monochromatic flux shortward of the Lyman edge. As discussed by \cite{Izotov18b}, this method therefore measures the total ionizing photon escape fraction. However, if strong free-bound emission is present, this method might also overestimate \fescabs. Comparing this method to the ``monochromatic'' method, \cite{Izotov18b} find good agreement overall and no systematic differences \citep[cf.][]{Izotov18a,Izotov18b,Izotov2021}. Similarly, for the larger sample analyzed here, we do not find systematic differences between the two methods, although the scatter is fairly large \citep{FluryI}, and possibly driven by uncertainties related to the reddening corrections (both in the optical and UV). Indeed, as shown above (see Fig.\ \ref{fig:FESC_comparison}), the \fescabs\ values derived here solely from the UV spectrum depend quite strongly on the assumed attenuation law, making this probably the largest source of uncertainty of our analysis.

Finally, as it is known that theoretical SEDs of star-forming galaxies have a significant lack of ionizing photons at energies above 54 eV, which are required to explain nebular lines of He~{\sc ii} frequently observed in low-metallicity SF galaxies \citep[see e.g.,][]{Stasinska2015,Schaerer2019}, one might wonder whether or not these photons could alter our analysis and play a role in the ionizing photon budget of hydrogen. The answer appears to be no, primarily because the number of photons emitted above the \ion{He}{ii} ionization potential ($\lambda <228$ \AA) is small compared to those ionizing hydrogen. Indeed, their relative ratio, $Q_2/Q_O$, can be determined empirically from the observed \ion{He}{ii} and H recombination lines and is typically $\sim (0.2-2) \times 10^{-2}$ in low-$z$ galaxies with nebular \ion{He}{ii}$\lambda 4686$ detections \citep[see][]{Schaerer2019}. Their contribution to the ionization of hydrogen is therefore negligible. From model SEDs, it is well known that the bulk of the H-ionizing photons are emitted relatively close to the Lyman edge; for example $\sim 60\%-70$ \% of the photons in the LyC are emitted longward of the \ion{He}{i} ionization potential (24.6eV, or 504\AA) at young ages \citep[see e.g.,][]{Nakajima2018,Xiao2018}. Although all photons with energies $>13.6$ eV can ionize hydrogen, the majority are emitted at relatively low energies. In any case, methods using H$\beta$ (or other H recombination lines) to determine the LyC escape fraction inherently probe the {total} H-ionizing photon budget.  
\section{Summary and Conclusions}\label{sec:conclusions}
The {\em Low-Redshift Lyman Continuum Survey} \citep[LzLCS,][]{FluryI} recently observed 66 low-$z$ ($z \sim 0.3$) star-forming galaxies with HST/COS to measure the LyC in a large sample of galaxies  for the first time. We augmented the LzLCS data, adding 23 previous LyC observations taken with COS \citep{Izotov16a, Izotov16b, Izotov18a, Izotov18b}, \citet{Wang2019}, and \citet{Izotov2021}, and the available LyC, UV, and optical spectra analyzed in a homogenous manner. The sample provides unique insight into the LyC emission, the LyC escape fraction, and many related properties; it covers galaxies with a wide range of physical properties (UV magnitude, metallicity, and others) allowing us to empirically study how the escape of LyC emission varies from galaxy-to-galaxy and, in particular, how this phenomenon is related to the ISM properties of these sources. 

To study the ISM properties of these low-redshift galaxies, we analyzed the COS UV spectra in the framework of a picket-fence model with a uniform dust screen. First, we fit the UV stellar continuum (see Sect.\ \ref{sub:method_fitting}) to constrain the massive star content of these galaxies. From our nonparametric fits, we derived direct properties such us the stellar ages and metallicities, UV dust attenuations (\ebv), intrinsic UV magnitudes (\mfuv), and absolute LyC escape fractions (\fescabs). The stellar population properties will be discussed in a separate paper (Chisholm et al., in prep.).

The derived LyC escape fractions range up to $60\%$~($85\%$) for the LzLCS (published) sample. Approximately 45\% of the sample is not detected in the LyC, with $1\sigma$ upper limits of $\fescabs <0.01$. The absolute escape fraction strongly depends on the dust-attenuation parameter (\ebv), with the strongest LyC emitters only found with low UV attenuation. We also study the impact of the assumed attenuation law on \fescabs: adopting the SMC extinction law instead of the attenuation law from \cite{R16} leads to an increase in \fescabs\ by $\sim 10\%-50$\%, and the largest differences appear for galaxies with the lowest LyC escape fractions (Sect.\ \ref{sec:fit_results}). 

With the stellar continuum fits in hand, we then measured the equivalent widths ($W_{\lambda}$) and residual fluxes ($R_{\lambda}$) of different interstellar absorption lines of \hi\ (Lyman series) and metallic low ionization state (LIS) lines, such as \siia, \siib, \oi, and \cii\ (see Sect.\ \ref{sub:method_abs}). With the picket-fence model, the residual fluxes allow us to infer a geometrical covering fraction of the UV source by the gas. Our analysis of the absorption lines and other galaxy properties has led us to the following main conclusions: 

\begin{enumerate}

    \item[$-$] Lyman continuum emitters (leakers) have weak but likely saturated \hi\ and LIS lines (Sect.\ \ref{sub:results_lowews}), and changes in $W_{\lambda}$ are mainly driven by the neutral gas covering fraction (\cfhi).  Metallicity plays a secondary role (Sect.\ \ref{sub:results_HILIS}). 
    
    \item[$-$] We find that leakers have high \hi\ and LIS residual fluxes ($R_{\lambda}$) and low UV attenuations (\auv). This indicates that leakers tend to host a dust-poor ISM, with low \hi\ column density channels intermixed between higher column density regions (Sect.\ \ref{sub:results_HILIS}). Our large sample confirms the finding that these channels enable the escape of ionizing photons, as shown in other recent studies \citep{Reddy2016b, Gazagnes2018, Steidel2018}.
    
    \item[$-$] Leakers have large \lya\ equivalent widths (\ewlya), and this correlates strongly with the \hi\ and LIS equivalent widths, showing that the absorption lines trace the same gas that regulates the radiative transfer of \lya\ photons. \ewlya\ is therefore partly driven by the neutral gas covering fraction \cfhi\ (Sect.\ \ref{sub:results_Lya}), as also found in earlier studies \citep[e.g.,][]{Gazagnes2020, Pahl2020}.
    
    \item[$-$] We find \hi\ and LIS covering fractions varying from $\sim 0.5$ to $1$ and $\sim 0.2$ to $1,$ respectively. As in \cite{Reddy2016b} and \cite{Gazagnes2018}, the LIS covering fraction is on average smaller than that of \hi, and the two quantities are strongly correlated. This probably indicates that \hi\ and metals are not evenly distributed within the host galaxy, but that they trace each other (Sect.\ \ref{sub:results_cfs}). 
    
    \item[$-$] The LyC photon escape fraction (\fescabs) strongly correlates with both \hi\ and LIS equivalent widths and residual fluxes of the lines (Sect.\ \ref{sub:results_cf_fesc}), and also with the UV attenuation (Sect.\ \ref{sub:results_ebv_fesc}). We derive empirical calibrations to estimate \fescabs\ from UV absorption line measurements.
    
    \item[$-$] Finally, we show that simultaneous UV absorption line and attenuation measurements can, in general, predict the escape fractions of galaxies (Sect.\ \ref{sub:discussion_fescs}). We apply our methods to literature measurements of UV LIS lines in $z \sim 4-6$ galaxies, estimating  modestly low LyC escape fractions of $\fescabs \la 0.1$ for relatively bright, $\mobs \la -21$, high-redshift galaxies (Sect.\ \ref{sub:discussion_highz}).
    
\end{enumerate}

In short, our detailed quantitative analysis of the UV spectra of 89 low-$z$ galaxies shows that LyC leakage is a complex phenomenon (see Sect.\ \ref{sub:discussion_ism}), which mainly depends on the gas distribution (covering fractions) and dust content (UV attenuation) of the host galaxy. We demonstrate that the \hi\ Lyman series lines shortward of \lya\ and metallic absorption lines can be used to statistically predict the escape fraction of ionizing photons, as proposed by \cite{Chisholm2018}. Although several uncertainties and open questions remain, our study demonstrates how LIS UV absorption lines can be used to study and constrain the escape of ionizing photons. These diagnostics will be invaluable and key in upcoming studies of galaxies up to the Epoch of Reionization, with the advent of the James Webb Space Telescope (JWST) and other 30-meter class telescopes.

\begin{acknowledgements}
    A.S.L. and D.S. acknowledge support from Swiss National Science Foundation. S.R.F and A.E.J. acknowledge support from NASA through grant HST-GO-15626 from the Space Telescope Science Institute. R.A. acknowledges support from ANID  Fondecyt Regular 1202007. M.T. acknowledges support from the NWO grant 016.VIDI.189.162 (``ODIN''). 
    This research is based on observations made with the NASA/ESA Hubble Space Telescope obtained from the Space Telescope Science Institute, which is operated by the Association of Universities for Research in Astronomy, Inc., under NASA contract NAS 5–26555. These observations are associated with program GO 15626 (P.I. Jaskot). Additional work was obtained from the data archive at the Space Telescope Science Institute from HST proposals 13744, 14635, 15341, and 15639. 
    
\end{acknowledgements}

%
%

\bibliographystyle{aa}
\bibliography{references}

\newcommand{\noop}[1]{}
\begin{thebibliography}{174}
\expandafter\ifx\csname natexlab\endcsname\relax\def\natexlab#1{#1}\fi

\bibitem[{{Akritas} \& {Siebert}(1996)}]{ATS1996}
{Akritas}, M.~G. \& {Siebert}, J. 1996, \mnras, 278, 919

\bibitem[{{Alavi} {et~al.}(2020){Alavi}, {Colbert}, {Teplitz}, {Siana},
  {Scarlata}, {Rutkowski}, {Mehta}, {Henry}, {Dai}, {Haardt}, \&
  {Bagley}}]{Alavi2020}
{Alavi}, A., {Colbert}, J., {Teplitz}, H.~I., {et~al.} 2020, \apj, 904, 59

\bibitem[{{Baldwin} {et~al.}(1981){Baldwin}, {Phillips}, \&
  {Terlevich}}]{BPT1981}
{Baldwin}, J.~A., {Phillips}, M.~M., \& {Terlevich}, R. 1981, \pasp, 93, 5

\bibitem[{{Bassett} {et~al.}(2019){Bassett}, {Ryan-Weber}, {Cooke}, {Diaz},
  {Nanayakkara}, {Yuan}, {Spitler}, {Me{\v{s}}tri{\'c}}, {Garel}, {Sawicki},
  {Gwyn}, \& {Golob}}]{Basset2019}
{Bassett}, R., {Ryan-Weber}, E.~V., {Cooke}, J., {et~al.} 2019, \mnras, 483,
  5223

\bibitem[{{Becker} {et~al.}(2001){Becker}, {Fan}, {White}, {Strauss},
  {Narayanan}, {Lupton}, {Gunn}, {Annis}, {Bahcall}, {Brinkmann}, {Connolly},
  {Csabai}, {Czarapata}, {Doi}, {Heckman}, {Hennessy}, {Ivezi{\'c}}, {Knapp},
  {Lamb}, {McKay}, {Munn}, {Nash}, {Nichol}, {Pier}, {Richards}, {Schneider},
  {Stoughton}, {Szalay}, {Thakar}, \& {York}}]{Becker2001}
{Becker}, R.~H., {Fan}, X., {White}, R.~L., {et~al.} 2001, \aj, 122, 2850

\bibitem[{{Belokurov} {et~al.}(2007){Belokurov}, {Evans}, {Moiseev}, {King},
  {Hewett}, {Pettini}, {Wyrzykowski}, {McMahon}, {Smith}, {Gilmore}, {Sanchez},
  {Udalski}, {Koposov}, {Zucker}, \& {Walcher}}]{Belokurov2007}
{Belokurov}, V., {Evans}, N.~W., {Moiseev}, A., {et~al.} 2007, \apjl, 671, L9

\bibitem[{{Bergvall} {et~al.}(2006){Bergvall}, {Zackrisson}, {Andersson},
  {Arnberg}, {Masegosa}, \& {{\"O}stlin}}]{Bergvall2006}
{Bergvall}, N., {Zackrisson}, E., {Andersson}, B.~G., {et~al.} 2006, \aap, 448,
  513

\bibitem[{{Bian} \& {Fan}(2020)}]{Bian2020}
{Bian}, F. \& {Fan}, X. 2020, \mnras, 493, L65

\bibitem[{{Bian} {et~al.}(2017){Bian}, {Fan}, {McGreer}, {Cai}, \&
  {Jiang}}]{Bian2017}
{Bian}, F., {Fan}, X., {McGreer}, I., {Cai}, Z., \& {Jiang}, L. 2017, \apjl,
  837, L12

\bibitem[{{Blanton} {et~al.}(2017){Blanton}, {Bershady}, {Abolfathi},
  {Albareti}, {Allende Prieto}, {Almeida}, {Alonso-Garc{\'\i}a}, {Anders},
  {Anderson}, {Andrews}, {Aquino-Ort{\'\i}z}, {Arag{\'o}n-Salamanca},
  {Argudo-Fern{\'a}ndez}, {Armengaud}, {Aubourg}, {Avila-Reese}, {Badenes},
  {Bailey}, {Barger}, {Barrera-Ballesteros}, {Bartosz}, {Bates}, {Baumgarten},
  {Bautista}, {Beaton}, {Beers}, {Belfiore}, {Bender}, {Berlind}, {Bernardi},
  {Beutler}, {Bird}, {Bizyaev}, {Blanc}, {Blomqvist}, {Bolton}, {Boquien},
  {Borissova}, {van den Bosch}, {Bovy}, {Brandt}, {Brinkmann}, {Brownstein},
  {Bundy}, {Burgasser}, {Burtin}, {Busca}, {Cappellari}, {Delgado Carigi},
  {Carlberg}, {Carnero Rosell}, {Carrera}, {Chanover}, {Cherinka}, {Cheung},
  {G{\'o}mez Maqueo Chew}, {Chiappini}, {Choi}, {Chojnowski}, {Chuang},
  {Chung}, {Cirolini}, {Clerc}, {Cohen}, {Comparat}, {da Costa}, {Cousinou},
  {Covey}, {Crane}, {Croft}, {Cruz-Gonzalez}, {Garrido Cuadra}, {Cunha},
  {Damke}, {Darling}, {Davies}, {Dawson}, {de la Macorra}, {Dell'Agli}, {De
  Lee}, {Delubac}, {Di Mille}, {Diamond-Stanic}, {Cano-D{\'\i}az}, {Donor},
  {Downes}, {Drory}, {du Mas des Bourboux}, {Duckworth}, {Dwelly}, {Dyer},
  {Ebelke}, {Eigenbrot}, {Eisenstein}, {Emsellem}, {Eracleous}, {Escoffier},
  {Evans}, {Fan}, {Fern{\'a}ndez-Alvar}, {Fernandez-Trincado}, {Feuillet},
  {Finoguenov}, {Fleming}, {Font-Ribera}, {Fredrickson}, {Freischlad},
  {Frinchaboy}, {Fuentes}, {Galbany}, {Garcia-Dias},
  {Garc{\'\i}a-Hern{\'a}ndez}, {Gaulme}, {Geisler}, {Gelfand},
  {Gil-Mar{\'\i}n}, {Gillespie}, {Goddard}, {Gonzalez-Perez}, {Grabowski},
  {Green}, {Grier}, {Gunn}, {Guo}, {Guy}, {Hagen}, {Hahn}, {Hall}, {Harding},
  {Hasselquist}, {Hawley}, {Hearty}, {Gonzalez Hern{\'a}ndez}, {Ho}, {Hogg},
  {Holley-Bockelmann}, {Holtzman}, {Holzer}, {Huehnerhoff}, {Hutchinson},
  {Hwang}, {Ibarra-Medel}, {da Silva Ilha}, {Ivans}, {Ivory}, {Jackson},
  {Jensen}, {Johnson}, {Jones}, {J{\"o}nsson}, {Jullo}, {Kamble}, {Kinemuchi},
  {Kirkby}, {Kitaura}, {Klaene}, {Knapp}, {Kneib}, {Kollmeier}, {Lacerna},
  {Lane}, {Lang}, {Law}, {Lazarz}, {Lee}, {Le Goff}, {Liang}, {Li}, {Li},
  {Lian}, {Lima}, {Lin}, {Lin}, {Bertran de Lis}, {Liu}, {de Icaza Lizaola},
  {Long}, {Lucatello}, {Lundgren}, {MacDonald}, {Deconto Machado}, {MacLeod},
  {Mahadevan}, {Geimba Maia}, {Maiolino}, {Majewski}, {Malanushenko},
  {Malanushenko}, {Manchado}, {Mao}, {Maraston}, {Marques-Chaves}, {Masseron},
  {Masters}, {McBride}, {McDermid}, {McGrath}, {McGreer}, {Medina Pe{\~n}a},
  {Melendez}, {Merloni}, {Merrifield}, {Meszaros}, {Meza}, {Minchev},
  {Minniti}, {Miyaji}, {More}, {Mulchaey}, {M{\"u}ller-S{\'a}nchez}, {Muna},
  {Munoz}, {Myers}, {Nair}, {Nandra}, {Correa do Nascimento}, {Negrete},
  {Ness}, {Newman}, {Nichol}, {Nidever}, {Nitschelm}, {Ntelis}, {O'Connell},
  {Oelkers}, {Oravetz}, {Oravetz}, {Pace}, {Padilla}, {Palanque-Delabrouille},
  {Alonso Palicio}, {Pan}, {Parejko}, {Parikh}, {P{\^a}ris}, {Park}, {Patten},
  {Peirani}, {Pellejero-Ibanez}, {Penny}, {Percival}, {Perez-Fournon},
  {Petitjean}, {Pieri}, {Pinsonneault}, {Pisani}, {Poleski}, {Prada},
  {Prakash}, {Queiroz}, {Raddick}, {Raichoor}, {Barboza Rembold}, {Richstein},
  {Riffel}, {Riffel}, {Rix}, {Robin}, {Rockosi}, {Rodr{\'\i}guez-Torres},
  {Roman-Lopes}, {Rom{\'a}n-Z{\'u}{\~n}iga}, {Rosado}, {Ross}, {Rossi}, {Ruan},
  {Ruggeri}, {Rykoff}, {Salazar-Albornoz}, {Salvato}, {S{\'a}nchez}, {Aguado},
  {S{\'a}nchez-Gallego}, {Santana}, {Santiago}, {Sayres}, {Schiavon}, {da Silva
  Schimoia}, {Schlafly}, {Schlegel}, {Schneider}, {Schultheis}, {Schuster},
  {Schwope}, {Seo}, {Shao}, {Shen}, {Shetrone}, {Shull}, {Simon}, {Skinner},
  {Skrutskie}, {Slosar}, {Smith}, {Sobeck}, {Sobreira}, {Somers}, {Souto},
  {Stark}, {Stassun}, {Stauffer}, {Steinmetz}, {Storchi-Bergmann},
  {Streblyanska}, {Stringfellow}, {Su{\'a}rez}, {Sun}, {Suzuki}, {Szigeti},
  {Taghizadeh-Popp}, {Tang}, {Tao}, {Tayar}, {Tembe}, {Teske}, {Thakar},
  {Thomas}, {Thompson}, {Tinker}, {Tissera}, {Tojeiro}, {Hernandez Toledo}, {de
  la Torre}, {Tremonti}, {Troup}, {Valenzuela}, {Martinez Valpuesta},
  {Vargas-Gonz{\'a}lez}, {Vargas-Maga{\~n}a}, {Vazquez}, {Villanova}, {Vivek},
  {Vogt}, {Wake}, {Walterbos}, {Wang}, {Weaver}, {Weijmans}, {Weinberg},
  {Westfall}, {Whelan}, {Wild}, {Wilson}, {Wood-Vasey}, {Wylezalek}, {Xiao},
  {Yan}, {Yang}, {Ybarra}, {Y{\`e}che}, {Zakamska}, {Zamora}, {Zarrouk},
  {Zasowski}, {Zhang}, {Zhao}, {Zheng}, {Zheng}, {Zhou}, {Zhou}, {Zhu},
  {Zoccali}, \& {Zou}}]{SDSS15}
{Blanton}, M.~R., {Bershady}, M.~A., {Abolfathi}, B., {et~al.} 2017, \aj, 154,
  28

\bibitem[{{Borthakur} {et~al.}(2016){Borthakur}, {Heckman}, {Tumlinson},
  {Bordoloi}, {Kauffmann}, {Catinella}, {Schiminovich}, {Dav{\'e}}, {Moran}, \&
  {Saintonge}}]{Borthakur2016}
{Borthakur}, S., {Heckman}, T., {Tumlinson}, J., {et~al.} 2016, \apj, 833, 259

\bibitem[{{Bouchet} {et~al.}(1985){Bouchet}, {Lequeux}, {Maurice}, {Prevot}, \&
  {Prevot-Burnichon}}]{SMC85}
{Bouchet}, P., {Lequeux}, J., {Maurice}, E., {Prevot}, L., \&
  {Prevot-Burnichon}, M.~L. 1985, \aap, 149, 330

\bibitem[{{Bouwens} {et~al.}(2020){Bouwens}, {Gonz{\'a}lez-L{\'o}pez},
  {Aravena}, {Decarli}, {Novak}, {Stefanon}, {Walter}, {Boogaard}, {Carilli},
  {Dudzevi{\v{c}}i{\={u}}t{\.{e}}}, {Smail}, {Daddi}, {da Cunha}, {Ivison},
  {Nanayakkara}, {Cortes}, {Cox}, {Inami}, {Oesch}, {Popping}, {Riechers}, {van
  der Werf}, {Weiss}, {Fudamoto}, \& {Wagg}}]{Bouwens2020}
{Bouwens}, R., {Gonz{\'a}lez-L{\'o}pez}, J., {Aravena}, M., {et~al.} 2020,
  \apj, 902, 112

\bibitem[{{Bridge} {et~al.}(2010){Bridge}, {Teplitz}, {Siana}, {Scarlata},
  {Conselice}, {Ferguson}, {Brown}, {Salvato}, {Rudie}, {de Mello}, {Colbert},
  {Gardner}, {Giavalisco}, \& {Armus}}]{Bridge2010}
{Bridge}, C.~R., {Teplitz}, H.~I., {Siana}, B., {et~al.} 2010, \apj, 720, 465

\bibitem[{{Calzetti} {et~al.}(2000){Calzetti}, {Armus}, {Bohlin}, {Kinney},
  {Koornneef}, \& {Storchi-Bergmann}}]{Calzetti2000}
{Calzetti}, D., {Armus}, L., {Bohlin}, R.~C., {et~al.} 2000, \apj, 533, 682

\bibitem[{{Cardelli} {et~al.}(1989){Cardelli}, {Clayton}, \& {Mathis}}]{CCM89}
{Cardelli}, J.~A., {Clayton}, G.~C., \& {Mathis}, J.~S. 1989, \apj, 345, 245

\bibitem[{{Carr} {et~al.}(2021){Carr}, {Scarlata}, {Henry}, \&
  {Panagia}}]{Carr2021}
{Carr}, C., {Scarlata}, C., {Henry}, A., \& {Panagia}, N. 2021, \apj, 906, 104

\bibitem[{{Cen}(2020)}]{Cen2020}
{Cen}, R. 2020, \apjl, 889, L22

\bibitem[{{Charlot} \& {Fall}(2000)}]{CF00}
{Charlot}, S. \& {Fall}, S.~M. 2000, \apj, 539, 718

\bibitem[{{Chisholm} {et~al.}(2018){Chisholm}, {Gazagnes}, {Schaerer},
  {Verhamme}, {Rigby}, {Bayliss}, {Sharon}, {Gladders}, \&
  {Dahle}}]{Chisholm2018}
{Chisholm}, J., {Gazagnes}, S., {Schaerer}, D., {et~al.} 2018, \aap, 616, A30

\bibitem[{{Chisholm} {et~al.}(2017){Chisholm}, {Orlitov{\'a}}, {Schaerer},
  {Verhamme}, {Worseck}, {Izotov}, {Thuan}, \& {Guseva}}]{Chisholm2017}
{Chisholm}, J., {Orlitov{\'a}}, I., {Schaerer}, D., {et~al.} 2017, \aap, 605,
  A67

\bibitem[{{Chisholm} {et~al.}(2020){Chisholm}, {Prochaska}, {Schaerer},
  {Gazagnes}, \& {Henry}}]{Chisholm2020}
{Chisholm}, J., {Prochaska}, J.~X., {Schaerer}, D., {Gazagnes}, S., \& {Henry},
  A. 2020, \mnras, 498, 2554

\bibitem[{{Chisholm} {et~al.}(2019){Chisholm}, {Rigby}, {Bayliss}, {Berg},
  {Dahle}, {Gladders}, \& {Sharon}}]{Chisholm2019}
{Chisholm}, J., {Rigby}, J.~R., {Bayliss}, M., {et~al.} 2019, \apj, 882, 182

\bibitem[{{Chisholm} {et~al.}(2016){Chisholm}, {Tremonti}, {Leitherer}, {Chen},
  \& {Wofford}}]{Chisholm2016}
{Chisholm}, J., {Tremonti}, C.~A., {Leitherer}, C., {Chen}, Y., \& {Wofford},
  A. 2016, \mnras, 457, 3133

\bibitem[{{Clarke} \& {Oey}(2002)}]{ClarkeOey2002}
{Clarke}, C. \& {Oey}, M.~S. 2002, \mnras, 337, 1299

\bibitem[{{Dayal} {et~al.}(2020){Dayal}, {Volonteri}, {Choudhury}, {Schneider},
  {Trebitsch}, {Gnedin}, {Atek}, {Hirschmann}, \& {Reines}}]{Dayal2020}
{Dayal}, P., {Volonteri}, M., {Choudhury}, T.~R., {et~al.} 2020, \mnras, 495,
  3065

\bibitem[{{de Barros} {et~al.}(2016){de Barros}, {Vanzella}, {Amor{\'\i}n},
  {Castellano}, {Siana}, {Grazian}, {Suh}, {Balestra}, {Vignali}, {Verhamme},
  {Zamorani}, {Mignoli}, {Hasinger}, {Comastri}, {Pentericci},
  {P{\'e}rez-Montero}, {Fontana}, {Giavalisco}, \& {Gilli}}]{deBarros2016}
{de Barros}, S., {Vanzella}, E., {Amor{\'\i}n}, R., {et~al.} 2016, \aap, 585,
  A51

\bibitem[{{Dessauges-Zavadsky} {et~al.}(2010){Dessauges-Zavadsky}, {D'Odorico},
  {Schaerer}, {Modigliani}, {Tapken}, \& {Vernet}}]{Mirka2010}
{Dessauges-Zavadsky}, M., {D'Odorico}, S., {Schaerer}, D., {et~al.} 2010, \aap,
  510, A26

\bibitem[{{Dijkstra} {et~al.}(2016){Dijkstra}, {Gronke}, \&
  {Venkatesan}}]{Dijkstra2016}
{Dijkstra}, M., {Gronke}, M., \& {Venkatesan}, A. 2016, \apj, 828, 71

\bibitem[{{Draine}(2011)}]{Draine}
{Draine}, B.~T. 2011, {Physics of the Interstellar and Intergalactic Medium}

\bibitem[{{Du} {et~al.}(2018){Du}, {Shapley}, {Reddy}, {Jones}, {Stark},
  {Steidel}, {Strom}, {Rudie}, {Erb}, {Ellis}, \& {Pettini}}]{Du2018}
{Du}, X., {Shapley}, A.~E., {Reddy}, N.~A., {et~al.} 2018, \apj, 860, 75

\bibitem[{{Du} {et~al.}(2021){Du}, {Shapley}, {Topping}, {Reddy}, {Sanders},
  {Coil}, {Kriek}, {Mobasher}, \& {Siana}}]{Du2021}
{Du}, X., {Shapley}, A.~E., {Topping}, M.~W., {et~al.} 2021, \apj, 920, 95

\bibitem[{{Eldridge} {et~al.}(2017){Eldridge}, {Stanway}, {Xiao}, {McClelland},
  {Taylor}, {Ng}, {Greis}, \& {Bray}}]{BPASS17}
{Eldridge}, J.~J., {Stanway}, E.~R., {Xiao}, L., {et~al.} 2017, \pasa, 34, e058

\bibitem[{{Erb}(2015)}]{Erb2015}
{Erb}, D.~K. 2015, \nat, 523, 169

\bibitem[{{Faisst}(2016)}]{Faisst2016}
{Faisst}, A.~L. 2016, \apj, 829, 99

\bibitem[{{Ferland} {et~al.}(2017){Ferland}, {Chatzikos}, {Guzm{\'a}n},
  {Lykins}, {van Hoof}, {Williams}, {Abel}, {Badnell}, {Keenan}, {Porter}, \&
  {Stancil}}]{Ferland2017}
{Ferland}, G.~J., {Chatzikos}, M., {Guzm{\'a}n}, F., {et~al.} 2017, \rmxaa, 53,
  385

\bibitem[{{Finkelstein} {et~al.}(2019){Finkelstein}, {D'Aloisio},
  {Paardekooper}, {Ryan}, {Behroozi}, {Finlator}, {Livermore}, {Upton
  Sanderbeck}, {Dalla Vecchia}, \& {Khochfar}}]{Finkelstein2019}
{Finkelstein}, S.~L., {D'Aloisio}, A., {Paardekooper}, J.-P., {et~al.} 2019,
  \apj, 879, 36

\bibitem[{{Fitzpatrick}(1999)}]{F99}
{Fitzpatrick}, E.~L. 1999, \pasp, 111, 63

\bibitem[{{Fletcher} {et~al.}(2019){Fletcher}, {Tang}, {Robertson}, {Nakajima},
  {Ellis}, {Stark}, \& {Inoue}}]{Fletcher2019}
{Fletcher}, T.~J., {Tang}, M., {Robertson}, B.~E., {et~al.} 2019, \apj, 878, 87

\bibitem[{{Flury} {et~al.}(\noop{9992} 2022b){Flury}, {Ferguson}, {Worseck},
  {Makan}, {Chisholm}, {Saldana-Lopez}, {Schaerer}, {McCandliss}, {Wang}, M.,
  {Heckman}, {Zhiyuan}, \& {Giavalisco}}]{FluryII}
{Flury}, S.~R.~{Jaskot}, A.~E., {Ferguson}, H.~C., {Worseck}, G., {et~al.}
  \noop{9992} 2022b, \apj

\bibitem[{{Flury} {et~al.}(2022){Flury}, {Jaskot}, {Ferguson}, {Worseck},
  {Makan}, {Chisholm}, {Saldana-Lopez}, {Schaerer}, {McCandless}, {Wang},
  {Ford}, {Heckman}, {Ji}, {Giavalisco}, {Amorin}, {Atek}, {Blaizot},
  {Borthakur}, {Carr}, {Castellano}, {Cristiani}, {de Barros}, {Dickinson},
  {Finkelstein}, {Fleming}, {Fontanot}, {Garel}, {Grazian}, {Hayes}, {Henry},
  {Mauerhofer}, {Micheva}, {Oey}, {Ostlin}, {Papovich}, {Pentericci},
  {Ravindranath}, {Rosdahl}, {Rutkowski}, {Santini}, {Scarlata}, {Teplitz},
  {Thuan}, {Trebitsch}, {Vanzella}, {Verhamme}, \& {Xu}}]{FluryI}
{Flury}, S.~R., {Jaskot}, A.~E., {Ferguson}, H.~C., {et~al.} 2022, arXiv
  e-prints, arXiv:2201.11716

\bibitem[{{Flury} \& {Moran}(2020)}]{Flury2020}
{Flury}, S.~R. \& {Moran}, E.~C. 2020, \mnras, 496, 2191

\bibitem[{{Fontanot} {et~al.}(2014){Fontanot}, {Cristiani}, {Pfrommer},
  {Cupani}, \& {Vanzella}}]{Fontanot2014}
{Fontanot}, F., {Cristiani}, S., {Pfrommer}, C., {Cupani}, G., \& {Vanzella},
  E. 2014, \mnras, 438, 2097

\bibitem[{{Gazagnes} {et~al.}(2020){Gazagnes}, {Chisholm}, {Schaerer},
  {Verhamme}, \& {Izotov}}]{Gazagnes2020}
{Gazagnes}, S., {Chisholm}, J., {Schaerer}, D., {Verhamme}, A., \& {Izotov}, Y.
  2020, \aap, 639, A85

\bibitem[{{Gazagnes} {et~al.}(2018){Gazagnes}, {Chisholm}, {Schaerer},
  {Verhamme}, {Rigby}, \& {Bayliss}}]{Gazagnes2018}
{Gazagnes}, S., {Chisholm}, J., {Schaerer}, D., {et~al.} 2018, \aap, 616, A29

\bibitem[{{Giallongo} {et~al.}(2015){Giallongo}, {Grazian}, {Fiore}, {Fontana},
  {Pentericci}, {Vanzella}, {Dickinson}, {Kocevski}, {Castellano}, {Cristiani},
  {Ferguson}, {Finkelstein}, {Grogin}, {Hathi}, {Koekemoer}, {Newman}, \&
  {Salvato}}]{Giallongo2015}
{Giallongo}, E., {Grazian}, A., {Fiore}, F., {et~al.} 2015, \aap, 578, A83

\bibitem[{{Gnedin} {et~al.}(2008){Gnedin}, {Kravtsov}, \& {Chen}}]{Gnedin2008}
{Gnedin}, N.~Y., {Kravtsov}, A.~V., \& {Chen}, H. 2008, \apj, 672, 765

\bibitem[{{Grazian} {et~al.}(2020){Grazian}, {Giallongo}, {Fiore}, {Boutsia},
  {Civano}, {Cristiani}, {Cupani}, {Dickinson}, {Fontanot}, {Menci}, \&
  {Romano}}]{Grazian2020}
{Grazian}, A., {Giallongo}, E., {Fiore}, F., {et~al.} 2020, \apj, 897, 94

\bibitem[{{Grazian} {et~al.}(2016){Grazian}, {Giallongo}, {Gerbasi}, {Fiore},
  {Fontana}, {Le F{\`e}vre}, {Pentericci}, {Vanzella}, {Zamorani}, {Cassata},
  {Garilli}, {Le Brun}, {Maccagni}, {Tasca}, {Thomas}, {Zucca}, {Amor{\'\i}n},
  {Bardelli}, {Cassar{\`a}}, {Castellano}, {Cimatti}, {Cucciati}, {Durkalec},
  {Giavalisco}, {Hathi}, {Ilbert}, {Lemaux}, {Paltani}, {Ribeiro}, {Schaerer},
  {Scodeggio}, {Sommariva}, {Talia}, {Tresse}, {Vergani}, {Bonchi}, {Boutsia},
  {Capak}, {Charlot}, {Contini}, {de la Torre}, {Dunlop}, {Fotopoulou},
  {Guaita}, {Koekemoer}, {L{\'o}pez-Sanjuan}, {Mellier}, {Merlin}, {Paris},
  {Pforr}, {Pilo}, {Santini}, {Scoville}, {Taniguchi}, \& {Wang}}]{Grazian2016}
{Grazian}, A., {Giallongo}, E., {Gerbasi}, R., {et~al.} 2016, \aap, 585, A48

\bibitem[{{Grazian} {et~al.}(2017){Grazian}, {Giallongo}, {Paris}, {Boutsia},
  {Dickinson}, {Santini}, {Windhorst}, {Jansen}, {Cohen}, {Ashcraft},
  {Scarlata}, {Rutkowski}, {Vanzella}, {Cusano}, {Cristiani}, {Giavalisco},
  {Ferguson}, {Koekemoer}, {Grogin}, {Castellano}, {Fiore}, {Fontana},
  {Marchi}, {Pedichini}, {Pentericci}, {Amor{\'\i}n}, {Barro}, {Bonchi},
  {Bongiorno}, {Faber}, {Fumana}, {Galametz}, {Guaita}, {Kocevski}, {Merlin},
  {Nonino}, {O'Connell}, {Pilo}, {Ryan}, {Sani}, {Speziali}, {Testa}, {Weiner},
  \& {Yan}}]{Grazian2017}
{Grazian}, A., {Giallongo}, E., {Paris}, D., {et~al.} 2017, \aap, 602, A18

\bibitem[{{Green}(2018)}]{Green2018}
{Green}, G.~M. 2018, The Journal of Open Source Software, 3, 695

\bibitem[{{Grimes} {et~al.}(2009){Grimes}, {Heckman}, {Aloisi}, {Calzetti},
  {Leitherer}, {Martin}, {Meurer}, {Sembach}, \& {Strickland}}]{Grimes2009}
{Grimes}, J.~P., {Heckman}, T., {Aloisi}, A., {et~al.} 2009, \apjs, 181, 272

\bibitem[{{Gronke} {et~al.}(2016){Gronke}, {Dijkstra}, {McCourt}, \&
  {Oh}}]{Gronke2016}
{Gronke}, M., {Dijkstra}, M., {McCourt}, M., \& {Oh}, S.~P. 2016, \apjl, 833,
  L26

\bibitem[{{Guaita} {et~al.}(2016){Guaita}, {Pentericci}, {Grazian}, {Vanzella},
  {Nonino}, {Giavalisco}, {Zamorani}, {Bongiorno}, {Cassata}, {Castellano},
  {Garilli}, {Gawiser}, {Le Brun}, {Le F{\`e}vre}, {Lemaux}, {Maccagni},
  {Merlin}, {Santini}, {Tasca}, {Thomas}, {Zucca}, {De Barros}, {Hathi},
  {Amorin}, {Bardelli}, \& {Fontana}}]{Guaita2016}
{Guaita}, L., {Pentericci}, L., {Grazian}, A., {et~al.} 2016, \aap, 587, A133

\bibitem[{{Gunn} \& {Peterson}(1965)}]{GP1965}
{Gunn}, J.~E. \& {Peterson}, B.~A. 1965, \apj, 142, 1633

\bibitem[{{Guseva} {et~al.}(2020){Guseva}, {Izotov}, {Schaerer},
  {V{\'\i}lchez}, {Amor{\'\i}n}, {P{\'e}rez-Montero}, {Iglesias-P{\'a}ramo},
  {Verhamme}, {Kehrig}, \& {Ramambason}}]{Guseva2020}
{Guseva}, N.~G., {Izotov}, Y.~I., {Schaerer}, D., {et~al.} 2020, \mnras, 497,
  4293

\bibitem[{{Harikane} {et~al.}(2020){Harikane}, {Laporte}, {Ellis}, \&
  {Matsuoka}}]{Harikane2020}
{Harikane}, Y., {Laporte}, N., {Ellis}, R.~S., \& {Matsuoka}, Y. 2020, \apj,
  902, 117

\bibitem[{{Heckman} {et~al.}(2011){Heckman}, {Borthakur}, {Overzier},
  {Kauffmann}, {Basu-Zych}, {Leitherer}, {Sembach}, {Martin}, {Rich},
  {Schiminovich}, \& {Seibert}}]{Heckman2011}
{Heckman}, T.~M., {Borthakur}, S., {Overzier}, R., {et~al.} 2011, \apj, 730, 5

\bibitem[{{Heckman} {et~al.}(2001){Heckman}, {Sembach}, {Meurer}, {Leitherer},
  {Calzetti}, \& {Martin}}]{Heckman2001}
{Heckman}, T.~M., {Sembach}, K.~R., {Meurer}, G.~R., {et~al.} 2001, \apj, 558,
  56

\bibitem[{{Henry} {et~al.}(2018){Henry}, {Berg}, {Scarlata}, {Verhamme}, \&
  {Erb}}]{Henry2018}
{Henry}, A., {Berg}, D.~A., {Scarlata}, C., {Verhamme}, A., \& {Erb}, D. 2018,
  \apj, 855, 96

\bibitem[{{Henry} {et~al.}(2015){Henry}, {Scarlata}, {Martin}, \&
  {Erb}}]{Henry2015}
{Henry}, A., {Scarlata}, C., {Martin}, C.~L., \& {Erb}, D. 2015, \apj, 809, 19

\bibitem[{{Hogarth} {et~al.}(2020){Hogarth}, {Amor{\'\i}n}, {V{\'\i}lchez},
  {H{\"a}gele}, {Cardaci}, {P{\'e}rez-Montero}, {Firpo}, {Jaskot}, \&
  {Ch{\'a}vez}}]{Hogarth2020}
{Hogarth}, L., {Amor{\'\i}n}, R., {V{\'\i}lchez}, J.~M., {et~al.} 2020, \mnras,
  494, 3541

\bibitem[{{Inoue}(2010)}]{Inoue2010}
{Inoue}, A.~K. 2010, \mnras, 401, 1325

\bibitem[{{Inoue} {et~al.}(2014){Inoue}, {Shimizu}, {Iwata}, \&
  {Tanaka}}]{Inoue2014}
{Inoue}, A.~K., {Shimizu}, I., {Iwata}, I., \& {Tanaka}, M. 2014, \mnras, 442,
  1805

\bibitem[{{Izotov} {et~al.}(2016{\natexlab{a}}){Izotov}, {Orlitov{\'a}},
  {Schaerer}, {Thuan}, {Verhamme}, {Guseva}, \& {Worseck}}]{Izotov16a}
{Izotov}, Y.~I., {Orlitov{\'a}}, I., {Schaerer}, D., {et~al.}
  2016{\natexlab{a}}, \nat, 529, 178

\bibitem[{{Izotov} {et~al.}(2016{\natexlab{b}}){Izotov}, {Schaerer}, {Thuan},
  {Worseck}, {Guseva}, {Orlitov{\'a}}, \& {Verhamme}}]{Izotov16b}
{Izotov}, Y.~I., {Schaerer}, D., {Thuan}, T.~X., {et~al.} 2016{\natexlab{b}},
  \mnras, 461, 3683

\bibitem[{{Izotov} {et~al.}(2018{\natexlab{a}}){Izotov}, {Schaerer}, {Worseck},
  {Guseva}, {Thuan}, {Verhamme}, {Orlitov{\'a}}, \& {Fricke}}]{Izotov18a}
{Izotov}, Y.~I., {Schaerer}, D., {Worseck}, G., {et~al.} 2018{\natexlab{a}},
  \mnras, 474, 4514

\bibitem[{{Izotov} {et~al.}(2020){Izotov}, {Schaerer}, {Worseck}, {Verhamme},
  {Guseva}, {Thuan}, {Orlitov{\'a}}, \& {Fricke}}]{Izotov2020}
{Izotov}, Y.~I., {Schaerer}, D., {Worseck}, G., {et~al.} 2020, \mnras, 491, 468

\bibitem[{{Izotov} \& {Thuan}(1999)}]{IzotovThuan1999}
{Izotov}, Y.~I. \& {Thuan}, T.~X. 1999, \apj, 511, 639

\bibitem[{{Izotov} {et~al.}(2017){Izotov}, {Thuan}, \& {Guseva}}]{Izotov2017}
{Izotov}, Y.~I., {Thuan}, T.~X., \& {Guseva}, N.~G. 2017, \mnras, 471, 548

\bibitem[{{Izotov} {et~al.}(1994){Izotov}, {Thuan}, \&
  {Lipovetsky}}]{Izotov1994}
{Izotov}, Y.~I., {Thuan}, T.~X., \& {Lipovetsky}, V.~A. 1994, \apj, 435, 647

\bibitem[{{Izotov} {et~al.}(2021){Izotov}, {Worseck}, {Schaerer}, {Guseva},
  {Chisholm}, {Thuan}, {Fricke}, \& {Verhamme}}]{Izotov2021}
{Izotov}, Y.~I., {Worseck}, G., {Schaerer}, D., {et~al.} 2021, \mnras, 503,
  1734

\bibitem[{{Izotov} {et~al.}(2018{\natexlab{b}}){Izotov}, {Worseck}, {Schaerer},
  {Guseva}, {Thuan}, {Fricke}, \& {Orlitov{\'a}}}]{Izotov18b}
{Izotov}, Y.~I., {Worseck}, G., {Schaerer}, D., {et~al.} 2018{\natexlab{b}},
  \mnras, 478, 4851

\bibitem[{{Japelj} {et~al.}(2017){Japelj}, {Vanzella}, {Fontanot}, {Cristiani},
  {Caminha}, {Tozzi}, {Balestra}, {Rosati}, \& {Meneghetti}}]{Japelj2017}
{Japelj}, J., {Vanzella}, E., {Fontanot}, F., {et~al.} 2017, \mnras, 468, 389

\bibitem[{{Jaskot} {et~al.}(2019){Jaskot}, {Dowd}, {Oey}, {Scarlata}, \&
  {McKinney}}]{Jaskot2019}
{Jaskot}, A.~E., {Dowd}, T., {Oey}, M.~S., {Scarlata}, C., \& {McKinney}, J.
  2019, \apj, 885, 96

\bibitem[{{Jaskot} \& {Oey}(2013)}]{JaskotOey2013}
{Jaskot}, A.~E. \& {Oey}, M.~S. 2013, \apj, 766, 91

\bibitem[{{Jaskot} \& {Oey}(2014)}]{Jaskot2014}
{Jaskot}, A.~E. \& {Oey}, M.~S. 2014, \apjl, 791, L19

\bibitem[{{Ji} {et~al.}(2020){Ji}, {Giavalisco}, {Vanzella}, {Siana},
  {Pentericci}, {Jaskot}, {Liu}, {Nonino}, {Ferguson}, {Castellano},
  {Mannucci}, {Schaerer}, {Fynbo}, {Papovich}, {Carnall}, {Amorin}, {Simons},
  {Hathi}, {Cullen}, \& {McLeod}}]{Ji2020}
{Ji}, Z., {Giavalisco}, M., {Vanzella}, E., {et~al.} 2020, \apj, 888, 109

\bibitem[{{Jones} {et~al.}(2012){Jones}, {Stark}, \& {Ellis}}]{Jones2012}
{Jones}, T., {Stark}, D.~P., \& {Ellis}, R.~S. 2012, \apj, 751, 51

\bibitem[{{Jones} {et~al.}(2013){Jones}, {Ellis}, {Schenker}, \&
  {Stark}}]{Jones2013}
{Jones}, T.~A., {Ellis}, R.~S., {Schenker}, M.~A., \& {Stark}, D.~P. 2013,
  \apj, 779, 52

\bibitem[{{Kakiichi} \& {Gronke}(2021)}]{Kakiichi2021}
{Kakiichi}, K. \& {Gronke}, M. 2021, \apj, 908, 30

\bibitem[{{Kelly}(2007)}]{Kelly2007}
{Kelly}, B.~C. 2007, \apj, 665, 1489

\bibitem[{{Kennicutt} \& {De Los Reyes}(2021)}]{Kennicutt2021}
{Kennicutt}, Robert~C., J. \& {De Los Reyes}, M. A.~C. 2021, \apj, 908, 61

\bibitem[{{Kimm} {et~al.}(2019){Kimm}, {Blaizot}, {Garel}, {Michel-Dansac},
  {Katz}, {Rosdahl}, {Verhamme}, \& {Haehnelt}}]{Kimm2019}
{Kimm}, T., {Blaizot}, J., {Garel}, T., {et~al.} 2019, \mnras, 486, 2215

\bibitem[{{Kroupa}(2001)}]{Kroupa2001}
{Kroupa}, P. 2001, \mnras, 322, 231

\bibitem[{{Kulkarni} {et~al.}(2017){Kulkarni}, {Choudhury}, {Puchwein}, \&
  {Haehnelt}}]{Kulkarni2017}
{Kulkarni}, G., {Choudhury}, T.~R., {Puchwein}, E., \& {Haehnelt}, M.~G. 2017,
  \mnras, 469, 4283

\bibitem[{{Lebouteiller} {et~al.}(2006){Lebouteiller}, {Kunth}, {Lequeux},
  {Aloisi}, {D{\'e}sert}, {H{\'e}brard}, {Lecavelier Des {\'E}tangs}, \&
  {Vidal-Madjar}}]{Lebouteiller2006}
{Lebouteiller}, V., {Kunth}, D., {Lequeux}, J., {et~al.} 2006, \aap, 459, 161

\bibitem[{{Lecavelier des Etangs} {et~al.}(2004){Lecavelier des Etangs},
  {D{\'e}sert}, {Kunth}, {Vidal-Madjar}, {Callejo}, {Ferlet}, {H{\'e}brard}, \&
  {Lebouteiller}}]{LdE2004}
{Lecavelier des Etangs}, A., {D{\'e}sert}, J.~M., {Kunth}, D., {et~al.} 2004,
  \aap, 413, 131

\bibitem[{{Leethochawalit} {et~al.}(2016){Leethochawalit}, {Jones}, {Ellis},
  {Stark}, \& {Zitrin}}]{Leethochawalit2016}
{Leethochawalit}, N., {Jones}, T.~A., {Ellis}, R.~S., {Stark}, D.~P., \&
  {Zitrin}, A. 2016, \apj, 831, 152

\bibitem[{{Leitet} {et~al.}(2013){Leitet}, {Bergvall}, {Hayes}, {Linn{\'e}}, \&
  {Zackrisson}}]{Leitet2013}
{Leitet}, E., {Bergvall}, N., {Hayes}, M., {Linn{\'e}}, S., \& {Zackrisson}, E.
  2013, \aap, 553, A106

\bibitem[{{Leitherer} {et~al.}(2014){Leitherer}, {Ekstr{\"o}m}, {Meynet},
  {Schaerer}, {Agienko}, \& {Levesque}}]{Leitherer2014}
{Leitherer}, C., {Ekstr{\"o}m}, S., {Meynet}, G., {et~al.} 2014, \apjs, 212, 14

\bibitem[{{Leitherer} {et~al.}(2016){Leitherer}, {Hernandez}, {Lee}, \&
  {Oey}}]{Leitherer2016}
{Leitherer}, C., {Hernandez}, S., {Lee}, J.~C., \& {Oey}, M.~S. 2016, \apj,
  823, 64

\bibitem[{{Leitherer} {et~al.}(2010){Leitherer}, {Ortiz Ot{\'a}lvaro},
  {Bresolin}, {Kudritzki}, {Lo Faro}, {Pauldrach}, {Pettini}, \&
  {Rix}}]{Leitherer2010}
{Leitherer}, C., {Ortiz Ot{\'a}lvaro}, P.~A., {Bresolin}, F., {et~al.} 2010,
  \apjs, 189, 309

\bibitem[{{Leitherer} {et~al.}(2011{\natexlab{a}}){Leitherer}, {Schaerer},
  {Goldader}, {Gonzalez-Delgado}, {Robert}, {Foo Kune}, {de Mello}, {Devost},
  {Heckman}, {Aloisi}, {Martins}, \& {Vazquez}}]{Leitherer2011b}
{Leitherer}, C., {Schaerer}, D., {Goldader}, J., {et~al.} 2011{\natexlab{a}},
  {Starburst99: Synthesis Models for Galaxies with Active Star Formation}

\bibitem[{{Leitherer} {et~al.}(2011{\natexlab{b}}){Leitherer}, {Tremonti},
  {Heckman}, \& {Calzetti}}]{Leitherer2011a}
{Leitherer}, C., {Tremonti}, C.~A., {Heckman}, T.~M., \& {Calzetti}, D.
  2011{\natexlab{b}}, \aj, 141, 37

\bibitem[{{Lusso} {et~al.}(2015){Lusso}, {Worseck}, {Hennawi}, {Prochaska},
  {Vignali}, {Stern}, \& {O'Meara}}]{Lusso2015}
{Lusso}, E., {Worseck}, G., {Hennawi}, J.~F., {et~al.} 2015, \mnras, 449, 4204

\bibitem[{{Ma} {et~al.}(2020){Ma}, {Quataert}, {Wetzel}, {Hopkins},
  {Faucher-Gigu{\`e}re}, \& {Kere{\v{s}}}}]{Ma2020}
{Ma}, X., {Quataert}, E., {Wetzel}, A., {et~al.} 2020, \mnras, 498, 2001

\bibitem[{{Madau} \& {Haardt}(2015)}]{Madau2015}
{Madau}, P. \& {Haardt}, F. 2015, \apjl, 813, L8

\bibitem[{{Makan} {et~al.}(2021){Makan}, {Worseck}, {Davies}, {Hennawi},
  {Prochaska}, \& {Richter}}]{Makan2021}
{Makan}, K., {Worseck}, G., {Davies}, F.~B., {et~al.} 2021, \apj, 912, 38

\bibitem[{{Malkan} \& {Malkan}(2021)}]{Malkan2021}
{Malkan}, M.~A. \& {Malkan}, B.~K. 2021, \apj, 909, 92

\bibitem[{{Marchi} {et~al.}(2017){Marchi}, {Pentericci}, {Guaita}, {Ribeiro},
  {Castellano}, {Schaerer}, {Hathi}, {Lemaux}, {Grazian}, {Le F{\`e}vre},
  {Garilli}, {Maccagni}, {Amorin}, {Bardelli}, {Cassata}, {Fontana},
  {Koekemoer}, {Le Brun}, {Tasca}, {Thomas}, {Vanzella}, {Zamorani}, \&
  {Zucca}}]{Marchi2017}
{Marchi}, F., {Pentericci}, L., {Guaita}, L., {et~al.} 2017, \aap, 601, A73

\bibitem[{{Marchi} {et~al.}(2018){Marchi}, {Pentericci}, {Guaita}, {Schaerer},
  {Verhamme}, {Castellano}, {Ribeiro}, {Garilli}, {F{\`e}vre}, {Amorin},
  {Bardelli}, {Cassata}, {Durkalec}, {Grazian}, {Hathi}, {Lemaux}, {Maccagni},
  {Vanzella}, \& {Zucca}}]{Marchi2018Lyalpha-Lyman-c}
{Marchi}, F., {Pentericci}, L., {Guaita}, L., {et~al.} 2018, \aap, 614, A11

\bibitem[{{Matsuoka} {et~al.}(2018){Matsuoka}, {Strauss}, {Kashikawa}, {Onoue},
  {Iwasawa}, {Tang}, {Lee}, {Imanishi}, {Nagao}, {Akiyama}, {Asami}, {Bosch},
  {Furusawa}, {Goto}, {Gunn}, {Harikane}, {Ikeda}, {Izumi}, {Kawaguchi},
  {Kato}, {Kikuta}, {Kohno}, {Komiyama}, {Lupton}, {Minezaki}, {Miyazaki},
  {Murayama}, {Niida}, {Nishizawa}, {Noboriguchi}, {Oguri}, {Ono}, {Ouchi},
  {Price}, {Sameshima}, {Schulze}, {Shirakata}, {Silverman}, {Sugiyama},
  {Tait}, {Takada}, {Takata}, {Tanaka}, {Toba}, {Utsumi}, {Wang}, \&
  {Yamashita}}]{Matsuoka2018}
{Matsuoka}, Y., {Strauss}, M.~A., {Kashikawa}, N., {et~al.} 2018, \apj, 869,
  150

\bibitem[{{Mauerhofer} {et~al.}(2021){Mauerhofer}, {Verhamme}, {Blaizot},
  {Garel}, {Kimm}, {Michel-Dansac}, \& {Rosdahl}}]{Mauerhofer2020}
{Mauerhofer}, V., {Verhamme}, A., {Blaizot}, J., {et~al.} 2021, \aap, 646, A80

\bibitem[{{McCandliss} \& {O'Meara}(2017)}]{McCandliss2017}
{McCandliss}, S.~R. \& {O'Meara}, J.~M. 2017, \apj, 845, 111

\bibitem[{{McKinney} {et~al.}(2019){McKinney}, {Jaskot}, {Oey}, {Yun}, {Dowd},
  \& {Lowenthal}}]{McKinney2019}
{McKinney}, J.~H., {Jaskot}, A.~E., {Oey}, M.~S., {et~al.} 2019, \apj, 874, 52

\bibitem[{{Me{\v{s}}tri{\'c}} {et~al.}(2021){Me{\v{s}}tri{\'c}}, {Ryan-Weber},
  {Cooke}, {Bassett}, {Prichard}, \& {Rafelski}}]{Mestric2021}
{Me{\v{s}}tri{\'c}}, U., {Ryan-Weber}, E.~V., {Cooke}, J., {et~al.} 2021,
  \mnras, 508, 4443

\bibitem[{{Meynet} {et~al.}(1994){Meynet}, {Maeder}, {Schaller}, {Schaerer}, \&
  {Charbonnel}}]{Meynet1994}
{Meynet}, G., {Maeder}, A., {Schaller}, G., {Schaerer}, D., \& {Charbonnel}, C.
  1994, \aaps, 103, 97

\bibitem[{{Micheva} {et~al.}(2017{\natexlab{a}}){Micheva}, {Iwata}, \&
  {Inoue}}]{Micheva2017a}
{Micheva}, G., {Iwata}, I., \& {Inoue}, A.~K. 2017{\natexlab{a}}, \mnras, 465,
  302

\bibitem[{{Micheva} {et~al.}(2017{\natexlab{b}}){Micheva}, {Iwata}, {Inoue},
  {Matsuda}, {Yamada}, \& {Hayashino}}]{Micheva2017b}
{Micheva}, G., {Iwata}, I., {Inoue}, A.~K., {et~al.} 2017{\natexlab{b}},
  \mnras, 465, 316

\bibitem[{{Morrissey} {et~al.}(2007){Morrissey}, {Conrow}, {Barlow}, {Small},
  {Seibert}, {Wyder}, {Budav{\'a}ri}, {Arnouts}, {Friedman}, {Forster},
  {Martin}, {Neff}, {Schiminovich}, {Bianchi}, {Donas}, {Heckman}, {Lee},
  {Madore}, {Milliard}, {Rich}, {Szalay}, {Welsh}, \& {Yi}}]{Morrissey2007}
{Morrissey}, P., {Conrow}, T., {Barlow}, T.~A., {et~al.} 2007, \apjs, 173, 682

\bibitem[{{Mostardi} {et~al.}(2015){Mostardi}, {Shapley}, {Steidel}, {Trainor},
  {Reddy}, \& {Siana}}]{Mostardi2015}
{Mostardi}, R.~E., {Shapley}, A.~E., {Steidel}, C.~C., {et~al.} 2015, \apj,
  810, 107

\bibitem[{{Muratov} {et~al.}(2015){Muratov}, {Kere{\v{s}}},
  {Faucher-Gigu{\`e}re}, {Hopkins}, {Quataert}, \& {Murray}}]{Muratov2015}
{Muratov}, A.~L., {Kere{\v{s}}}, D., {Faucher-Gigu{\`e}re}, C.-A., {et~al.}
  2015, \mnras, 454, 2691

\bibitem[{{Naidu} {et~al.}(2018){Naidu}, {Forrest}, {Oesch}, {Tran}, \&
  {Holden}}]{Naidu2018}
{Naidu}, R.~P., {Forrest}, B., {Oesch}, P.~A., {Tran}, K.-V.~H., \& {Holden},
  B.~P. 2018, \mnras, 478, 791

\bibitem[{{Naidu} {et~al.}(2020){Naidu}, {Tacchella}, {Mason}, {Bose}, {Oesch},
  \& {Conroy}}]{Naidu2020}
{Naidu}, R.~P., {Tacchella}, S., {Mason}, C.~A., {et~al.} 2020, \apj, 892, 109

\bibitem[{{Nakajima} {et~al.}(2016){Nakajima}, {Ellis}, {Iwata}, {Inoue},
  {Kusakabe}, {Ouchi}, \& {Robertson}}]{Nakajima2016}
{Nakajima}, K., {Ellis}, R.~S., {Iwata}, I., {et~al.} 2016, \apjl, 831, L9

\bibitem[{{Nakajima} {et~al.}(2020){Nakajima}, {Ellis}, {Robertson}, {Tang}, \&
  {Stark}}]{Nakajima2020}
{Nakajima}, K., {Ellis}, R.~S., {Robertson}, B.~E., {Tang}, M., \& {Stark},
  D.~P. 2020, \apj, 889, 161

\bibitem[{{Nakajima} \& {Ouchi}(2014)}]{Nakajima2014}
{Nakajima}, K. \& {Ouchi}, M. 2014, \mnras, 442, 900

\bibitem[{{Nakajima} {et~al.}(2018){Nakajima}, {Schaerer}, {Le F{\`e}vre},
  {Amor{\'\i}n}, {Talia}, {Lemaux}, {Tasca}, {Vanzella}, {Zamorani},
  {Bardelli}, {Grazian}, {Guaita}, {Hathi}, {Pentericci}, \&
  {Zucca}}]{Nakajima2018}
{Nakajima}, K., {Schaerer}, D., {Le F{\`e}vre}, O., {et~al.} 2018, \aap, 612,
  A94

\bibitem[{{Newville} {et~al.}(2016){Newville}, {Stensitzki}, {Allen}, {Rawlik},
  {Ingargiola}, \& {Nelson}}]{lmfit}
{Newville}, M., {Stensitzki}, T., {Allen}, D.~B., {et~al.} 2016, {Lmfit:
  Non-Linear Least-Square Minimization and Curve-Fitting for Python}

\bibitem[{{Orlitov{\'a}} {et~al.}(2018){Orlitov{\'a}}, {Verhamme}, {Henry},
  {Scarlata}, {Jaskot}, {Oey}, \& {Schaerer}}]{Orlitova2018}
{Orlitov{\'a}}, I., {Verhamme}, A., {Henry}, A., {et~al.} 2018, \aap, 616, A60

\bibitem[{{{\"O}stlin} {et~al.}(2021){{\"O}stlin}, {Rivera-Thorsen}, {Menacho},
  {Hayes}, {Runnholm}, {Micheva}, {Oey}, {Adamo}, {Bik}, {Cannon}, {Gronke},
  {Kunth}, {Laursen}, {Mas-Hesse}, {Melinder}, {Messa}, {Sirressi}, \&
  {Smith}}]{Ostlin2021}
{{\"O}stlin}, G., {Rivera-Thorsen}, T.~E., {Menacho}, V., {et~al.} 2021, \apj,
  912, 155

\bibitem[{{Pahl} {et~al.}(2020){Pahl}, {Shapley}, {Faisst}, {Capak}, {Du},
  {Reddy}, {Laursen}, \& {Topping}}]{Pahl2020}
{Pahl}, A.~J., {Shapley}, A., {Faisst}, A.~L., {et~al.} 2020, \mnras, 493, 3194

\bibitem[{{Pahl} {et~al.}(2021){Pahl}, {Shapley}, {Steidel}, {Chen}, \&
  {Reddy}}]{Pahl2021}
{Pahl}, A.~J., {Shapley}, A., {Steidel}, C.~C., {Chen}, Y., \& {Reddy}, N.~A.
  2021, \mnras, 505, 2447

\bibitem[{{Pauldrach} {et~al.}(2001){Pauldrach}, {Hoffmann}, \&
  {Lennon}}]{Pauldrach2001}
{Pauldrach}, A.~W.~A., {Hoffmann}, T.~L., \& {Lennon}, M. 2001, \aap, 375, 161

\bibitem[{{Planck Collaboration} {et~al.}(2016){Planck Collaboration}, {Adam},
  {Aghanim}, {Ashdown}, {Aumont}, {Baccigalupi}, {Ballardini}, {Banday},
  {Barreiro}, {Bartolo}, {Basak}, {Battye}, {Benabed}, {Bernard}, {Bersanelli},
  {Bielewicz}, {Bock}, {Bonaldi}, {Bonavera}, {Bond}, {Borrill}, {Bouchet},
  {Boulanger}, {Bucher}, {Burigana}, {Calabrese}, {Cardoso}, {Carron},
  {Chiang}, {Colombo}, {Combet}, {Comis}, {Couchot}, {Coulais}, {Crill},
  {Curto}, {Cuttaia}, {Davis}, {de Bernardis}, {de Rosa}, {de Zotti},
  {Delabrouille}, {Di Valentino}, {Dickinson}, {Diego}, {Dor{\'e}}, {Douspis},
  {Ducout}, {Dupac}, {Elsner}, {En{\ss}lin}, {Eriksen}, {Falgarone}, {Fantaye},
  {Finelli}, {Forastieri}, {Frailis}, {Fraisse}, {Franceschi}, {Frolov},
  {Galeotta}, {Galli}, {Ganga}, {G{\'e}nova-Santos}, {Gerbino}, {Ghosh},
  {Gonz{\'a}lez-Nuevo}, {G{\'o}rski}, {Gruppuso}, {Gudmundsson}, {Hansen},
  {Helou}, {Henrot-Versill{\'e}}, {Herranz}, {Hivon}, {Huang}, {Ili{\'c}},
  {Jaffe}, {Jones}, {Keih{\"a}nen}, {Keskitalo}, {Kisner}, {Knox},
  {Krachmalnicoff}, {Kunz}, {Kurki-Suonio}, {Lagache}, {L{\"a}hteenm{\"a}ki},
  {Lamarre}, {Langer}, {Lasenby}, {Lattanzi}, {Lawrence}, {Le Jeune},
  {Levrier}, {Lewis}, {Liguori}, {Lilje}, {L{\'o}pez-Caniego}, {Ma},
  {Mac{\'\i}as-P{\'e}rez}, {Maggio}, {Mangilli}, {Maris}, {Martin},
  {Mart{\'\i}nez-Gonz{\'a}lez}, {Matarrese}, {Mauri}, {McEwen}, {Meinhold},
  {Melchiorri}, {Mennella}, {Migliaccio}, {Miville-Desch{\^e}nes}, {Molinari},
  {Moneti}, {Montier}, {Morgante}, {Moss}, {Naselsky}, {Natoli}, {Oxborrow},
  {Pagano}, {Paoletti}, {Partridge}, {Patanchon}, {Patrizii}, {Perdereau},
  {Perotto}, {Pettorino}, {Piacentini}, {Plaszczynski}, {Polastri}, {Polenta},
  {Puget}, {Rachen}, {Racine}, {Reinecke}, {Remazeilles}, {Renzi}, {Rocha},
  {Rossetti}, {Roudier}, {Rubi{\~n}o-Mart{\'\i}n}, {Ruiz-Granados}, {Salvati},
  {Sandri}, {Savelainen}, {Scott}, {Sirri}, {Sunyaev}, {Suur-Uski}, {Tauber},
  {Tenti}, {Toffolatti}, {Tomasi}, {Tristram}, {Trombetti}, {Valiviita}, {Van
  Tent}, {Vielva}, {Villa}, {Vittorio}, {Wandelt}, {Wehus}, {White}, {Zacchei},
  \& {Zonca}}]{Planck2016}
{Planck Collaboration}, {Adam}, R., {Aghanim}, N., {et~al.} 2016, \aap, 596,
  A108

\bibitem[{{Prevot} {et~al.}(1984){Prevot}, {Lequeux}, {Maurice}, {Prevot}, \&
  {Rocca-Volmerange}}]{SMC84}
{Prevot}, M.~L., {Lequeux}, J., {Maurice}, E., {Prevot}, L., \&
  {Rocca-Volmerange}, B. 1984, \aap, 132, 389

\bibitem[{{Puschnig} {et~al.}(2017){Puschnig}, {Hayes}, {{\"O}stlin},
  {Rivera-Thorsen}, {Melinder}, {Cannon}, {Menacho}, {Zackrisson}, {Bergvall},
  \& {Leitet}}]{Puschnig2017}
{Puschnig}, J., {Hayes}, M., {{\"O}stlin}, G., {et~al.} 2017, \mnras, 469, 3252

\bibitem[{{Ramambason} {et~al.}(2020){Ramambason}, {Schaerer}, {Stasi{\'n}ska},
  {Izotov}, {Guseva}, {V{\'\i}lchez}, {Amor{\'\i}n}, \&
  {Morisset}}]{Ramambason2020}
{Ramambason}, L., {Schaerer}, D., {Stasi{\'n}ska}, G., {et~al.} 2020, \aap,
  644, A21

\bibitem[{{Reddy} {et~al.}(2015){Reddy}, {Kriek}, {Shapley}, {Freeman},
  {Siana}, {Coil}, {Mobasher}, {Price}, {Sanders}, \& {Shivaei}}]{Reddy2015}
{Reddy}, N.~A., {Kriek}, M., {Shapley}, A.~E., {et~al.} 2015, \apj, 806, 259

\bibitem[{{Reddy} {et~al.}(2016{\natexlab{a}}){Reddy}, {Steidel}, {Pettini}, \&
  {Bogosavljevi{\'c}}}]{R16}
{Reddy}, N.~A., {Steidel}, C.~C., {Pettini}, M., \& {Bogosavljevi{\'c}}, M.
  2016{\natexlab{a}}, \apj, 828, 107

\bibitem[{{Reddy} {et~al.}(2016{\natexlab{b}}){Reddy}, {Steidel}, {Pettini},
  {Bogosavljevi{\'c}}, \& {Shapley}}]{Reddy2016b}
{Reddy}, N.~A., {Steidel}, C.~C., {Pettini}, M., {Bogosavljevi{\'c}}, M., \&
  {Shapley}, A.~E. 2016{\natexlab{b}}, \apj, 828, 108

\bibitem[{{Reddy} {et~al.}(2021){Reddy}, {Topping}, {Shapley}, {Steidel},
  {Sanders}, {Du}, {Coil}, {Mobasher}, \& {Price}}]{Reddy2021}
{Reddy}, N.~A., {Topping}, M.~W., {Shapley}, A.~E., {et~al.} 2021, arXiv
  e-prints, arXiv:2108.05363

\bibitem[{{Rivera-Thorsen} {et~al.}(2019){Rivera-Thorsen}, {Dahle}, {Chisholm},
  {Florian}, {Gronke}, {Rigby}, {Gladders}, {Mahler}, {Sharon}, \&
  {Bayliss}}]{RT2019}
{Rivera-Thorsen}, T.~E., {Dahle}, H., {Chisholm}, J., {et~al.} 2019, Science,
  366, 738

\bibitem[{{Rivera-Thorsen} {et~al.}(2017){Rivera-Thorsen}, {Dahle}, {Gronke},
  {Bayliss}, {Rigby}, {Simcoe}, {Bordoloi}, {Turner}, \&
  {Furesz}}]{Ribera-Thorsen2017}
{Rivera-Thorsen}, T.~E., {Dahle}, H., {Gronke}, M., {et~al.} 2017, \aap, 608,
  L4

\bibitem[{{Rivera-Thorsen} {et~al.}(2015){Rivera-Thorsen}, {Hayes},
  {{\"O}stlin}, {Duval}, {Orlitov{\'a}}, {Verhamme}, {Mas-Hesse}, {Schaerer},
  {Cannon}, {Ot{\'\i}-Floranes}, {Sandberg}, {Guaita}, {Adamo}, {Atek},
  {Herenz}, {Kunth}, {Laursen}, \& {Melinder}}]{RiberaThorsen2015}
{Rivera-Thorsen}, T.~E., {Hayes}, M., {{\"O}stlin}, G., {et~al.} 2015, \apj,
  805, 14

\bibitem[{{Robertson} {et~al.}(2015){Robertson}, {Ellis}, {Furlanetto}, \&
  {Dunlop}}]{Robertson2015}
{Robertson}, B.~E., {Ellis}, R.~S., {Furlanetto}, S.~R., \& {Dunlop}, J.~S.
  2015, \apjl, 802, L19

\bibitem[{{Robertson} {et~al.}(2013){Robertson}, {Furlanetto}, {Schneider},
  {Charlot}, {Ellis}, {Stark}, {McLure}, {Dunlop}, {Koekemoer}, {Schenker},
  {Ouchi}, {Ono}, {Curtis-Lake}, {Rogers}, {Bowler}, \&
  {Cirasuolo}}]{Robertson2013}
{Robertson}, B.~E., {Furlanetto}, S.~R., {Schneider}, E., {et~al.} 2013, \apj,
  768, 71

\bibitem[{{Rosdahl} {et~al.}(2018){Rosdahl}, {Katz}, {Blaizot}, {Kimm},
  {Michel-Dansac}, {Garel}, {Haehnelt}, {Ocvirk}, \& {Teyssier}}]{Rosdahl2018}
{Rosdahl}, J., {Katz}, H., {Blaizot}, J., {et~al.} 2018, \mnras, 479, 994

\bibitem[{{Rutkowski} {et~al.}(2017){Rutkowski}, {Scarlata}, {Henry}, {Hayes},
  {Mehta}, {Hathi}, {Cohen}, {Windhorst}, {Koekemoer}, {Teplitz}, {Haardt}, \&
  {Siana}}]{Rutkowski2017}
{Rutkowski}, M.~J., {Scarlata}, C., {Henry}, A., {et~al.} 2017, \apjl, 841, L27

\bibitem[{{Saha} {et~al.}(2020){Saha}, {Tandon}, {Simmonds}, {Verhamme},
  {Paswan}, {Schaerer}, {Rutkowski}, {Borgohain}, {Elmegreen}, {Inoue},
  {Combes}, {Elmegreen}, \& {Paalvast}}]{Saha2020}
{Saha}, K., {Tandon}, S.~N., {Simmonds}, C., {et~al.} 2020, Nature Astronomy,
  4, 1185

\bibitem[{{Salim} {et~al.}(2018){Salim}, {Boquien}, \& {Lee}}]{Salim2018}
{Salim}, S., {Boquien}, M., \& {Lee}, J.~C. 2018, \apj, 859, 11

\bibitem[{{Scarlata} {et~al.}(2009){Scarlata}, {Colbert}, {Teplitz}, {Panagia},
  {Hayes}, {Siana}, {Rau}, {Francis}, {Caon}, {Pizzella}, \&
  {Bridge}}]{Scarlata2009}
{Scarlata}, C., {Colbert}, J., {Teplitz}, H.~I., {et~al.} 2009, \apjl, 704, L98

\bibitem[{{Scarlata} \& {Panagia}(2015)}]{Scarlata2015}
{Scarlata}, C. \& {Panagia}, N. 2015, \apj, 801, 43

\bibitem[{{Schaerer} {et~al.}(2013){Schaerer}, {de Barros}, \&
  {Sklias}}]{Schaerer2013}
{Schaerer}, D., {de Barros}, S., \& {Sklias}, P. 2013, \aap, 549, A4

\bibitem[{{Schaerer} {et~al.}(2019){Schaerer}, {Fragos}, \&
  {Izotov}}]{Schaerer2019}
{Schaerer}, D., {Fragos}, T., \& {Izotov}, Y.~I. 2019, \aap, 622, L10

\bibitem[{{Schaerer} {et~al.}(2016){Schaerer}, {Izotov}, {Verhamme},
  {Orlitov{\'a}}, {Thuan}, {Worseck}, \& {Guseva}}]{Schaerer2016}
{Schaerer}, D., {Izotov}, Y.~I., {Verhamme}, A., {et~al.} 2016, \aap, 591, L8

\bibitem[{{Shapley} {et~al.}(2003){Shapley}, {Steidel}, {Pettini}, \&
  {Adelberger}}]{Shapley2003}
{Shapley}, A.~E., {Steidel}, C.~C., {Pettini}, M., \& {Adelberger}, K.~L. 2003,
  \apj, 588, 65

\bibitem[{{Shapley} {et~al.}(2006){Shapley}, {Steidel}, {Pettini},
  {Adelberger}, \& {Erb}}]{Shapley2006}
{Shapley}, A.~E., {Steidel}, C.~C., {Pettini}, M., {Adelberger}, K.~L., \&
  {Erb}, D.~K. 2006, \apj, 651, 688

\bibitem[{{Shapley} {et~al.}(2016){Shapley}, {Steidel}, {Strom},
  {Bogosavljevi{\'c}}, {Reddy}, {Siana}, {Mostardi}, \& {Rudie}}]{Shapley2016}
{Shapley}, A.~E., {Steidel}, C.~C., {Strom}, A.~L., {et~al.} 2016, \apjl, 826,
  L24

\bibitem[{{Shivaei} {et~al.}(2020){Shivaei}, {Reddy}, {Rieke}, {Shapley},
  {Kriek}, {Battisti}, {Mobasher}, {Sanders}, {Fetherolf}, {Azadi}, {Coil},
  {Freeman}, {de Groot}, {Leung}, {Price}, {Siana}, \& {Zick}}]{Shivaei2020}
{Shivaei}, I., {Reddy}, N., {Rieke}, G., {et~al.} 2020, \apj, 899, 117

\bibitem[{{Siana} {et~al.}(2015){Siana}, {Shapley}, {Kulas}, {Nestor},
  {Steidel}, {Teplitz}, {Alavi}, {Brown}, {Conselice}, {Ferguson}, {Dickinson},
  {Giavalisco}, {Colbert}, {Bridge}, {Gardner}, \& {de Mello}}]{Siana2015}
{Siana}, B., {Shapley}, A.~E., {Kulas}, K.~R., {et~al.} 2015, \apj, 804, 17

\bibitem[{{Smith} {et~al.}(2015){Smith}, {Safranek-Shrader}, {Bromm}, \&
  {Milosavljevi{\'c}}}]{Smith2015}
{Smith}, A., {Safranek-Shrader}, C., {Bromm}, V., \& {Milosavljevi{\'c}}, M.
  2015, \mnras, 449, 4336

\bibitem[{{Smith} {et~al.}(2020){Smith}, {Windhorst}, {Cohen}, {Koekemoer},
  {Jansen}, {White}, {Borthakur}, {Hathi}, {Jiang}, {Rutkowski}, {Ryan},
  {Inoue}, {O'Connell}, {MacKenty}, {Conselice}, \& {Silk}}]{Smith2020}
{Smith}, B.~M., {Windhorst}, R.~A., {Cohen}, S.~H., {et~al.} 2020, \apj, 897,
  41

\bibitem[{{Stasi{\'n}ska} {et~al.}(2015){Stasi{\'n}ska}, {Izotov}, {Morisset},
  \& {Guseva}}]{Stasinska2015}
{Stasi{\'n}ska}, G., {Izotov}, Y., {Morisset}, C., \& {Guseva}, N. 2015, \aap,
  576, A83

\bibitem[{{Steidel} {et~al.}(2018){Steidel}, {Bogosavljevi{\'c}}, {Shapley},
  {Reddy}, {Rudie}, {Pettini}, {Trainor}, \& {Strom}}]{Steidel2018}
{Steidel}, C.~C., {Bogosavljevi{\'c}}, M., {Shapley}, A.~E., {et~al.} 2018,
  \apj, 869, 123

\bibitem[{{Stevans} {et~al.}(2014){Stevans}, {Shull}, {Danforth}, \&
  {Tilton}}]{Stevans2014}
{Stevans}, M.~L., {Shull}, J.~M., {Danforth}, C.~W., \& {Tilton}, E.~M. 2014,
  \apj, 794, 75

\bibitem[{{Sugahara} {et~al.}(2019){Sugahara}, {Ouchi}, {Harikane},
  {Bouch{\'e}}, {Mitchell}, \& {Blaizot}}]{Sugahara2019}
{Sugahara}, Y., {Ouchi}, M., {Harikane}, Y., {et~al.} 2019, \apj, 886, 29

\bibitem[{{Trainor} {et~al.}(2015){Trainor}, {Steidel}, {Strom}, \&
  {Rudie}}]{Trainor2015}
{Trainor}, R.~F., {Steidel}, C.~C., {Strom}, A.~L., \& {Rudie}, G.~C. 2015,
  \apj, 809, 89

\bibitem[{{Trainor} {et~al.}(2019){Trainor}, {Strom}, {Steidel}, {Rudie},
  {Chen}, \& {Theios}}]{Trainor2019}
{Trainor}, R.~F., {Strom}, A.~L., {Steidel}, C.~C., {et~al.} 2019, \apj, 887,
  85

\bibitem[{{Trebitsch} {et~al.}(2017){Trebitsch}, {Blaizot}, {Rosdahl},
  {Devriendt}, \& {Slyz}}]{Trebitsch2017}
{Trebitsch}, M., {Blaizot}, J., {Rosdahl}, J., {Devriendt}, J., \& {Slyz}, A.
  2017, \mnras, 470, 224

\bibitem[{{Vanzella} {et~al.}(2020){Vanzella}, {Caminha}, {Calura}, {Cupani},
  {Meneghetti}, {Castellano}, {Rosati}, {Mercurio}, {Sani}, {Grillo}, {Gilli},
  {Mignoli}, {Comastri}, {Nonino}, {Cristiani}, {Giavalisco}, \&
  {Caputi}}]{Vanzella2020}
{Vanzella}, E., {Caminha}, G.~B., {Calura}, F., {et~al.} 2020, \mnras, 491,
  1093

\bibitem[{{Vanzella} {et~al.}(2015){Vanzella}, {de Barros}, {Castellano},
  {Grazian}, {Inoue}, {Schaerer}, {Guaita}, {Zamorani}, {Giavalisco}, {Siana},
  {Pentericci}, {Giallongo}, {Fontana}, \& {Vignali}}]{Vanzella2015}
{Vanzella}, E., {de Barros}, S., {Castellano}, M., {et~al.} 2015, \aap, 576,
  A116

\bibitem[{{Vanzella} {et~al.}(2016){Vanzella}, {de Barros}, {Vasei}, {Alavi},
  {Giavalisco}, {Siana}, {Grazian}, {Hasinger}, {Suh}, {Cappelluti}, {Vito},
  {Amorin}, {Balestra}, {Brusa}, {Calura}, {Castellano}, {Comastri}, {Fontana},
  {Gilli}, {Mignoli}, {Pentericci}, {Vignali}, \& {Zamorani}}]{Vanzella2016}
{Vanzella}, E., {de Barros}, S., {Vasei}, K., {et~al.} 2016, \apj, 825, 41

\bibitem[{{Vanzella} {et~al.}(2012){Vanzella}, {Guo}, {Giavalisco}, {Grazian},
  {Castellano}, {Cristiani}, {Dickinson}, {Fontana}, {Nonino}, {Giallongo},
  {Pentericci}, {Galametz}, {Faber}, {Ferguson}, {Grogin}, {Koekemoer},
  {Newman}, \& {Siana}}]{Vanzella2012}
{Vanzella}, E., {Guo}, Y., {Giavalisco}, M., {et~al.} 2012, \apj, 751, 70

\bibitem[{{Vanzella} {et~al.}(2018){Vanzella}, {Nonino}, {Cupani},
  {Castellano}, {Sani}, {Mignoli}, {Calura}, {Meneghetti}, {Gilli}, {Comastri},
  {Mercurio}, {Caminha}, {Caputi}, {Rosati}, {Grillo}, {Cristiani}, {Balestra},
  {Fontana}, \& {Giavalisco}}]{Vanzella2018}
{Vanzella}, E., {Nonino}, M., {Cupani}, G., {et~al.} 2018, \mnras, 476, L15

\bibitem[{{Vanzella} {et~al.}(2010){Vanzella}, {Siana}, {Cristiani}, \&
  {Nonino}}]{Vanzella2010}
{Vanzella}, E., {Siana}, B., {Cristiani}, S., \& {Nonino}, M. 2010, \mnras,
  404, 1672

\bibitem[{{Vasei} {et~al.}(2016){Vasei}, {Siana}, {Shapley}, {Quider}, {Alavi},
  {Rafelski}, {Steidel}, {Pettini}, \& {Lewis}}]{Vasei2016}
{Vasei}, K., {Siana}, B., {Shapley}, A.~E., {et~al.} 2016, \apj, 831, 38

\bibitem[{{Verhamme} {et~al.}(2015){Verhamme}, {Orlitov{\'a}}, {Schaerer}, \&
  {Hayes}}]{Verhamme2015}
{Verhamme}, A., {Orlitov{\'a}}, I., {Schaerer}, D., \& {Hayes}, M. 2015, \aap,
  578, A7

\bibitem[{{Verhamme} {et~al.}(2017){Verhamme}, {Orlitov{\'a}}, {Schaerer},
  {Izotov}, {Worseck}, {Thuan}, \& {Guseva}}]{Verhamme2017}
{Verhamme}, A., {Orlitov{\'a}}, I., {Schaerer}, D., {et~al.} 2017, \aap, 597,
  A13

\bibitem[{{Wang} {et~al.}(2021){Wang}, {Heckman}, {Amor{\'\i}n}, {Borthakur},
  {Chisholm}, {Ferguson}, {Flury}, {Giavalisco}, {Grazian}, {Hayes}, {Henry},
  {Jaskot}, {Ji}, {Makan}, {McCandliss}, {Oey}, {{\"O}stlin}, {Saldana-Lopez},
  {Schaerer}, {Thuan}, {Worseck}, \& {Xu}}]{Wang2021}
{Wang}, B., {Heckman}, T.~M., {Amor{\'\i}n}, R., {et~al.} 2021, \apj, 916, 3

\bibitem[{{Wang} {et~al.}(2019){Wang}, {Heckman}, {Leitherer}, {Alexandroff},
  {Borthakur}, \& {Overzier}}]{Wang2019}
{Wang}, B., {Heckman}, T.~M., {Leitherer}, C., {et~al.} 2019, \apj, 885, 57

\bibitem[{{Xiao} {et~al.}(2018){Xiao}, {Stanway}, \& {Eldridge}}]{Xiao2018}
{Xiao}, L., {Stanway}, E.~R., \& {Eldridge}, J.~J. 2018, \mnras, 477, 904

\bibitem[{{Zackrisson} {et~al.}(2013){Zackrisson}, {Inoue}, \&
  {Jensen}}]{Zackarisson2013}
{Zackrisson}, E., {Inoue}, A.~K., \& {Jensen}, H. 2013, \apj, 777, 39

\end{thebibliography}

\appendix
\section{Fits and absorption lines results}\label{appA}
Summary tables with the main products from our UV--SED fits (+errors) and the absorption line measurements (+errors) are shown in this Appendix for both the LzLCS and literature samples. Tables \ref{tab:fits_results1} and \ref{tab:fits_results2} include the \ebv\, average stellar age and metallicities, inferred \fescabs, and intrinsic (dust-free) and observed UV absolute magnitudes (in AB system). Tables \ref{tab:lines_results1} and \ref{tab:lines_results2} include the average \hi\ and LIS equivalent widths, residual fluxes, and derived \feschi\ and \fesclis. Sources are identified and sorted by its coordinates, and the strong/weak/nonleaker categories are differentiated according to the LzLCS criteria.

\begin{table*}[b]
\tiny
\centering
\caption{UV--SED fits results for the LzLCS sample. Sources are sorted by Right-Ascension coordinate.}
\begin{tabular}{c|cccccccc}
object & LyC type & $z$ & $E_{\mathrm{B-V}}$ (mag.) & Age (Myr) & Z (Z$_{\odot}$) & $f\mathrm{_{esc}^{~abs}}$ & $\mathrm{M_{1500}^{int.}~(AB)}$ & $\mathrm{M_{1500}^{obs.}~(AB)}$ \\
\midrule
J003601+003307 & {\tiny non-leaker} & 0.3479 & 0.013$~\pm~$0.033 & 3.61$~\pm~$1.40 & 0.14$~\pm~$0.08 & $\leq~$0.029 & -18.64$~\pm~$0.37 & -18.53$~\pm~$0.10 \\
J004743+015440 & {\tiny weak} & 0.3535 & 0.134$~\pm~$0.041 & 1.17$~\pm~$1.12 & 0.24$~\pm~$0.09 & 0.013$^{+0.021}_{-0.003}$ & -21.90$~\pm~$0.45 & -20.73$~\pm~$0.10 \\
J011309+000223 & {\tiny weak} & 0.3062 & 0.136$~\pm~$0.049 & 3.97$~\pm~$1.30 & 0.50$~\pm~$0.19 & 0.022$^{+0.016}_{-0.012}$ & -21.46$~\pm~$0.54 & -20.27$~\pm~$0.12 \\
J012217+052044 & {\tiny weak} & 0.3656 & 0.075$~\pm~$0.040 & 3.79$~\pm~$1.73 & 0.05$~\pm~$0.05 & 0.038$^{+0.046}_{-0.016}$ & -20.79$~\pm~$0.44 & -20.14$~\pm~$0.10 \\
J012910+145935 & {\tiny non-leaker} & 0.2800 & 0.158$~\pm~$0.036 & 6.32$~\pm~$1.26 & 0.20$~\pm~$0.12 & $\leq~$0.007 & -21.82$~\pm~$0.40 & -20.44$~\pm~$0.09 \\
J072326+414608 & {\tiny non-leaker} & 0.2969 & 0.184$~\pm~$0.034 & 3.19$~\pm~$1.03 & 0.22$~\pm~$0.10 & $\leq~$0.004 & -21.51$~\pm~$0.37 & -19.91$~\pm~$0.08 \\
J080425+472607 & {\tiny strong} & 0.3565 & 0.062$~\pm~$0.057 & 6.85$~\pm~$1.93 & 0.10$~\pm~$0.11 & 0.584$^{+0.416}_{-0.370}$ & -19.00$~\pm~$0.64 & -18.46$~\pm~$0.15 \\
J081112+414146 & {\tiny weak} & 0.3331 & 0.182$~\pm~$0.022 & 5.27$~\pm~$0.85 & 0.08$~\pm~$0.05 & 0.020$^{+0.012}_{-0.005}$ & -21.11$~\pm~$0.23 & -19.52$~\pm~$0.05 \\
J081409+211459 & {\tiny non-leaker} & 0.2272 & 0.259$~\pm~$0.016 & 6.39$~\pm~$0.75 & 0.23$~\pm~$0.06 & $\leq~$0.007 & -23.19$~\pm~$0.17 & -20.93$~\pm~$0.04 \\
J082652+182052 & {\tiny non-leaker} & 0.2972 & 0.191$~\pm~$0.049 & 7.12$~\pm~$1.37 & 0.16$~\pm~$0.07 & $\leq~$0.009 & -20.79$~\pm~$0.54 & -19.12$~\pm~$0.12 \\
\dots & \dots & \dots & \dots & \dots & \dots & \dots & \dots & \dots \\
\bottomrule
\end{tabular}
\label{tab:fits_results1}\\
\tablefoot{
Column 1: object name. Column 2: object type (strong, weak or nonleaker). Column 3: redshift. Column 4: UV dust-attenuation parameter (\ebv, in mag.). Column 5: light-weighted stellar age (in Myr). Column 6: light-weighted stellar metallicity (in solar units, Z$_{\odot}$). Column 6: LyC absolute photon escape fraction (\fescabs). Column 7: intrinsic (dust corrected, \mfuv). Column 8: observed UV absolute magnitudes (\mobs). 
}
\end{table*}

\begin{table*}[b]
\tiny
\centering
\caption{Absorption lines results for the LzLCS sample. Sources are sorted by Right-Ascension coordinate.}
\begin{tabular}{c|cccccccc}
object & LyC type & $z$ & $W{\mathrm{_{HI}~(\AA)}}$ & $W{\mathrm{_{LIS}~(\AA)}}$ & $\mathrm{R_{HI}}$ & $\mathrm{R_{LIS}}$ & $f\mathrm{_{esc}^{~abs,~HI}}$ & $f\mathrm{_{esc}^{~abs,~LIS}}$ \\
\midrule
J003601+003307 & {\tiny non-leaker} & 0.3479 & 2.51$~\pm~$0.23 & 0.97$~\pm~$0.88 & 0.17$~\pm~$0.07 & 0.28$~\pm~$0.29 & 0.153$~\pm~$0.099 & 0.044$~\pm~$0.151 \\
J004743+015440 & {\tiny weak} & 0.3535 & 3.33$~\pm~$0.17 & 2.16$~\pm~$0.80 & 0.01$~\pm~$0.03 & 0.26$~\pm~$0.26 & 0.001$~\pm~$0.008 & 0.008$~\pm~$0.032 \\
J011309+000223 & {\tiny weak} & 0.3062 & 2.42$~\pm~$0.34 & 0.58$~\pm~$0.70 & 0.19$~\pm~$0.11 & 0.66$~\pm~$0.24 & 0.055$~\pm~$0.050 & 0.048$~\pm~$0.055 \\
J012217+052044 & {\tiny weak} & 0.3656 & 2.97$~\pm~$0.16 & -0.10$~\pm~$0.76 & 0.23$~\pm~$0.05 & 0.54$~\pm~$0.24 & 0.093$~\pm~$0.059 & 0.078$~\pm~$0.098 \\
J012910+145935 & {\tiny non-leaker} & 0.2800 & 2.40$~\pm~$0.30 & 1.87$~\pm~$0.49 & 0.14$~\pm~$0.09 & 0.36$~\pm~$0.15 & 0.015$~\pm~$0.016 & 0.001$~\pm~$0.014 \\
J072326+414608 & {\tiny non-leaker} & 0.2969 & 1.83$~\pm~$0.27 & 0.82$~\pm~$0.48 & 0.13$~\pm~$0.06 & 0.56$~\pm~$0.16 & 0.014$~\pm~$0.010 & 0.021$~\pm~$0.017 \\
J080425+472607 & {\tiny strong} & 0.3565 & 0.87$~\pm~$0.28 & 0.38$~\pm~$1.46 & 0.30$~\pm~$0.10 & 0.72$~\pm~$0.51 & 0.135$~\pm~$0.186 & 0.136$~\pm~$0.298 \\
J081112+414146 & {\tiny weak} & 0.3331 & 1.77$~\pm~$0.16 & 0.65$~\pm~$0.38 & 0.27$~\pm~$0.05 & 0.77$~\pm~$0.15 & 0.033$~\pm~$0.011 & 0.039$~\pm~$0.017 \\
J081409+211459 & {\tiny non-leaker} & 0.2272 & 4.20$~\pm~$0.11 & 1.28$~\pm~$0.24 & 0.21$~\pm~$0.04 & 0.76$~\pm~$0.10 & 0.010$~\pm~$0.003 & 0.015$~\pm~$0.004 \\
J082652+182052 & {\tiny non-leaker} & 0.2972 & 3.09$~\pm~$0.41 & 2.71$~\pm~$0.61 & 0.00$~\pm~$0.07 & 0.27$~\pm~$0.16 & 0.003$~\pm~$0.010 & -- \\
\dots & \dots & \dots & \dots & \dots & \dots & \dots & \dots & \dots \\
\bottomrule
\end{tabular}
\label{tab:lines_results1}\\
\tablefoot{
Column 1: object name. Column 2: object type (strong, weak or non-leaker). Column 3: redshift. Columns 4 and 5: weighted-average \hi\ and LIS equivalent widths (\ewhi, \ewlis). Columns 6 and 7: weighted-average \hi\ and LIS residual fluxes (\cfhi, \cflis). Columns 8 and 9: Lyman series and LIS derived absolute photon escape fraction (\feschi, \fesclis). 
}
\end{table*}

\begin{table*}[b]
\centering
\tiny
\caption{UV--SED fits results for the literature sample$^1$. Columns are the same as in Table \ref{tab:fits_results1}. Sources are sorted by year of publication and Right-Ascension coordinate.}
\begin{tabular}{c|cccccccc}
object & LyC type & $z$ & $E_{\mathrm{B-V}}$ (mag.) & Age (Myr) & Z (Z$_{\odot}$) & $f\mathrm{_{esc}^{~abs}}$ & $\mathrm{M_{1500}^{int.}~(AB)}$ & $\mathrm{M_{1500}^{obs.}~(AB)}$ \\
\midrule
J0925+1403 & {\tiny strong} & 0.3013 & 0.106$~\pm~$0.024 & 4.32$~\pm~$0.94 & 0.43$~\pm~$0.09 & 0.092$^{+0.019}_{-0.034}$ & -20.74$~\pm~$0.26 & -19.81$~\pm~$0.05 \\
J1152+3400 & {\tiny strong} & 0.3419 & 0.095$~\pm~$0.029 & 6.85$~\pm~$1.28 & 0.11$~\pm~$0.08 & 0.177$^{+0.195}_{-0.063}$ & -21.51$~\pm~$0.32 & -20.68$~\pm~$0.07 \\
J1333+6246 & {\tiny strong} & 0.3181 & 0.097$~\pm~$0.044 & 7.00$~\pm~$1.17 & 0.10$~\pm~$0.04 & 0.128$^{+0.126}_{-0.062}$ & -20.60$~\pm~$0.49 & -19.75$~\pm~$0.11 \\
J1442-0209 & {\tiny strong} & 0.2937 & 0.091$~\pm~$0.028 & 5.79$~\pm~$1.08 & 0.16$~\pm~$0.04 & 0.120$^{+0.064}_{-0.050}$ & -20.94$~\pm~$0.30 & -20.14$~\pm~$0.06 \\
J1503+3644 & {\tiny weak} & 0.3557 & 0.165$~\pm~$0.026 & 1.69$~\pm~$0.53 & 0.49$~\pm~$0.18 & 0.033$^{+0.019}_{-0.006}$ & -21.52$~\pm~$0.28 & -20.08$~\pm~$0.06 \\
\dots & \dots & \dots & \dots & \dots & \dots & \dots & \dots & \dots \\
\bottomrule
\end{tabular}
\label{tab:fits_results2}\\
\tablefoot{
$^1$ References: \citet{Izotov16a, Izotov16b, Izotov18a, Izotov18b, Izotov2021, Wang2019}.
}
\end{table*}

\begin{table*}[b]
\tiny
\centering
\caption{Absorption lines results for the literature sample$^1$. Columns are the same as in Table \ref{tab:lines_results1}. Sources are sorted by year of publication and Right-Ascension coordinate.}
\begin{tabular}{c|cccccccc}
object & LyC type & $z$ & $W{\mathrm{_{HI}~(\AA)}}$ & $W{\mathrm{_{LIS}~(\AA)}}$ & $\mathrm{R_{HI}}$ & $\mathrm{R_{LIS}}$ & $f\mathrm{_{esc}^{~abs,~HI}}$ & $f\mathrm{_{esc}^{~abs,~LIS}}$ \\
\midrule
J0925+1403 & {\tiny strong} & 0.3013 & 0.41$~\pm~$0.48 & 0.82$~\pm~$0.35 & 0.33$~\pm~$0.19 & 0.70$~\pm~$0.11 & 0.092$~\pm~$0.062 & 0.083$~\pm~$0.032 \\
J1152+3400 & {\tiny strong} & 0.3419 & 2.21$~\pm~$0.30 & 0.58$~\pm~$0.51 & 0.17$~\pm~$0.10 & 0.66$~\pm~$0.17 & 0.052$~\pm~$0.043 & 0.090$~\pm~$0.054 \\
J1333+6246 & {\tiny strong} & 0.3181 & 2.47$~\pm~$0.26 & 0.92$~\pm~$0.66 & 0.09$~\pm~$0.06 & 0.54$~\pm~$0.21 & 0.044$~\pm~$0.039 & 0.075$~\pm~$0.073 \\
J1442-0209 & {\tiny strong} & 0.2937 & 1.61$~\pm~$0.29 & 0.10$~\pm~$0.44 & 0.26$~\pm~$0.10 & 0.70$~\pm~$0.16 & 0.121$~\pm~$0.053 & 0.105$~\pm~$0.052 \\
J1503+3644 & {\tiny weak} & 0.3557 & 1.53$~\pm~$0.21 & 0.73$~\pm~$0.50 & 0.30$~\pm~$0.06 & 0.52$~\pm~$0.14 & 0.043$~\pm~$0.017 & 0.021$~\pm~$0.014 \\
\dots & \dots & \dots & \dots & \dots & \dots & \dots & \dots & \dots \\
\bottomrule
\end{tabular}
\label{tab:lines_results2}\\
\tablefoot{
$^1$ References: \citet{Izotov16a, Izotov16b, Izotov18a, Izotov18b, Izotov2021, Wang2019}.
}\\
--\\

{Tables \ref{tab:fits_results1} to \ref{tab:lines_results2} are part of the LzLCS science products and will be publicly available in a dedicated website.}

\end{table*}

\section{Absorption lines -- saturation}\label{appB}
The ratio of two absorption lines of the same element and ionization state can indicate whether the species in question is saturated or not. Let $f_i,\lambda_i$ be the oscillator strength and central rest-frame wavelength of the line in question, where the  2(1) subscript represents the stronger (weaker) transition. If both transitions are saturated, the equivalent width ratio $W_2/W_1$ can be approximated by \citep{Draine}
\begin{equation}
    \dfrac{W_2}{W_1} \approx \left[1+\dfrac{\ln (f_2 \lambda_2 / f_1 \lambda_1)}{\ln (\tau_1 / \ln2)}\right]^{(1/2)}
,\end{equation}

\noindent where $\tau_1$ is the optical depth of the stronger transition, being $\tau_1 \gg 1$ the optically thick limit. If, on the opposite, both lines behave under the optically thin regime ($\tau_1 \ll 1$), the equivalent width doublet-ratio is simply:
\begin{equation}
    \dfrac{W_2}{W_1} = \dfrac{f_2 \lambda_2}{f_1 \lambda_1}
.\end{equation}

In practice, we assume $\tau_1 > 1$ for saturation, and a smooth transition between the two regimes. To test the saturation of \hi\ in our galaxy spectra, we compare the equivalent width of \lyb\ to \lyd. The use of lines pairs which are close in wavelength ensures that both lines are in the same optical regime simultaneously. 

\begin{figure}
    \centering
    \includegraphics[width=0.90\columnwidth, page=1]{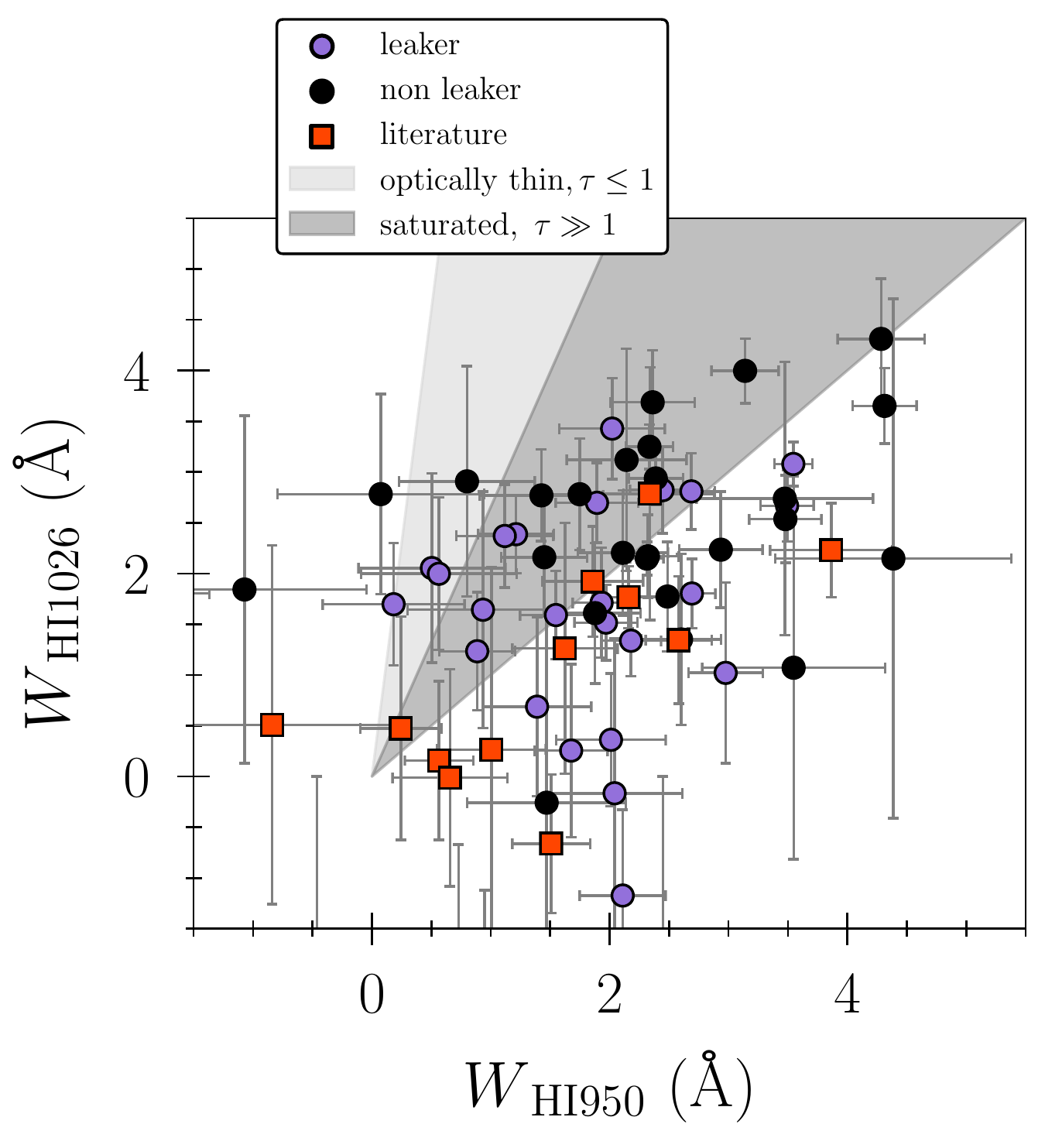}
\caption{Absorption line \hi\ equivalent width comparisons for studying the condition of saturation. The optically thin (light-gray) and optically thick (dark-gray) regions are delimited using the approximations given in \citet{Draine}. As in the entire document, purple symbols represent the leakers while the black symbols show the nonleakers. Red squares correspond to the \citet{Izotov16a, Izotov16b, Izotov18a, Izotov18b, Izotov2021} and \citet{Wang2019} galaxies.}
\label{fig:saturation_plot}
\end{figure}

The observations are shown in \autoref{fig:saturation_plot}, together with the regions delimiting the optically thick (saturation) and optically thin (linear) regimes, respectively, following the equations above. As stressed in the text, most of the sources are compatible with saturation for \hi\ within the 1$\sigma$ errors, and only a low number of them fall in the optically thin limit. Several sources are found to show a higher equivalent width in \lyd\ than \lyb, which is not expected from simple curve-of-growth analysis. The exact reasons for this behavior are not known but could be due to multiple effects, such as systematic errors in the continuum placement in the blue part of the spectrum, blends with other lines such as \ion{O}{i}, and other causes. In any case, in most of the cases, they are compatible with the saturated region at 1$\sigma$. Spectra with higher resolution and S/N, such as the subset of galaxies analyzed by \citet{Gazagnes2018}, would be required to better examine the question of saturation.

\section{Absorption lines -- systematic errors}\label{appC}
\begin{figure}
    \centering
    \includegraphics[width=0.90\columnwidth]{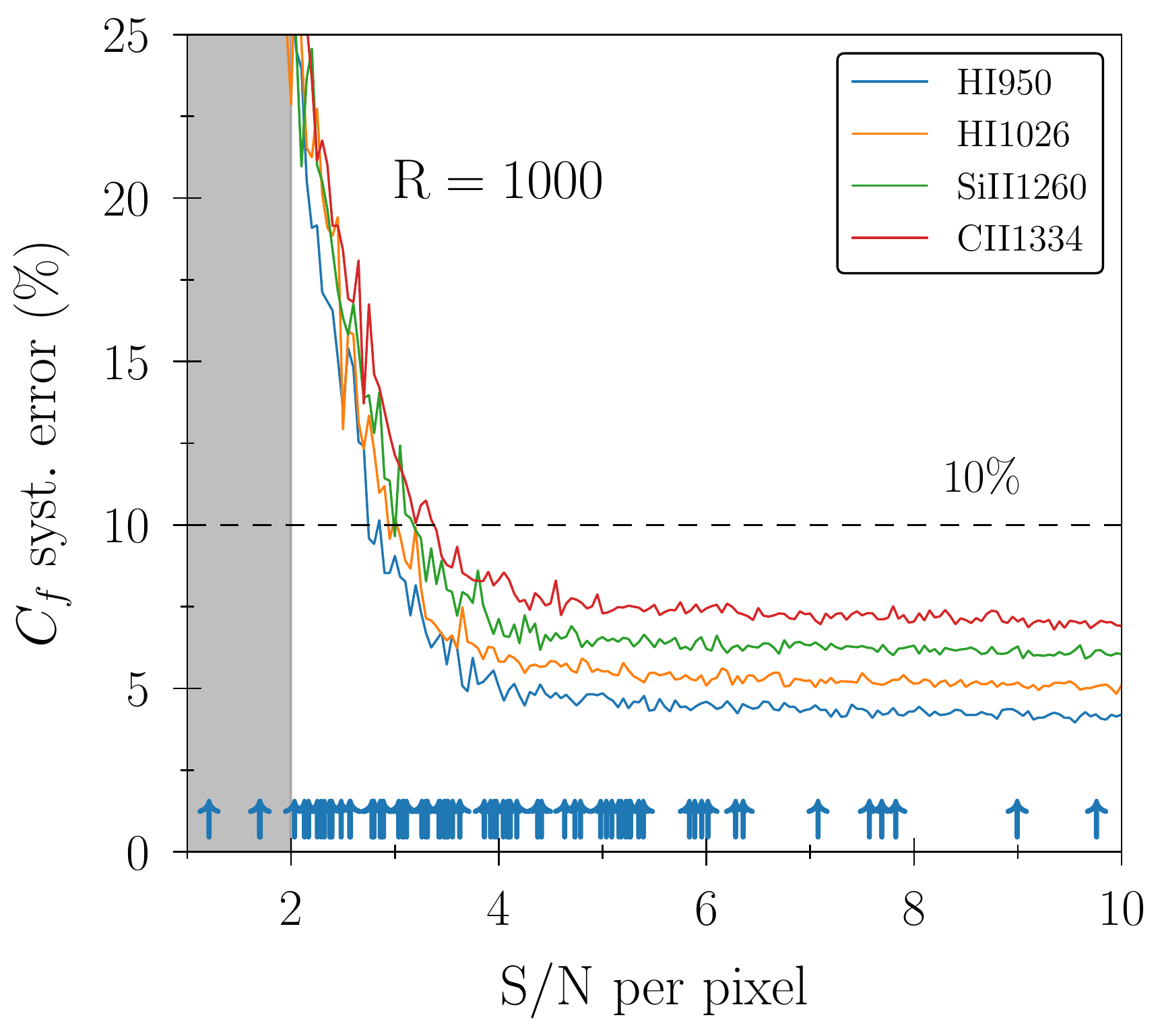}
\caption{Estimation of the systematic error on the observed covering fraction as a function of the spectral S/N per pixel, for different simulated lines (see legend). A resolution of $R=1000$ is assumed. The upward arrows at the bottom correspond to the measured S/N for the LzLCS sample around the \ion{H}{I}$\lambda$950 line. Horizontal dashed line limits the $10\%$ representative value.}
\label{fig:syst_err}
\end{figure}

Here, we carefully account for the impact of the resolution and S/N in the measured covering fraction. The spectrograph resolution tends to make the absorption lines wider but less deep. Together with the noise and sensitivity --and other possible instrumental effects-- this can lead to a systematically overestimated measurement of the residual flux. 

We use absorption line simulations to study how the resolution ($R$) and S/N both impact our measurements of the residual flux. We first simulate Ly$\delta$, Ly$\beta$, \siib,\ and \cii\ line intensities ($I_{\lambda}$) in a foreground dust-screen picket-fence model:
\begin{equation}
    I_{\lambda} = C_f(\lambda) \times \exp{(-\sigma_{\lambda}N_{\lambda})} + 1-C_f(\lambda)
,\end{equation}

\noindent where $C_f(\lambda)$ represents the covering fraction of the line in question, i.e., the fraction of sight lines for which the transition is optically thick around the host galaxy. $C_f(\lambda)$ can be inferred from the line depth as $C_f(\lambda) = 1-R(\lambda)$ for the dust-screen picket-fence model. The $\tau_{\lambda}=\sigma_{\lambda} N_{\lambda}$ product is usually known as the optical depth of the line, and the $\lambda$ subscripts belong to the transition in question. The line cross-section, $\sigma_{\lambda}$, which shapes the absorption profile, is given by a usual Voigt function so that:
\begin{equation}
    \sigma_{\lambda} = \dfrac{\sqrt{\pi} e^2 f_{\lambda}}{m_e c} \dfrac{\lambda}{b}~{\rm Voigt(\lambda, A_{\lambda}, b)}
.\end{equation}

For the Voigt($\lambda$,A$_{\lambda}$,b) function, we use the numerical approximation described in \citet{Smith2015}. The Doppler broadening parameter is input through $b$ (km~s$^{-1}$). Finally, oscillator strengths $f_{\lambda}$ and Einstein coefficients A$_{\lambda}$ are taken from the NIST database \footnote{NIST stands for National Institute of Standards and Technology, whose website is: \url{https://physics.nist.gov/PhysRefData/ASD/lines_form.html}.}. We do not consider any velocity gradient effect (outflows) in the mock realizations. 

For our purposes, we assume a Doppler parameter of $b=125$~km/s, and typical column densities in SF galaxies of $N_{\lambda}=10^{19}, 10^{19}, 10^{17}, 10^{17}$ cm$^{-2}$ for Ly$\delta$, Ly$\beta$, \siib,\ and \cii\ lines, respectively. We also explore a wide range of $(N,b)$ parameters and conclude that results do not change drastically as soon as the lines are in the saturation region of the curve of growth. The covering fraction is fixed to $C_f(\lambda)=0.85$ throughout the processes. 

The mock lines are convolved with a Gaussian kernel to match the COS/G140L nominal resolution ($R=1000$) and interpolated into the rest-frame COS/G140L wavelength spacing (0.5621/(1+z)~\AA. where $z=0.3$). A set of Gaussian S/N are then randomly added to every pixel. Finally, the residual flux is measured in the same way as in the original real spectra, and related to the covering fraction by Eq. \ref{eq:cfA} (uniform dust-screen). This process is repeated 500 times for every S/N value, and the median covering fraction is compared to the input value. 

This method is summarized in \autoref{fig:syst_err}, which shows the covering fraction relative difference (in percentage) as a function of the input S/N per pixel in the simulations. The four colored lines corresponds to the different lines considered. The upward arrows at the bottom correspond to the measured S/N for the LzLCS sample around the \lyd\ line, for comparison (the S/N has been measured at the continuum level close to the line in question). In general, the observed continuum S/N increases at shorter wavelengths, i.e., it is larger for Ly$\delta$ and Ly$\beta$, but systematically lower for \siib\ and \cii. 

In all cases, and as indicated in the text, the systematic error on the covering fraction spans between 5\%\ and 20$\%$ for these four representative lines (with a typical value of 10$\%$). These values are within the original error bars we are reporting for single covering fraction measurements. We therefore leave our initial lines measurements and errors  unaltered, and we prefer not to correct the $C_f(\lambda)$ values by this systematic offset.

\section{The \textsc{linmix} package}\label{appD}
The \textsc{linmix} estimator \citep{Kelly2007} is a Python-based Bayesian fitting code that allows the user to model a two-dimensional data set $(x,y)$ with a linear regression, accounting for errors on both variables and intrinsic random scatter, and with the capability of including censored (upper or lower limits) data. 

\textsc{linmix} assumes that measured data points $(x,y)$ can be sampled from a two-dimensional Gaussian distribution $P_1$ with mean $(\xi, \eta)$, and covariance matrix determined from measurement uncertainties. It also assumes that the measurement errors in the $x$- and $y$-directions are normally distributed and uncorrelated. 

The ``true'' value of the $\eta$ variable is drawn from another Gaussian distribution $P_2$, with mean $\alpha + \beta \eta$, and variance $\sigma_y^2$, where $\beta$ describes the slope, $\alpha$ the intercept of the straight line, and $\sigma_y^2$ is the intrinsic $y-$dispersion (also normally distributed). Finally, the ``true'' value of the independent variable $\xi$ is sampled from a weighted sum of $K$ Gaussian distributions $P_3$ (we choose $K=2$). The distributions $P_1, P_2, P_3$ are convolved hierarchically to compute the full likelihood of obtaining the observed data $(x,y)$ given the parameters of the fit: $\alpha, \beta, \sigma^2_y$. 

Assuming uniform prior distributions for the three parameters: $\alpha, \beta \in (-\infty, \infty), ~\sigma_y^2 \in [0,\infty)$; a Markov Chain Monte-Carlo (MCMC) algorithm is used to sample from the posterior distributions until the convergence of the chains ($\times$4 chains, 5~000 to 10~000 iterations each). For a practical example, see e.g., \citet{Kennicutt2021}. A \textsc{python} version of the \textsc{linmix} code can be found in \small{\url{https://linmix.readthedocs.io/en/latest/index.html}}.

\section{Main results at a glance}\label{appE}
This Appendix is dedicated to illustrating the global behavior of the absolute LyC photon escape fraction (\fescabs) with the main UV spectral properties measured in this work, namely: \rhi, \rlis, \ewhi, \ewlis\ and \ebv. These results are presented in \autoref{fig:fesc_cfHILIS}, \ref{fig:fesc_ewHILIS}, and \ref{fig:fesc_EBV}, but here we focus on the median values of the working sample as a take- home message. To do this, we compute the median value of \rhi, \rlis, \ewhi, \ewlis\ and \ebv\ in three regimes of \fescabs, specifically at $\fescabs \in [0\%-5\%), [5\%-10\%), [10\%-100\%)$. We additionally apply a bootstrap + Monte Carlo algorithm to the number of sources at each \fescabs\ interval allowing us to compute the 0.16 and 0.84 quantiles at each bin. The results are summarized in \autoref{tab:summary_table}. 

\begin{figure*}
    \centering
    \includegraphics[width=0.8\textwidth]{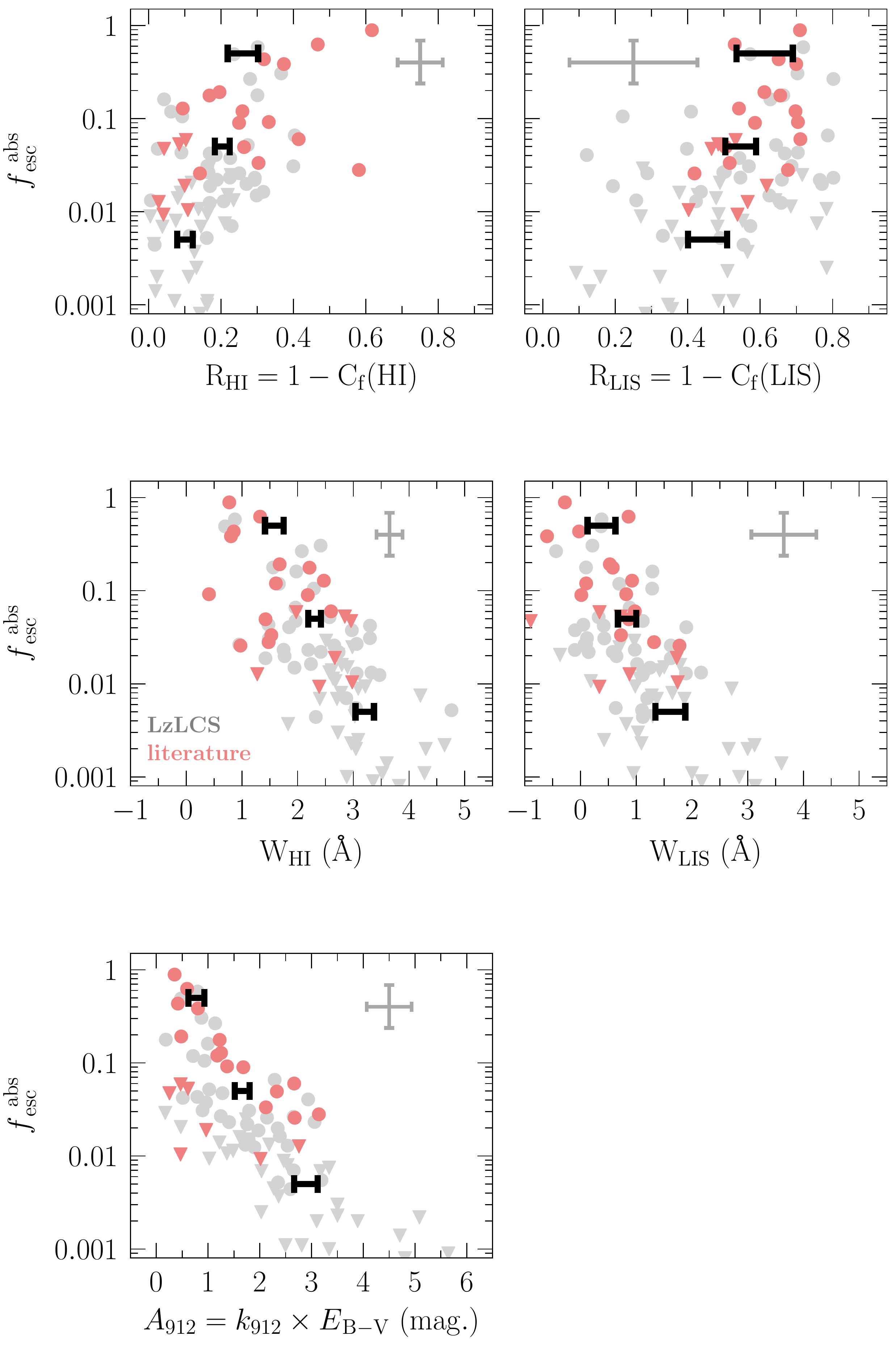}
\caption{Summary plot with the correlations between the LyC absolute photon escape fraction (\fescabs) and the main UV spectral properties derived in this work: \rhi, \rlis, \ewhi, \ewlis\ and \ebv\ ({\em from top to bottom and left to right}, respectively). Black thick error bars comprise the running median values and 0.16 and 0.84 quantiles of the parameters as a function of \fescabs. The individual measurements are shown in the background through gray (LzLCS) and red (literature) symbols. Error bars at the top of the panels represent the median 1$\sigma$ uncertainty of each parameter, based on individual errors.}
\label{fig:summary_plots}
\end{figure*}

In \autoref{fig:summary_plots}, the running medians are plotted as a function of \fescabs\ (in black), together with the individual LzLCS+literature measurements (gray and red dots in the background). The reader may find a detailed discussion about these results in Sect.\ \ref{sec:results_lya_cf_ebv_fesc}. A proper analysis of the median results of the sample will be investigated in future LzLCS publications, by the use of stacked spectra.

\begin{table}[H]
\centering
\tiny
\caption{Running median results for \rhi, \rlis, \ewhi, \ewlis\ (in \AA) and \ebv\ (in mag.) in three \fescabs\ intervals (including 0.16 and 0.84 quantiles).}
\begin{tabular}{cccc}
\toprule
\multicolumn{1}{c}{} & \multicolumn{3}{c}{LyC absolute photon escape fraction (\fescabs)} \\
\midrule
 & $0-5~(\%)$ & $5-10~(\%)$ & $10-100~(\%)$ \\ 
\cmidrule{2-4} 
\rhi\ & 0.10 (0.08,~0.12) & 0.20 (0.18,~0.22) & 0.26 (0.21,~0.30) \\
\rlis\ & 0.46 (0.40,~0.51) & 0.54 (0.50,~0.58) & 0.61 (0.54,~0.69) \\
\ewhi\ & 3.23 (3.06,~3.39) & 2.30 (2.19,~2.41) & 1.58 (1.41,1.75) \\
\ewlis\ & 1.61 (1.35,~1.87) & 0.85 (0.69,~1.01) & 0.37 (0.11,~0.62) \\
\auv\ & 2.88 (2.66,~3.11) & 1.67 (1.53,~1.81) & 0.77 (0.63,~0.93) \\
\bottomrule
\end{tabular}
\label{tab:summary_table}
\end{table}

\end{document}